\pdfoutput=1
\PassOptionsToPackage{dvipsnames}{xcolor}
\documentclass[letterpaper,DIV=12]{scrartcl}

\usepackage{import}
\usepackage[utf8]{inputenc}

\PassOptionsToPackage{hyphens}{url}%
\usepackage{hyperref}
\usepackage{graphicx}

\usepackage{amsmath}
\usepackage{amsthm}
\usepackage{amssymb}
\usepackage{amsfonts}
\usepackage{mathtools}

\usepackage{shortsym}

\usepackage{microtype}

\usepackage{IEEEtrantools}
\allowdisplaybreaks

\usepackage[shortlabels,inline]{enumitem}
\setlist[itemize]{leftmargin=1.25em}
\setlist[enumerate]{leftmargin=1.75em}
\usepackage[normalem]{ulem}

\usepackage{import}
\theoremstyle{plain}
\newtheorem{lemma}{Lemma}
\newtheorem*{theorem*}{Theorem}
\newtheorem*{conjecture*}{Conjecture}

\theoremstyle{definition}
\makeatletter
\renewcommand*\env@matrix[1][*\c@MaxMatrixCols c]{%
    \hskip -\arraycolsep
    \let\@ifnextchar\new@ifnextchar
    \array{#1}}
\makeatother

\DeclarePairedDelimiter{\abs}{\lvert}{\rvert}
\DeclarePairedDelimiter{\len}{\lvert}{\rvert}
\DeclarePairedDelimiter{\norm}{\lVert}{\rVert}
\DeclarePairedDelimiter{\floor}{\lfloor}{\rfloor}
\DeclarePairedDelimiter{\ceil}{\lceil}{\rceil}

\makeatletter
\let\oldabs\abs
\def\abs{\@ifstar{\oldabs}{\oldabs*}}
\let\oldlen\len
\def\len{\@ifstar{\oldlen}{\oldlen*}}
\let\oldnorm\norm
\def\norm{\@ifstar{\oldnorm}{\oldnorm*}}
\let\oldfloor\floor
\def\floor{\@ifstar{\oldfloor}{\oldfloor*}}
\let\oldceil\ceil
\def\ceil{\@ifstar{\oldceil}{\oldceil*}}
\makeatother

\usepackage{pifont}

\usepackage{dsfont}
\newcommand{\Ione}{{\mathds{1}}}

\newcommand{\getsRandom}{\ensuremath{\overset{\text{\$}}{\gets}}}

\usepackage{xspace}

\newcommand{\cf}[0]{cf.\xspace}

\usepackage[dvipsnames]{xcolor}

\definecolor{myA16zGrayLight}{RGB}{235,235,235}     %
\definecolor{myA16zGrayMedium}{RGB}{196,196,196}    %
\definecolor{myA16zGrayDark}{RGB}{44,34,34}         %
\definecolor{myA16zLavender}{RGB}{208,161,255}      %
\definecolor{myA16zMagenta}{RGB}{195,70,206}        %
\definecolor{myA16zMulberry}{RGB}{113,24,88}        %
\definecolor{myA16zLemonChiffon}{RGB}{250,234,157}  %
\definecolor{myA16zAmber}{RGB}{230,154,48}          %
\definecolor{myA16zRust}{RGB}{174,59,10}            %
\definecolor{myA16zLime}{RGB}{197,222,107}          %
\definecolor{myA16zAquamarine}{RGB}{82,216,145}     %
\definecolor{myA16zPine}{RGB}{60,87,44}             %
\definecolor{myA16zPacific}{RGB}{145,224,235}       %
\definecolor{myA16zTeal}{RGB}{36,197,201}           %
\definecolor{myA16zAzure}{RGB}{18,51,90}            %

\definecolor{myTechnionDeepBlue}{HTML}{002147}           %
\definecolor{myTechnionGoldenOchre}{HTML}{D59F0F}        %
\definecolor{myTechnionBlack}{HTML}{000000}              %
\definecolor{myTechnionWhite}{HTML}{FFFFFF}              %

\definecolor{myTechnionRed}{HTML}{E31D1A}             %
\definecolor{myTechnionPink}{HTML}{EA094B}           %
\definecolor{myTechnionPurple1}{HTML}{AE3B72}         %
\definecolor{myTechnionPurple2}{HTML}{4D4084}         %
\definecolor{myTechnionBlue1}{HTML}{216093}            %
\definecolor{myTechnionBlue2}{HTML}{5686DA}           %
\definecolor{myTechnionTeal}{HTML}{32B1CA}            %
\definecolor{myTechnionGreen1}{HTML}{EA094B}           %
\definecolor{myTechnionGreen2}{HTML}{A3D65C}           %
\definecolor{myTechnionGreen3}{HTML}{94D60A}           %
\definecolor{myTechnionYellow}{HTML}{FDD700}          %
\definecolor{myTechnionOrange}{HTML}{FF6B00}         %
\definecolor{myTechnionBrown}{HTML}{97775C}          %
\definecolor{myTechnionBeige}{HTML}{D9D1C3}          %
\definecolor{myTechnionGray1}{HTML}{A2A9AE}            %
\definecolor{myTechnionGray2}{HTML}{5A6771}            %

\definecolor{mySuCardinalRed}{HTML}{8c1515}
\definecolor{mySuCardinalRedLight}{HTML}{B83A4B}
\definecolor{mySuCardinalRedDark}{HTML}{820000}
\definecolor{mySuWhite}{HTML}{ffffff}
\definecolor{mySuCoolGrey}{HTML}{53565A}
\definecolor{mySuBlack}{HTML}{2e2d29}
\definecolor{mySuBlack100}{HTML}{2e2d29}
\definecolor{mySuBlack90}{HTML}{43423E}
\definecolor{mySuBlack80}{HTML}{585754}
\definecolor{mySuBlack70}{HTML}{6D6C69}
\definecolor{mySuBlack60}{HTML}{767674}
\definecolor{mySuBlack50}{HTML}{979694}
\definecolor{mySuBlack40}{HTML}{ABABA9}
\definecolor{mySuBlack30}{HTML}{C0C0BF}
\definecolor{mySuBlack20}{HTML}{D5D5D4}
\definecolor{mySuBlack10}{HTML}{EAEAEA}

\definecolor{mySuPaloAlto}{HTML}{175E54}
\definecolor{mySuPaloAltoLight}{HTML}{2D716F}
\definecolor{mySuPaloAltoDark}{HTML}{014240}
\definecolor{mySuPaloVerde}{HTML}{279989}
\definecolor{mySuPaloVerdeLight}{HTML}{59B3A9}
\definecolor{mySuPaloVerdeDark}{HTML}{017E7C}
\definecolor{mySuOlive}{HTML}{8F993E}
\definecolor{mySuOliveLight}{HTML}{A6B168}
\definecolor{mySuOliveDark}{HTML}{7A863B}
\definecolor{mySuBay}{HTML}{6FA287}
\definecolor{mySuBayLight}{HTML}{8AB8A7}
\definecolor{mySuBayDark}{HTML}{417865}
\definecolor{mySuSky}{HTML}{4298B5}
\definecolor{mySuSkyLight}{HTML}{67AFD2}
\definecolor{mySuSkyDark}{HTML}{016895}
\definecolor{mySuLagunita}{HTML}{007C92}
\definecolor{mySuLagunitaLight}{HTML}{009AB4}
\definecolor{mySuLagunitaDark}{HTML}{006B81}
\definecolor{mySuPoppy}{HTML}{E98300}
\definecolor{mySuPoppyLight}{HTML}{F9A44A}
\definecolor{mySuPoppyDark}{HTML}{D1660F}
\definecolor{mySuSpirited}{HTML}{E04F39}
\definecolor{mySuSpiritedLight}{HTML}{F4795B}
\definecolor{mySuSpiritedDark}{HTML}{C74632}
\definecolor{mySuIlluminating}{HTML}{FEDD5C}
\definecolor{mySuIlluminatingLight}{HTML}{FFE781}
\definecolor{mySuIlluminatingDark}{HTML}{FEC51D}
\definecolor{mySuPlum}{HTML}{620059}
\definecolor{mySuPlumLight}{HTML}{734675}
\definecolor{mySuPlumDark}{HTML}{350D36}
\definecolor{mySuBrick}{HTML}{651C32}
\definecolor{mySuBrickLight}{HTML}{7F2D48}
\definecolor{mySuBrickDark}{HTML}{42081B}
\definecolor{mySuArchway}{HTML}{5D4B3C}
\definecolor{mySuArchwayLight}{HTML}{766253}
\definecolor{mySuArchwayDark}{HTML}{2F2424}
\definecolor{mySuStone}{HTML}{7F7776}
\definecolor{mySuStoneLight}{HTML}{D4D1D1}
\definecolor{mySuStoneDark}{HTML}{544948}
\definecolor{mySuFog}{HTML}{DAD7CB}
\definecolor{mySuFogLight}{HTML}{F4F4F4}
\definecolor{mySuFogDark}{HTML}{B6B1A9}

\definecolor{mySuDigitalRed}{HTML}{B1040E}
\definecolor{mySuDigitalRedLight}{HTML}{E50808}
\definecolor{mySuDigitalRedDark}{HTML}{820000}
\definecolor{mySuDigitalBlue}{HTML}{006CB8}
\definecolor{mySuDigitalBlueLight}{HTML}{6FC3FF}
\definecolor{mySuDigitalBlueDark}{HTML}{00548f}
\definecolor{mySuDigitalGreen}{HTML}{008566}
\definecolor{mySuDigitalGreenLight}{HTML}{1AECBA}
\definecolor{mySuDigitalGreenDark}{HTML}{006F54}

\definecolor{myParula1Blue}{RGB}{0,114,189}
\definecolor{myParula2Orange}{RGB}{217,83,25}
\definecolor{myParula3Yellow}{RGB}{237,177,32}
\definecolor{myParula4Purple}{RGB}{126,47,142}
\definecolor{myParula5Green}{RGB}{119,172,48}
\definecolor{myParula6LightBlue}{RGB}{77,190,238}
\definecolor{myParula7Red}{RGB}{162,20,47}

\usepackage{tikz}
\usetikzlibrary{calc}
\usetikzlibrary{fpu}
\usetikzlibrary{arrows.meta}
\usetikzlibrary{patterns}
\usetikzlibrary{positioning}
\usetikzlibrary{decorations.pathreplacing}
\usetikzlibrary{decorations.pathmorphing}
\usetikzlibrary{calligraphy}
\usetikzlibrary{shapes.misc}
\usetikzlibrary{shapes.symbols}
\usetikzlibrary{shapes.arrows}
\usetikzlibrary{spy}
\usetikzlibrary{shapes.geometric}
\usetikzlibrary{intersections}
\usetikzlibrary{fit}

\usepackage{pgfplots}
\pgfplotsset{compat=1.17}
\usepgfplotslibrary{fillbetween}

\pgfplotsset{
    discard if not/.style 2 args={
            x filter/.code={
                    \edef\tempa{\thisrow{#1}}
                    \edef\tempb{#2}
                    \ifx\tempa\tempb
                    \else
                        
                    \fi
                }
        },
}

\pgfplotsset{
    mysimpleplot/.style = {
            every axis plot/.prefix style={thick},
            width=1.05\linewidth,
            height=0.75\linewidth,
            title style={font=\scriptsize,align=center},
            legend cell align=left,
            legend style={font=\scriptsize},
            legend columns=3,
            legend style={
                    at={(0.5,1)},
                    yshift=0.3em,
                    anchor=south,
                    draw=none,
                    /tikz/every even column/.append style={
                            column sep=0.3em
                        },
                    cells={
                            align=left
                        }
                },
            grid=both,
            minor tick num=9,
            major grid style={solid,very thin,draw=gray!50},
            minor grid style={solid,ultra thin,draw=gray!20},
            label style={font=\scriptsize,align=center},
            tick label style={font=\scriptsize},
        },
}

\pgfplotsset{
    mysimplefig1plot/.style = {
        mysimpleplot,
        xlabel={$\netX$},
        ylabel={$\tALident/n$},
        xmin=0.0, xmax=0.5,
        ymin=0.0, ymax=0.35,
        height=0.75\linewidth,
        width=\linewidth,
        yticklabel style={
                /pgf/number format/fixed,
                /pgf/number format/precision=2
            },
        scaled y ticks=false,
        xtick={0,0.1,0.2,0.25,0.33333,0.5},
        xticklabels={0,1/10,1/5,1/4,1/3,1/2},
        yticklabels={0,1/10,1/6,1/5,1/4,1/3},
        ytick={0,0.1,0.16666,0.2,0.25,0.33333},
    }
}

\tikzset{myparula11/.style={color=myParula1Blue,solid,mark=+,mark options={solid}}}
\tikzset{myparula12/.style={color=myParula1Blue,densely dashed,mark=x,mark options={solid}}}
\tikzset{myparula13/.style={color=myParula1Blue,densely dotted,mark=o,mark options={solid}}}
\tikzset{myparula14/.style={color=myParula1Blue,dashdotted,mark=triangle,mark options={solid}}}
\tikzset{myparula15/.style={color=myParula1Blue,dashdotdotted,mark=square,mark options={solid}}}

\tikzset{myparula21/.style={color=myParula2Orange,solid,mark=+,mark options={solid}}}
\tikzset{myparula22/.style={color=myParula2Orange,densely dashed,mark=x,mark options={solid}}}
\tikzset{myparula23/.style={color=myParula2Orange,densely dotted,mark=o,mark options={solid}}}
\tikzset{myparula24/.style={color=myParula2Orange,dashdotted,mark=triangle,mark options={solid}}}
\tikzset{myparula25/.style={color=myParula2Orange,dashdotdotted,mark=square,mark options={solid}}}

\tikzset{myparula31/.style={color=myParula3Yellow,solid,mark=+,mark options={solid}}}
\tikzset{myparula32/.style={color=myParula3Yellow,densely dashed,mark=x,mark options={solid}}}
\tikzset{myparula33/.style={color=myParula3Yellow,densely dotted,mark=o,mark options={solid}}}
\tikzset{myparula34/.style={color=myParula3Yellow,dashdotted,mark=triangle,mark options={solid}}}
\tikzset{myparula35/.style={color=myParula3Yellow,dashdotdotted,mark=square,mark options={solid}}}

\tikzset{myparula41/.style={color=myParula4Purple,solid,mark=+,mark options={solid}}}
\tikzset{myparula42/.style={color=myParula4Purple,densely dashed,mark=x,mark options={solid}}}
\tikzset{myparula43/.style={color=myParula4Purple,densely dotted,mark=o,mark options={solid}}}
\tikzset{myparula44/.style={color=myParula4Purple,dashdotted,mark=triangle,mark options={solid}}}
\tikzset{myparula45/.style={color=myParula4Purple,dashdotdotted,mark=square,mark options={solid}}}

\tikzset{myparula51/.style={color=myParula5Green,solid,mark=+,mark options={solid}}}
\tikzset{myparula52/.style={color=myParula5Green,densely dashed,mark=x,mark options={solid}}}
\tikzset{myparula53/.style={color=myParula5Green,densely dotted,mark=o,mark options={solid}}}
\tikzset{myparula54/.style={color=myParula5Green,dashdotted,mark=triangle,mark options={solid}}}
\tikzset{myparula55/.style={color=myParula5Green,dashdotdotted,mark=square,mark options={solid}}}

\tikzset{myparula61/.style={color=myParula6LightBlue,solid,mark=+,mark options={solid}}}
\tikzset{myparula62/.style={color=myParula6LightBlue,densely dashed,mark=x,mark options={solid}}}
\tikzset{myparula63/.style={color=myParula6LightBlue,densely dotted,mark=o,mark options={solid}}}
\tikzset{myparula64/.style={color=myParula6LightBlue,dashdotted,mark=triangle,mark options={solid}}}
\tikzset{myparula65/.style={color=myParula6LightBlue,dashdotdotted,mark=square,mark options={solid}}}

\tikzset{myparula71/.style={color=myParula7Red,solid,mark=+,mark options={solid}}}
\tikzset{myparula72/.style={color=myParula7Red,densely dashed,mark=x,mark options={solid}}}
\tikzset{myparula73/.style={color=myParula7Red,densely dotted,mark=o,mark options={solid}}}
\tikzset{myparula74/.style={color=myParula7Red,dashdotted,mark=triangle,mark options={solid}}}
\tikzset{myparula75/.style={color=myParula7Red,dashdotdotted,mark=square,mark options={solid}}}

\usepackage{multirow}
\usepackage{booktabs}   %
\usepackage{makecell}
\usepackage{threeparttable}   %

\usepackage{algorithm}
\usepackage{algorithmicx}
\usepackage[noend]{algpseudocode}

\makeatletter
\AddToHook{env/algorithmic/begin}{\def\@currentcounter{ALG@line}}
\renewcommand{\theHALG@line}{\thealgorithm.\arabic{ALG@line}}
\makeatother

\algnewcommand{\LineComment}[1]{\State {\textcolor{gray}{/\!/ #1}}}

\newcommand{\alglocref}[2]{\cref{#1}, \cref{#2}}

\algrenewcommand{\alglinenumber}[1]{\scriptsize\textcolor{gray}{\texttt{#1}}}
\algrenewcommand{\algorithmicindent}{1em}

\algnewcommand{\algfontsize}[0]{\footnotesize}
\AddToHook{env/algorithmic/begin}{\algfontsize}

\algnewcommand{\algorithmicswitch}{\textbf{switch}}
\algdef{SE}[SWITCH]{Switch}{EndSwitch}[1]{\algorithmicswitch\ #1\ \algorithmicdo}{\algorithmicend\ \algorithmicswitch}%
\algtext*{EndSwitch}%

\algnewcommand{\algorithmiccase}{\textbf{case}}
\algdef{SE}[CASE]{Case}{EndCase}[1]{\algorithmiccase\ #1}{\algorithmicend\ \algorithmiccase}%
\algtext*{EndCase}%

\algnewcommand{\algorithmicon}{\textbf{on}}
\algdef{SE}[ON]{On}{EndOn}[1]{\algorithmicon\ #1\ \algorithmicdo}{\algorithmicend\ \algorithmicon}%
\algtext*{EndOn}%

\algnewcommand{\algorithmicat}{\textbf{at}}
\algdef{SE}[AT]{At}{EndAt}[1]{\algorithmicat\ #1\ \algorithmicdo}{\algorithmicend\ \algorithmicat}%
\algtext*{EndAt}%

\algnewcommand{\algorithmicrealfunction}{\textbf{function}}
\algdef{SE}[REALFUNCTION]{RealFunction}{EndRealFunction}[1]{\algorithmicrealfunction\ #1\ \algorithmicdo}{\algorithmicend\ \algorithmicrealfunction}%
\algtext*{EndRealFunction}%

\algnewcommand{\algorithmicthroughout}{\textbf{do throughout}}
\algdef{SE}[Throughout]{Throughout}{EndThroughout}[1]{\algorithmicthroughout\ #1\ \algorithmicdo}{\algorithmicend\ \algorithmicthroughout}%
\algtext*{EndThroughout}%

\algnewcommand{\algorithmictry}{\textbf{try}}
\algnewcommand{\algorithmiccatch}{\textbf{catch}}
\algdef{SE}[TRY]{Try}{EndTry}{\algorithmictry\ \algorithmicdo}{\algorithmicend\ \algorithmictry}
\algdef{C}[TRY]{TRY}{Catch}[1]{\algorithmiccatch\ #1\ \algorithmicdo}
\algtext*{EndTry}

\algrenewcommand{\algorithmicdo}{}
\algrenewcommand{\algorithmicthen}{}

\algnewcommand{\algorithmicgoto}{\textbf{goto}}%
\algnewcommand{\Goto}[1]{\algorithmicgoto~\ref{#1}}%

\algnewcommand{\algorithmicassert}{\textbf{assert}}%
\algnewcommand{\Assert}[1]{\algorithmicassert~{#1}}%

\algnewcommand{\algorithmicbreak}{\textbf{break}}%
\algnewcommand{\Break}[0]{\algorithmicbreak}%
\algnewcommand{\BreakOutOf}[1]{\algorithmicbreak~out~of~#1}%

\algnewcommand{\algorithmicwaiton}{\textbf{wait on}}%
\algnewcommand{\WaitOn}[1]{\algorithmicwaiton~{#1}}%

\algnewcommand{\InlineRequire}[1]{\textbf{require} {#1}}

\algblock{ManualIndent}{EndManualIndent}
\algnotext{ManualIndent}
\algnotext{EndManualIndent}

\algdef{SE}[GENERICBLOCK]{GenericBlock}{EndGenericBlock}[1]{#1}{}%
\algtext*{EndGenericBlock}%

\usepackage{fontawesome5}
\usepackage{noto-emoji-easy}
\usepackage{tango-icons}

\subimport{./lib/}{defer.tex}
\IfFileExists{version-string.txt}{%
  \newread\versionfile
  \openin\versionfile=version-string.txt
  \read\versionfile to \myVersionString
  \closein\versionfile
}{}

\usepackage[sort&compress,capitalize,nameinlink]{cleveref}

\crefalias{ALG@line}{line}

\crefname{figure}{Fig.}{Figs.}
\Crefname{figure}{Fig.}{Figs.}

\crefname{table}{Tab.}{Tabs.}
\Crefname{table}{Tab.}{Tabs.}

\crefname{section}{Sec.}{Secs.}
\Crefname{section}{Sec.}{Secs.}
\crefname{subsection}{Sec.}{Secs.}
\Crefname{subsection}{Sec.}{Secs.}
\crefname{subsubsection}{Sec.}{Secs.}
\Crefname{subsubsection}{Sec.}{Secs.}
\crefname{subsubsubsection}{Sec.}{Secs.}
\Crefname{subsubsubsection}{Sec.}{Secs.}
\crefname{appendix}{App.}{Apps.}
\Crefname{appendix}{App.}{Apps.}
\crefname{subappendix}{App.}{Apps.}
\Crefname{subappendix}{App.}{Apps.}
\crefname{subsubappendix}{App.}{Apps.}
\Crefname{subsubappendix}{App.}{Apps.}
\crefname{subsubsubappendix}{App.}{Apps.}
\Crefname{subsubsubappendix}{App.}{Apps.}

\crefname{algorithm}{Alg.}{Algs.}
\Crefname{algorithm}{Alg.}{Algs.}
\crefname{line}{ln.}{lns.}
\Crefname{line}{ln.}{lns.}

\crefname{proposition}{Prop.}{Props.}
\Crefname{proposition}{Prop.}{Props.}
\crefname{lemma}{Lem.}{Lems.}
\Crefname{lemma}{Lem.}{Lems.}
\crefname{theorem}{Thm.}{Thms.}
\Crefname{theorem}{Thm.}{Thms.}
\crefname{corollary}{Cor.}{Cors.}
\Crefname{corollary}{Cor.}{Cors.}
\crefname{definition}{Def.}{Defs.}
\Crefname{definition}{Def.}{Defs.}
\crefname{conjecture}{Conj.}{Conjs.}
\Crefname{conjecture}{Conj.}{Conjs.}
\crefname{remark}{Rem.}{Rems.}
\Crefname{remark}{Rem.}{Rems.}

\usepackage{subcaption}
\usepackage{amssymb}
\usepackage{amsmath}
\usepackage{amsthm}

\usepackage{hyperref}
\usepackage[sort&compress,capitalize,nameinlink]{cleveref}

\usepackage{ifthen}
\usepackage{xspace}

\usepackage{tikz}
\usetikzlibrary{trees,arrows.meta,positioning,calc}
\tikzset{>=Stealth}

\algnewcommand{\algorithmicupon}{\textbf{upon}}
\algdef{SE}[UPON]{Upon}{EndUpon}[1]{\algorithmicupon\ #1\ \algorithmicdo}{\algorithmicend\ \algorithmicupon}
\algtext*{EndUpon}

\usepackage{tablefootnote}
\usepackage[dvipsnames]{xcolor}
\usepackage{ragged2e}

\theoremstyle{plain}
\newtheorem{theorem}{Theorem}

\theoremstyle{definition}
\newtheorem{definition}{Definition}

\newcommand{\tx}{\mathsf{tx}}
\newcommand{\txs}{\mathsf{txs}}

\newcommand{\consistent}{\ensuremath{\asymp}}

\tikzset{mygame/.style={
    x=1.5cm,
    y=0.5cm,
}}

\usepackage{enumitem}

\setlist{nosep}  %
\setlist[itemize]{leftmargin=*}  %
\setlist[enumerate]{leftmargin=*}  %

\newcommand{\GST}[0]{\ensuremath{\mathsf{GST}}}

\newcommand{\negl}{\ensuremath{\operatorname{negl}}}
\newcommand{\poly}{\ensuremath{\operatorname{poly}}}

\newcommand{\Prob}[1]{\ensuremath{\mathbb{P}\left[#1\right]}}
\newcommand{\Expe}[1]{\ensuremath{\mathbb{E}\left[#1\right]}}

\newcommand{\Adv}{\ensuremath{\CA}}

\newcommand{\heccK}{\ensuremath{K}}
\newcommand{\heccN}{\ensuremath{N}}
\newcommand{\heccT}{\ensuremath{T}}
\newcommand{\heccF}{\ensuremath{\mathbb{F}}}

\newcommand{\heccLong}{\ensuremath{\mathsf{HECC}}}
\newcommand{\heccEnc}{\ensuremath{\mathsf{Enc}}}
\newcommand{\heccDec}{\ensuremath{\mathsf{Dec}}}
\newcommand{\heccLongEnc}{\ensuremath{\heccLong.\heccEnc}}
\newcommand{\heccLongDec}{\ensuremath{\heccLong.\heccDec}}

\newcommand{\heccMsg}{\ensuremath{\Bm}}
\newcommand{\heccRand}{\ensuremath{\Br}}
\newcommand{\heccShreds}{\ensuremath{\Bc}}
\newcommand{\heccShred}[1]{\ensuremath{c_{#1}}}

\newcommand{\sigLong}{\ensuremath{\mathsf{Sig}}}
\newcommand{\sigGen}{\ensuremath{\mathsf{Gen}}}
\newcommand{\sigSign}{\ensuremath{\mathsf{Sign}}}
\newcommand{\sigVerify}{\ensuremath{\mathsf{Verify}}}
\newcommand{\sigLongGen}{\ensuremath{\sigLong.\sigGen}}
\newcommand{\sigLongSign}{\ensuremath{\sigLong.\sigSign}}
\newcommand{\sigLongVerify}{\ensuremath{\sigLong.\sigVerify}}

\newcommand{\sigSK}{\ensuremath{\mathsf{sk}}}
\newcommand{\sigPK}{\ensuremath{\mathsf{pk}}}
\newcommand{\sigMsg}{\ensuremath{m}}
\newcommand{\sigSig}{\ensuremath{\sigma}}

\newcommand{\vcLong}{\ensuremath{\mathsf{VC}}}

\newcommand{\vcCommit}{\ensuremath{\mathsf{Commit}}}
\newcommand{\vcOpen}{\ensuremath{\mathsf{Open}}}
\newcommand{\vcVerify}{\ensuremath{\mathsf{Verify}}}

\newcommand{\vcLongCommit}{\ensuremath{\vcLong.\vcCommit}}
\newcommand{\vcLongOpen}{\ensuremath{\vcLong.\vcOpen}}
\newcommand{\vcLongVerify}{\ensuremath{\vcLong.\vcVerify}}

\newcommand{\vcVals}{\ensuremath{\Bv}}
\newcommand{\vcVal}{\ensuremath{v}}
\newcommand{\vcValSet}{\ensuremath{\CV}}
\newcommand{\vcRands}{\ensuremath{\Br}}
\newcommand{\vcRand}{\ensuremath{r}}
\newcommand{\vcRandSet}{\ensuremath{\CR}}
\newcommand{\vcCom}{\ensuremath{C}}
\newcommand{\vcWit}{\ensuremath{w}}
\newcommand{\vcIdx}{\ensuremath{i}}
\newcommand{\vcLen}{\ensuremath{L}}

\newcommand{\roH}{\ensuremath{\mathsf{H}}}

\newcommand{\PIsmr}{\ensuremath{\Pi_{\mathrm{ab}}}}
\newcommand{\PImcp}{\ensuremath{\Pi_{\mathrm{mcp}}}}

\newcommand{\Proposers}[1]{\ensuremath{\CN_{\mathrm{prop}}(#1)}}
\newcommand{\Relays}[1]{\ensuremath{\CN_{\mathrm{relay}}(#1)}}
\newcommand{\NumProposers}{\ensuremath{n_{\mathrm{prop}}}}
\newcommand{\NumRelays}{\ensuremath{n_{\mathrm{relay}}}}

\newcommand{\Leader}[1]{\ensuremath{\CL(#1)}}
\newcommand{\SlotTime}[1]{\ensuremath{T_{#1}}}

\newcommand{\smrInput}{\ensuremath{\mathsf{Input}}}
\newcommand{\smrOutput}{\ensuremath{\mathsf{Output}}}

\newcommand{\CPah}{\ensuremath{\CN}}
\newcommand{\CPa}{\ensuremath{\CN_{\mathrm{a}}}}
\newcommand{\CPh}{\ensuremath{\CN_{\mathrm{h}}}}

\newcommand{\MCPlong}{Multiple Concurrent Proposers\xspace}

\newcommand{\TheLOG}{\ensuremath{L_\infty^*}}
\newcommand{\ValencyLOG}[1]{\ensuremath{L_{#1}^{\times}}}

\newcommand{\timeA}{\ensuremath{t_{\mathrm{A}}}}
\newcommand{\timeB}{\ensuremath{t_{\mathrm{B}}}}
\newcommand{\bidA}{\ensuremath{b_{\mathrm{A}}}}
\newcommand{\bidB}{\ensuremath{b_{\mathrm{B}}}}
\newcommand{\valueA}{\ensuremath{v_{\timeA}}}
\newcommand{\valueB}{\ensuremath{v_{\timeB}}}
\newcommand{\valueTau}{\ensuremath{v_{\tau}}}
\newcommand{\valueT}[1]{\ensuremath{v_{#1}}}
\newcommand{\utilityA}{\ensuremath{u_{\mathrm{A}}}}

\usepackage{pdflscape}

\newdefergroup{uninterestingproofs}[notdeferred]
\newdefergroup{merkledetails}[notdeferred]

\title{\MCPlong: Why and How}
\author{%
Pranav Garimidi\\%
\small\href{mailto:pgarimidi@a16z.com}{\texttt{pgarimidi@a16z.com}}%
\and%
Joachim Neu\\%
\small\href{mailto:jneu@a16z.com}{\texttt{jneu@a16z.com}}%
\and%
Max Resnick\\%
\small\href{mailto:max.resnick@anza.xyz}{\texttt{max.resnick@anza.xyz}}%
}
\date{\small Version: \myVersionString}%

\begin{document}
\maketitle
\begin{abstract}
Traditional single-proposer blockchains suffer from miner extractable value (MEV), where validators exploit their serial monopoly on transaction inclusion and ordering to extract rents from users. 
While there have been many developments at the application layer to reduce the impact of MEV, these approaches largely require auctions as a subcomponent. Running auctions efficiently on chain requires two key properties of the underlying consensus protocol: \emph{selective-censorship resistance} and \emph{hiding}.
These properties guarantee that an adversary can neither selectively delay transactions
nor see their contents before they are confirmed.
We propose a \emph{multiple concurrent proposer} (MCP) protocol offering exactly these properties.

\end{abstract}

\section{Introduction}
\label{sec:introduction}

A fundamental design flaw in modern blockchain protocols is the creation of a serial monopoly over transaction inclusion. In prevailing single-proposer systems, a single node\footnote{We use ``node'' and ``validator'' interchangeably in this paper.} is granted temporary but absolute authority to determine the composition of the next block, allowing it to extract significant economic rents, commonly known as \emph{miner extractable value} (MEV)~\cite{Daian2019}. This monopolistic power to censor, reorder, and insert transactions creates profound market distortions, undermines the principle of fair access, and severely constrains the set of viable on-chain financial applications. While numerous countermeasures to MEV have been proposed, the underlying cause has not been satisfactorily addressed: a market structure that grants a single proposer monopolistic control over transaction ordering and inclusion.

This paper develops a framework for analyzing these market failures by reducing the broad and amorphous MEV landscape to its economic core. The most persistent and lucrative forms of MEV are: cross-exchange arbitrage, atomic arbitrage, and liquidations.\footnote{We note that sandwiching can be solved at the application layer even on single-proposer blockchains~\cite{sandwich-resistant-amm}.} We argue that these activities are best understood through the lens of auction theory; specifically, as high-frequency public-value auctions where a publicly observable signal evolves continuously, creating opportunities for bidders to compete and capture profits. 

Today, running an auction that starts and finishes in the span of a single block is essentially impossible, for two reasons. First, in many systems a single node controls which transactions are included in the next slot (which means, in particular, that node controls which auction bids are included)~\cite{FoxPR23}. Second, that same node sees transaction contents before all transactions (and therefore bids) are confirmed (and thus,  
while the node can still decide what bid of its own to place, if any). We show in \Cref{sec:auctions_today} that either of these issues makes running an auction on chain non-viable. Addressing this failure requires a protocol that can guarantee both \emph{selective-censorship resistance} and \emph{hiding}. We formally define both of these properties, and show that our \emph{multiple concurrent proposers} (MCP) protocol (\cref{sec:protocol}) satisfies them, ensuring immediate inclusion for transactions submitted to honest nodes, while concealing their contents until the consensus decision is final.
While short-term censorship resistance~\cite{Alpos_CRoverview} and hiding have been studied before, to our knowledge, we are the first to define the properties in a protocol-agnostic way, and to devise a protocol satisfying both properties simultaneously. 

The importance of censorship resistance and hiding for on-chain market structure has been established before~\cite{FoxPR23,moallemiPR25}; however, no major blockchain has incorporated these properties into its consensus protocol. A common critique of multiple concurrent proposer designs is that these properties cannot be achieved without introducing significant overhead. Our main contribution is to provide MCP as a light-weight gadget that fits on top of any standard PBFT-style consensus protocol. We prove that this construction is safe, live, and provides both censorship resistance and hiding, with an additional latency overhead of only a single network round-trip after optimizations compared to latency optimal protocols without the economic guarantees.

While many other solutions to MEV have been proposed, none of these approaches are without significant shortcomings (see \Cref{sec:relatedwork} for a detailed discussion). Some protocols have addressed hiding via encrypted mempools and some protocols have achieved approximations of short-term censorship resistance with DAG-based protocols, but no protocol has achieved the strict version of selective-censorship resistance we describe here. Thus, turning away from addressing MEV directly, protocols have been designed to reduce the negative externalities of MEV such as consensus instability or network centralization~\cite{Flashbots2020,bahrani2024centralization,Kiffer2023}. It should be noted that while approaches like proposer-builder separation in Ethereum have worked to reduce the adverse side effects of MEV at the protocol level, they have done so by facilitating MEV at the application layer, degrading the efficiency of on-chain financial applications.

The remainder of this paper proceeds as follows. 
In \cref{sec:auctions_today}, we discuss three examples that illustrate the market failures introduced by the leader's monopoly on transaction inclusion/ordering, motivating our definitions of censorship resistance and hiding,
before discussing related work in \cref{sec:relatedwork}.
In \cref{sec:model}, we recapitulate cryptographic primitives, and define the network, consensus, and economic models that we rely on throughout the paper. 
In \cref{sec:protocol}, we present the MCP protocol in detail. 
In \cref{sec:analysis}, we show that the MCP protocol, coupled with a safe and live underlying atomic broadcast primitive, is secure and achieves the desired censorship resistance and hiding properties.

\subsection{Three Examples}
\label{sec:auctions_today}

We present three stylized games for auctions. Our goal here is twofold: first, to provide the reader with an understanding of how specific aspects of an auction's market structure---such as the ability of one bidder to censor other bidders, see their bids, or bid with slightly more up-to-date information---can be more or less harmful to the efficiency of the system; and second, to explicitly state the market structure our protocol aims to achieve.

Three properties of single-proposer blockchains make running an auction in a single slot difficult:
\begin{enumerate}
    \item The proposer can censor bids.
    \item The proposer can see other bids and change his own bid based on that.
    \item The proposer has a latency advantage compared to other bidders.
\end{enumerate}
We begin with a scenario where all three effects are present. We then remove these, one by one, to isolate their impact and motivate our desired properties and the design of our protocol.

Two bidders, Alice and Bob, are competing in a first price auction for an item for which both bidders have a common value that continuously changes over time. At time $t$, the value of the item is observed to be $\valueT{t}$ by both bidders. We refer to this value as the signal at time $t$. The auction concludes at time $\tau$ at which point the item is transferred to the winning bidder. We assume the item's value follows a martingale, so that $E[\valueTau| \{ \valueT{s} \mid s \leq t\}] = \valueT{t}$ for all $t \leq \tau$. This means at any given time, both bidders' expected future value of the item at the end of the auction is equal to the latest value they have observed. 

We assume that bidders are risk neutral and have quasi-linear utilities, so that Alice's utility is $\utilityA(\bidA,\bidB) = \Ione_{\bidA > \bidB} (\valueTau - \bidA)$, with Bob's utility defined analogously. Alice bids first at time $\timeA$ observing the signal $\valueA$, and Bob bids after at time $\timeB > \timeA$ observing the updated signal $\valueB$. For tractability throughout, we assume that if Alice and Bob have the same bid, Bob wins the item.

Let us consider, as our first example, a game where Bob has every possible advantage. He bids after Alice, sees her bid, observes new information about the item's value, and can censor her bid entirely. 
\begin{center}
\noindent\fbox{\begin{minipage}{0.9\linewidth}
\RaggedRight
\paragraph{Scenario~1: Neither censorship resistance nor hiding}
\begin{enumerate}
    \item At time $\timeA$, Alice observes the signal $\valueA$ and submits a bid $\bidA$.
    \item At time $\timeB$, Bob observes Alice's bid $\bidA$, the new signal $\valueB$, and then chooses whether to censor Alice's bid and submits his own bid $\bidB$.
    \item If Bob censors, only his bid is considered. If not, both bids are. The bidder with the highest included bid wins, values the item at $\valueTau$, and pays their bid.
\end{enumerate}
\end{minipage}}
\end{center}
In this game, Bob's optimal strategy is trivial and devastating for the auction's fairness and efficiency. He can simply choose to censor Alice's bid, regardless of its value or the new signal $\valueB$, and submit a near-zero bid himself. By doing so, he guarantees a win at virtually no cost. Censorship alone is so powerful that it completely breaks the auction, regardless of whether Bob can see Alice's bid. 

In response to the problem of censorship, protocols suggesting multiple concurrent proposers have emerged to mitigate it~\cite{FoxPR23}. Let us assume such a system is implemented, successfully removing Bob's ability to selectively cancel Alice's bid. Is the resulting auction efficient? In the second example, we explore a market where censorship is resolved, but Bob still bids second, sees Alice's bid, and has more recent price information.

\begin{center}
\noindent\fbox{\begin{minipage}{0.9\linewidth}
\RaggedRight
\paragraph{Scenario~2: Censorship resistance but no hiding}
\begin{enumerate}
    \item At time $\timeA$, Alice observes the signal $\valueA$ and submits a bid $\bidA$.
    \item At time $\timeB$, Bob observes $\bidA$, the updated signal $\valueB$, and then places his  bid $\bidB$.
    \item The highest bidder wins, values the item at $\valueTau$ and pays their bid.
\end{enumerate}
\end{minipage}}
\end{center}
In this game, Bob has a very simple best response to Alice's bid $\bidA$ and the new signal $\valueB$:
\begin{equation}
    \bidB(\bidA,\valueB) = \begin{cases}
        \bidA, &  \valueB \geq \bidA\\
        0 & \valueB < \bidA
    \end{cases}
\end{equation}

Notice that when Bob plays this strategy, Alice is subject to extreme \emph{adverse selection}. Whenever the price moves favorably, such that she would have had a positive payoff ($\valueB > \bidA$), Bob outbids her and captures the surplus. Whenever the price moves unfavorably against her ($\valueB < \bidA$), Bob lets her win, forcing her to take the loss. Knowing this, the subgame perfect equilibrium of this game is for Alice to bid nothing at all, allowing Bob to win the item for free. Even without censorship, the side information (Bob seeing $\bidA$) is enough for the auction to unravel completely. The acute adverse selection in this second scenario stems from Bob's ability to ``peek'' at Alice's bid. 

As our third and final example, we consider what happens when this ability is removed, leaving only the unavoidable \emph{last look}~\cite{moallemiPR25} feature where Bob's only advantage is bidding with a more recent signal than Alice.
\begin{center}
\noindent\fbox{\begin{minipage}{0.9\linewidth}
\RaggedRight
\paragraph{Scenario~3: Censorship resistance and hiding}
\begin{enumerate}
    \item At time $\timeA$, Alice observes the signal $\valueA$ and submits a bid $\bidA$.
    \item At time $\timeB$, Bob observes the updated signal $\valueB$ (but does not observe $\bidA$), and then places his  bid $\bidB$.
    \item The highest bidder wins, values the item at $\valueTau$ and pays their bid.
\end{enumerate}
\end{minipage}}
\end{center}
In this game, the equilibrium from the second example, where the auction completely unraveled, is no longer possible. Bob cannot condition his strategy on Alice's bid because he cannot observe it. We note that the equilibrium strategy profile is still hard to determine as any pure strategy played by Alice would allow Bob to know exactly what Alice bid (even without explicitly seeing her bid) causing the same unraveling as the previous case. The exact mixed-strategy equilibrium played will depend on the process by which the asset's price is determined. But, prior work has shown that if the price is determined by a geometric Brownian motion process, then the revenue of the auction converges to $\valueA$ as Bob's time advantage (and thus his informational advantage) tends to zero~\cite{moallemiPR25}. This is a notably different, and normatively much more desirable, outcome than the complete unraveling in the earlier second scenario, which occurs even for the smallest information advantage, if Bob can see Alice's bid. The effect of last look, when isolated, is not nearly as detrimental in a system with censorship resistance and hiding, and can be further reduced by increasing the block production frequency.

Now that we have explored these motivating examples---arriving at a functional market structure by peeling away layers of asymmetric advantage---we can proceed with the main body of our work, which is to propose a protocol that provides selective-censorship resistance and hiding, and thus can run an auction implementing Scenario~3 rather than Scenarios~1 or~2.

\subsection{Related Work}
\label{sec:relatedwork}

We start by reviewing earlier MEV mitigation approaches and their challenges.\footnote{For comprehensive surveys of MEV countermeasures, see~\cite{yang2023sokmevcountermeasurestheory,HeimbachW22}.}
Broadly, there are three approaches to addressing MEV in the literature: 
fair-ordering consensus protocols~\cite{Kelkar2020,kursawe2020wendy,Zhang2021,Kelkar2023,Cachin2022,Cachin2024,FairPoS2022,Xue2024,Zhang2024}, 
transaction encryption~\cite{F3B2022,Momeni2023,Bormet2024,Choudhuri2024,ChainlinkFSS2021, Dahlia_MEVprotectionDAG,UniswapTEE2025}, and 
batching~\cite{Budish2015,Angeris2022Batch,UniswapTEE2025}.

\paragraph{Fair Ordering}

Fair ordering protocols try to extend consensus to achieve a strict \emph{order fairness} property, where if a majority of nodes observe transaction $\tx_1$ before transaction $\tx_2$ then $\tx_1$ should be sequenced before $\tx_2$ in the final output log. 
But, with realistic network delays among a globally distributed node set and some nodes acting adversarially, 
so-called Condorcet cycles arise, where a majority of nodes each may see a transaction $\tx_1$ before transaction $\tx_2$, $\tx_2$ before transaction $\tx_3$, and $\tx_3$ before $\tx_1$.
Because of these cycles it is theoretically impossible to achieve the full order fairness property~\cite{Kelkar2020}. 
The Aequitas and Themis
protocols~\cite{Kelkar2020,Kelkar2023} achieve a weaker \emph{batch order fairness} property (which is still stronger than the fairness properties provided by Wendy~\cite{kursawe2020wendy} and Pompe~\cite{Zhang2021}), where if ``many'' nodes have seen transaction $\tx_1$ before transaction $\tx_2$, then $\tx_2$ would at least not be in an earlier consensus batch than $\tx_1$.

Although batch order fairness is one of the strongest definitions of fairness that has been achieved in the literature, its usefulness in practice has remained limited since a motivated attacker can create large cycles by inserting transactions~\cite{vafadar2023condorcet}, at which point the batch order fairness guarantee becomes weak.
Recent developments in fair ordering protocols have mostly been performance improvements~\cite{Xue2024}, not protocols that offer stronger fairness guarantees than Aequitas/Themis. 
There has also been work on defining order fairness in the universal composability framework~\cite{UC_fairness}. 
The authors of~\cite{UC_fairness} give lower bounds on the degree to which order fairness can be achieved by any protocol along with a matching upper bound with a protocol using trusted enclaves.

\paragraph{Encrypted Mempools}

Transaction encryption approaches aim to use cryptography to conceal details of transactions while consensus is underway, so that the proposer and consensus leader have limited information about how to profitably reorder transactions or which transactions to censor to extract value. 
\emph{Threshold encryption}~\cite{Bormet2024,Bormet2025BEASTMEV,Ferveo,F3B2022,BonehLT25,Choudhuri2024,AgarwalFP24,BonehBNRS25} is the more popular cryptographic primitive chosen for this purpose, but \emph{time-based cryptography}~\cite{delay} and \emph{identity-based encryption}~\cite{Momeni2023,McFly} have also been considered.
Encryption-based approaches have several key shortcomings.
First, in order to prevent denial-of-service (DoS) attacks, transactions must contain at least some unencrypted metadata about fees and the transaction sender, which presents an attack angle for MEV extraction.
Second, since the proposer is still a monopolist on inclusion, they can simply choose only to include transactions that are not encrypted (or which are encrypted but accompanied by the unencrypted transaction through a side channel). 
Time-sensitive transactions have no choice but to send the unencrypted transaction to the leader, thereby bypassing the mechanism, or wait for the next consensus round (i.e., drop out of on-chain auctions).

\paragraph{Batching}

Batching approaches aim to remove the power of the proposer to decide the order of transactions in the block, by making execution semantics order-agnostic~\cite{Angeris2022Batch, ramseyerspeedex,ramseyer2024groundhog} (or fixing the ordering deterministically, based on the set of transactions included in a block~\cite{prioritytaxes,executionconsensusseperation}).
While batching is a promising way to diffuse the ordering power of the proposer, it does not address the power of the proposer over inclusion and exclusion, and in particular the proposer's power to exclude (``censor'') transactions is enough to comprehensively manipulate many of the most popular batch-based mechanisms such as auctions~\cite{FoxPR23}. Another line of work has focused on producing a protocol that has a good market structure for auctions without focusing on integrating it into the broader consensus considerations required on a blockchain~\cite{Bonneau_Riggs,white2024leaderless}.

\paragraph{DAG-Based Protocols}

There is a large body of DAG-based consensus protocols that also allow for multiple nodes to propose transactions in every round. However, these protocols do this to increase throughput rather than for economic concerns. Because of this, design choices are made which make these protocols more performant at the cost of opening up avenues for selective censorship. Many of these protocols (Bullshark~\cite{bullshark}, Tusk~\cite{Narwhal_Tusk}, Shoal~\cite{shoal}), only commit special \emph{anchor} blocks, similar to leaders in traditional consensus protocols, in every round and all other blocks are implicitly confirmed based on the blocks an anchor refers to. This gives anchors the power to selectively leave out certain blocks from the previous round. While these blocks may still be picked up by future anchors, this still gives the adversary the power to selectively move certain blocks of transactions to the next round instead of the current round. While not outright censorship, these short-term delays can be enough to be deal-breaking for certain applications, as outlined in \Cref{sec:auctions_today}. 

More recent works such as Shoal++~\cite{shoal++} and Sailfish~\cite{sailfish} allow for multiple anchor candidates (referred to as leaders in Sailfish) per round to reduce the latency advantage exclusively enjoyed by a single leader node. In theory, this would alleviate the censorship concerns as a single honest anchor in a round would be enough to commit all the honest proposals from the previous round. However, the way both protocols implement multiple anchor candidates still gives an adversary the power to selectively censor. Both protocols have a predetermined order over anchors in each round, and will only commit a prefix of the anchors in this ordering, e.g., if the second anchor in the order crashes, none of the subsequent anchors in that round will be eligible to be committed. The first anchor in this ordering can be thought of as a traditional leader with the subsequent anchors being used to decrease latency. However, in these protocols if the adversary controls the first anchor in a round and crashes the second anchor they can still carry out the same attacks as in the single anchor case.

Additionally, to decrease latency, these protocols are optimistically responsive, and will only wait to hear from $n-f$ nodes before moving to the next round. While this allows these protocols to operate at network speed, it may also result in honest proposals being inadvertently censored. In particular, Mysticeti~\cite{mysticeti}, unlike other protocols, allows blocks to be committed if enough other nodes voted for them, even if the anchor did not vote for them. This means that if a node gets its block to enough honest nodes on time, the anchor has no power to censor it. However, honest nodes only wait to hear from $2f+1$ other nodes before advancing to the next round, so even honest nodes that announce their blocks on time might still be missed by other honest nodes. If the protocol were changed so that, instead of being optimistically responsive, honest nodes waited until a fixed timeout (guaranteeing that they hear all other honest nodes' proposals) before moving to the next round, then this would not be an issue and the protocol would be selective-censorship resistant.  Furthermore, as described, none of these DAG-based protocols satisfy hiding, although hiding can be incorporated into these protocols, for instance using the techniques described in~\cite{Dahlia_MEVprotectionDAG}. 

BigDipper~\cite{BigDipper}, while not DAG-based, is another protocol that allows for multiple nodes to submit transactions for the same round explicitly with the goal of short-term censorship resistance. BigDipper works similarly to our protocol in that it modifies a standard consensus protocol by constraining the types of blocks a leader can propose to including enough mini-blocks in their proposal. However, BigDipper does not use relays like we do, and in turn does not provide hiding and allows for the leader to exclude up to $2f$ mini-blocks, potentially from honest proposers.

\paragraph{``Leaderless'' Consensus Protocols}
There are also so-called ``leaderless''\footnote{The component protocols of these protocols are still based on leaders.}
protocols like HoneyBadgerBFT~\cite{honeybadger} and DispersedLedger~\cite{DispersedLedger}, targeting the asynchronous common subset formulation of consensus, where every node can include transactions in a round.
But these protocols still fall short of achieving full selective-censorship resistance. HoneyBadger only requires nodes to confirm $n-f$ proposals before moving to the next round, potentially allowing for $f$ honest proposals to be censored. 
DispersedLedger improves on this by guaranteeing that every honest proposal will eventually be included, but it achieves this by allowing later rounds to  confirm missed blocks from earlier rounds. Thus, while an adversary cannot censor transactions, they can potentially cause short-term confirmation delays for targeted batches. Both these protocols use cryptography to guarantee hiding before blocks are confirmed.

\paragraph{Verifiable Secret Sharing}
As becomes clear in \cref{sec:protocol},
our MCP protocol builds on verifiable secret sharing techniques.
Secret sharing was first introduced by~\cite{shamirsecretsharing,Blakley1979}, as a protocol with which a dealer can share a secret with a committee, such that any group of ``many'' committee members can recover the secret, while any group of ``few'' committee members learn nothing about the secret.
Verifiable secret sharing~\cite{ChorGMA85} (VSS) ensures that the secrets recovered by different ``large'' groups of committee members agree.
Packed multi-secret secret sharing~\cite{FranklinY92} allows amortizing communication overhead if the dealer shares multiple secrets simultaneously.
In this taxonomy, a component of MCP implements packed VSS.
Starting with~\cite{Feldman1987,Pedersen1991},
many VSS schemes were proposed in the past four decades---see~\cite{CachinKLS02,KateZG10,BackesKP11,DasXR21,ShoupS24} and references therein.
The closest relative of our MCP protocol among VSS protocols
is the acknowledgment-based approach of~\cite{Das25}.
Earlier works combining VSS and consensus
include~\cite{BasuTAMRS19,BenhamoudaGGHKLRR20}.
In~\cite{BasuTAMRS19}, the focus is on hiding the state (rather than the transactions) of a replicated state machine. No censorship resistance is achieved.
Similarly,~\cite{BenhamoudaGGHKLRR20} uses a blockchain as ``control plane for a storage system'' ``of secret information''.

\section{Preliminaries \& Model}
\label{sec:model}

\subsection{Cryptographic Primitives}
\label{sec:model-crypto}

Our scheme employs existentially-unforgeable signatures
and vector commitments,
which we recall below.
As is customary in the literature, we consider a
\emph{probabilistic polynomial-time} (PPT) adversary~$\Adv$.
A function $f$ is \emph{negligible} if for every $c > 0$ there exists an $n_0$ such that for all $n \geq n_0$, $f(n) < 1/n^c$.
We denote $[N] \triangleq \{1, ..., N\}$.
We use a computational security parameter $\kappa$,
and a finite field $\IF$ with $|\IF| \geq 2^\kappa$.

\subsubsection{Signatures}

\begin{definition}[Signatures]
    \label{def:sig}
    A signature scheme $\sigLong$ consists of three algorithms:
    \begin{itemize}
        \item
              \textbf{Key generation $\sigLongGen\colon 1^\kappa \to (\sigSK, \sigPK)$.}
              Given security parameter $\kappa$,
              the key generation algorithm outputs a \emph{secret key} $\sigSK$ and a corresponding \emph{public key} $\sigPK$.

        \item
              \textbf{Signing $\sigLongSign\colon (\sigSK, \sigMsg) \to \sigSig$.}
              Given a secret key $\sigSK$ and a \emph{message} $\sigMsg$ as inputs, the signing algorithm outputs a \emph{signature} $\sigSig$.

        \item
              \textbf{Verification $\sigLongVerify\colon (\sigPK, \sigMsg, \sigSig) \to \{0, 1\}$.}
              Given a public key $\sigPK$, a message $\sigMsg$, and a signature $\sigSig$ as inputs, the verification algorithm outputs $1$ if the signature is valid, and $0$ otherwise.
    \end{itemize}
    A ``good'' signature scheme satisfies the following two properties:
    \begin{itemize}
        \item
              \textbf{Correctness.}
              For all key pairs $(\sigSK, \sigPK) \gets \sigLongGen(1^\kappa)$ and all messages $\sigMsg$,
              if $\sigSig \gets \sigLongSign(\sigSK, \sigMsg)$, then $\sigLongVerify(\sigPK, \sigMsg, \sigSig) = 1$.

        \item
              \textbf{Existential unforgeability.}
              Every PPT adversary $\Adv$ succeeds in the following game only with probability negligible in $\kappa$:
              Generate $(\sigSK, \sigPK) \gets \sigLongGen(1^\kappa)$ and give $\sigPK$ to adversary $\Adv$.
              The adversary $\Adv$ may make polynomially many signing queries $\sigMsg_i$ and receive signatures $\sigSig_i \gets \sigLongSign(\sigSK, \sigMsg_i)$.
              Finally, $\Adv$ outputs a forgery $(\sigMsg^*, \sigSig^*)$.
              The adversary succeeds iff $\sigLongVerify(\sigPK, \sigMsg^*, \sigSig^*) = 1$ and $\sigMsg^*$ was not queried from the signing oracle.
    \end{itemize}
\end{definition}
An example signature scheme satisfying \cref{def:sig} is the BLS signature scheme~\cite{boneh2003aggregate}.

\subsubsection{Vector Commitments}

\begin{definition}[Vector Commitments]
    \label{def:vc}
    For a fixed positive integer $\vcLen$,
    finite set $\vcValSet$ of potential \emph{values},
    and finite set $\vcRandSet$ of potential \emph{masking values},
    with $|\vcRandSet| \geq 2^\kappa$ where $\kappa$ is a security parameter,
    a vector commitment scheme $\vcLong$ consists of three algorithms:
    \begin{itemize}
        \item
            \textbf{Commitment $\vcLongCommit\colon (\vcVals, \vcRands) \to \vcCom$.}
            Given 
            a \emph{vector} $\vcVals = (\vcVal_1, ..., \vcVal_{\vcLen}) \in \vcValSet^{\vcLen}$
            and
            \emph{randomness} $\vcRands = (\vcRand_1, ..., \vcRand_{\vcLen}) \in \vcRandSet^{\vcLen}$
            as inputs,
            the algorithm outputs 
            a deterministic \emph{commitment} $\vcCom$.

        \item
            \textbf{Opening $\vcLongOpen\colon (\vcVals, \vcRands, \vcIdx) \to \vcWit$.}
            Given 
            a vector $\vcVals \in \vcValSet^{\vcLen}$,
            randomness $\vcRands \in \vcRandSet^{\vcLen}$,
            and an \emph{index} $\vcIdx \in [\vcLen]$ 
            as inputs,
            the algorithm outputs 
            a \emph{witness} $\vcWit$ for the value of $\vcVals$ at index $\vcIdx$.

        \item
            \textbf{Verification $\vcLongVerify\colon (\vcCom, \vcIdx, \vcVal, \vcRand, \vcWit) \to \{0, 1\}$.}
            Given 
            a commitment $\vcCom$, 
            an index $\vcIdx \in [\vcLen]$, 
            a claimed \emph{opening} $(\vcVal, \vcRand) \in \vcValSet \times \vcRandSet$, 
            and a witness $\vcWit$ 
            as inputs,
            the algorithm outputs 
            $1$ if the opening is valid, and $0$ otherwise.
    \end{itemize}
    A ``good'' vector commitment scheme satisfies the following properties:
    \begin{itemize}
        \item
            \textbf{Correctness.}
            For all vectors $\vcVals = (\vcVal_1, ..., \vcVal_{\vcLen}) \in \vcValSet^{\vcLen}$, all randomness $\vcRands = (\vcRand_1, ..., \vcRand_{\vcLen}) \in \vcRandSet^{\vcLen}$, and all indices $\vcIdx \in [\vcLen]$:
            if $\vcCom \gets \vcLongCommit(\vcVals, \vcRands)$ and $\vcWit \gets \vcLongOpen(\vcVals, \vcRands, \vcIdx)$,
            then $\vcLongVerify(\vcCom, \vcIdx, \vcVal_{\vcIdx}, \vcRand_{\vcIdx}, \vcWit) = 1$.

        \item
            \textbf{Binding.}
            Every PPT adversary $\Adv$ succeeds in the following game only with probability negligible in $\kappa$:
            The adversary $\Adv$ outputs $(\vcVals, \vcVals', \vcRands, \vcRands')$ where $\vcVals, \vcVals' \in \vcValSet^{\vcLen}$ and $\vcRands, \vcRands' \in \vcRandSet^{\vcLen}$.
            The adversary succeeds iff $\vcVals \neq \vcVals'$ or $\vcRands \neq \vcRands'$, and $\vcLongCommit(\vcVals, \vcRands) = \vcLongCommit(\vcVals', \vcRands')$.

        \item
            \textbf{Position binding.}
            Every PPT adversary $\Adv$ succeeds in the following game only with probability negligible in $\kappa$:
            The adversary $\Adv$ outputs $(\vcCom, \vcIdx, \vcVal, \vcVal', \vcRand, \vcRand', \vcWit, \vcWit')$ where $\vcIdx \in [\vcLen]$, $\vcVal, \vcVal' \in \vcValSet$, and $\vcRand, \vcRand' \in \vcRandSet$.
            The adversary succeeds iff $\vcVal \neq \vcVal'$ or $\vcRand \neq \vcRand'$, and $\vcLongVerify(\vcCom, \vcIdx, \vcVal, \vcRand, \vcWit) = 1$
            and $\vcLongVerify(\vcCom, \vcIdx, \vcVal', \vcRand', \vcWit') = 1$.

        \item
            \textbf{Hiding.}
            Every PPT adversary $\Adv$ succeeds in the following game only with probability negligible in $\kappa$:
            The adversary $\Adv$ provides $(\vcVals^{(0)}, \vcVals^{(1)})$.
            The challenger flips an unbiased coin $b \getsRandom \{0,1\}$, samples randomness $\vcRands \getsRandom \vcRandSet^{\vcLen}$ uniformly at random, and gives $\vcCom \gets \vcLongCommit(\vcVals^{(b)}, \vcRands)$ to $\Adv$.
            The adversary $\Adv$ may make polynomially many opening queries $\vcIdx_j$ for indices where $\vcVal^{(0)}_{\vcIdx_j} = \vcVal^{(1)}_{\vcIdx_j}$ and receive openings $(\vcRand^{(j)}, \vcWit^{(j)})$ where $\vcRand^{(j)} \gets \vcRand_{\vcIdx_j}$ and $\vcWit^{(j)} \gets \vcLongOpen(\vcVals^{(b)}, \vcRands, \vcIdx_j)$.
            Finally, $\Adv$ outputs a guess $\hat{b} \in \{0,1\}$.
            The adversary succeeds iff $\hat{b} = b$.
    \end{itemize}
\end{definition}

\begin{landscape}
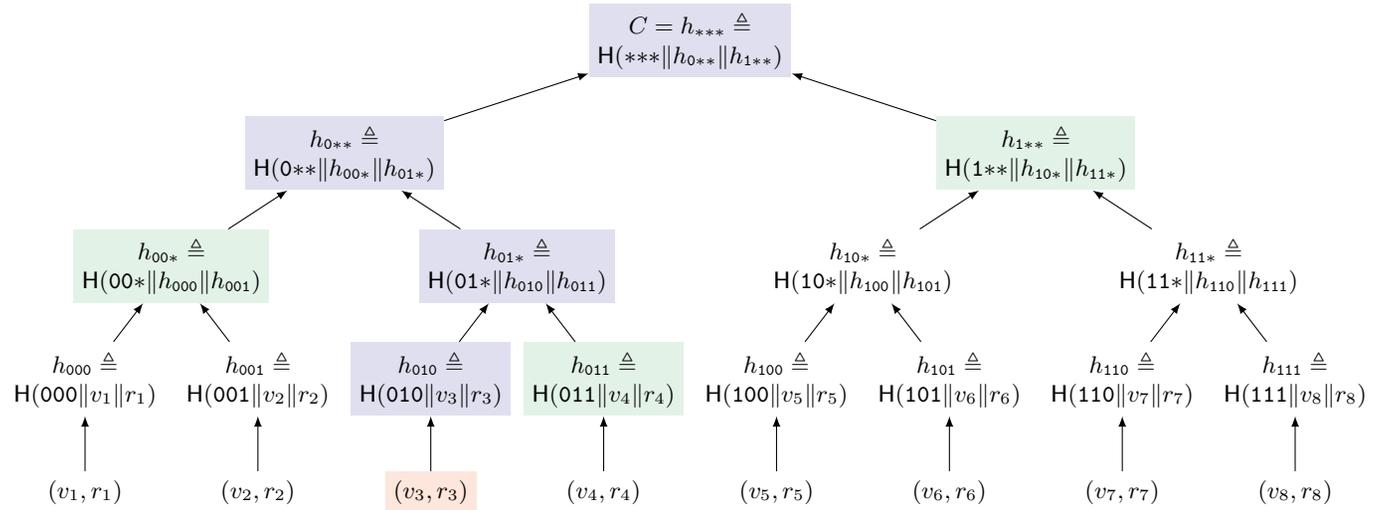
\begin{figure}[tb]%
    \centering%
    \begin{tikzpicture}[%
            x=2.3cm,
            y=1.5cm,
            node/.style={align=center},
            node_vr/.style={node},
            node_vr0/.style={node_vr},
            node_vr1/.style={node_vr},
            node_vr2/.style={node_vr,fill=Red!10},
            node_vr3/.style={node_vr},
            node_vr4/.style={node_vr},
            node_vr5/.style={node_vr},
            node_vr6/.style={node_vr},
            node_vr7/.style={node_vr},
            node_h/.style={node},
            node_h000/.style={node_h},
            node_h001/.style={node_h},
            node_h010/.style={node_h,fill=Blue!10},
            node_h011/.style={node_h,fill=Green!10},
            node_h100/.style={node_h},
            node_h101/.style={node_h},
            node_h110/.style={node_h},
            node_h111/.style={node_h},
            node_h00/.style={node_h,fill=Green!10},
            node_h01/.style={node_h,fill=Blue!10},
            node_h10/.style={node_h},
            node_h11/.style={node_h},
            node_h0/.style={node_h,fill=Blue!10},
            node_h1/.style={node_h,fill=Green!10},
            node_hroot/.style={node,fill=Blue!10},
        ]
        \footnotesize

        \foreach \i in {0,1,...,7} {
            \pgfmathtruncatemacro{\idx}{\i+1}
            \node [node_vr\i] (vr\i) at (\i, -1) {$(\vcVal_{\idx}, \vcRand_{\idx})$};
        }

        \foreach \i/\bi in {0/000, 1/001, 2/010, 3/011, 4/100, 5/101, 6/110, 7/111} {
            \pgfmathtruncatemacro{\idx}{\i+1}
            \node [node_h\bi] (h\bi) at (\i, 0) {$h_{\mathtt{\bi}} \triangleq$\\$\roH(\mathtt{\bi} \| \vcVal_{\idx} \| \vcRand_{\idx})$};
            \draw [-latex] (vr\i) -- (h\bi);
        }

        \foreach \i/\bi in {0/00, 1/01, 2/10, 3/11} {
            \node [node_h\bi] (h\bi) at (0.5+2*\i, 1) {$h_{\mathtt{\bi{*}}} \triangleq$\\$\roH(\mathtt{\bi{*}} \| h_{\mathtt{\bi0}} \| h_{\mathtt{\bi1}})$};
            \draw [-latex] (h\bi0) -- (h\bi);
            \draw [-latex] (h\bi1) -- (h\bi);
        }

        \foreach \i/\bi in {0/0, 1/1} {
            \node [node_h\bi] (h\bi) at (1.5+4*\i, 2) {$h_{\mathtt{\bi{*}{*}}} \triangleq$\\$\roH(\mathtt{\bi{*}{*}} \| h_{\mathtt{\bi0{*}}} \| h_{\mathtt{\bi1{*}}})$};
            \draw [-latex] (h\bi0) -- (h\bi);
            \draw [-latex] (h\bi1) -- (h\bi);
        }

        \node [node_hroot] (h) at (3.5, 3) {$\vcCom = h_{\mathtt{{*}{*}{*}}} \triangleq$\\$\roH(\mathtt{{*}{*}{*}} \| h_{\mathtt{0{*}{*}}} \| h_{\mathtt{1{*}{*}}})$};
        \draw [-latex] (h0) -- (h);
        \draw [-latex] (h1) -- (h);
        
    \end{tikzpicture}%
    \caption{%
        Illustration of hiding Merkle tree vector commitment scheme for $L=8$. The opening of commitment $\vcCom$ at index $3$ to $(v_3, r_3)$ (shaded red) is verified by recomputing the hashes (shaded blue) from the opening to the Merkle root using the opening and the Merkle path (shaded green) provided as witness.%
    }%
    \label{fig:hiding-merkle}%
\end{figure}%
\end{landscape}

\begin{algorithm}[tbp]
    \caption{Hiding Merkle tree (see \cref{fig:hiding-merkle}) vector commitment scheme (see \cref{def:vc}), for simplicity described for vectors of length $\vcLen = 2^d$ for some fixed $d \in \mathbb{N}$, using random oracle $\roH$}
    \label{alg:hiding-merkle}
    \begin{algorithmic}[1]
        
        \medskip
        \State Helper functions (see \cref{fig:hiding-merkle} for identification of hashes in the Merkle tree):
        \begin{itemize}
            \item $\operatorname{leaf}(\vcIdx)$: Returns the $d$-bit binary representation of $\vcIdx-1$ for $\vcIdx \in [\vcLen]$ as an identifier for the corresponding leaf hash in the Merkle tree. E.g., for $d=3$, $\operatorname{leaf}(3) = \mathtt{010}$.
            \item $\operatorname{ancestors}(b)$: For leaf hash identifier $b \in \{0,1\}^d$, returns the identifiers of all hashes on the path from the leaf to the root of the Merkle tree, including the leaf but excluding the root. E.g., for $d=3$, $\operatorname{ancestors}(\mathtt{010}) = [\mathtt{010}, \mathtt{01{*}}, \mathtt{0{*}{*}}]$.
            \item $\operatorname{parent}(b)$: For the identifier $b$ of a hash in the Merkle tree, returns the identifier of the parent node in the Merkle tree, by replacing the rightmost non-${*}$ bit in $b$ with ${*}$. E.g., for $d=3$, $\operatorname{parent}(\mathtt{01{*}}) = \mathtt{0{*}{*}}$.
            \item $\operatorname{sibling}(b)$: For the identifier $b$ of a hash in the Merkle tree, returns the identifier of the sibling in the Merkle tree, by flipping the rightmost non-${*}$ bit. E.g., for $d=3$, $\operatorname{sibling}(\mathtt{01{*}}) = \mathtt{00{*}}$.
        \end{itemize}
        
        \medskip
        \Procedure{$\operatorname{MerkleTree}$}{$\vcVals = (\vcVal_1, \ldots, \vcVal_\vcLen), \vcRands = (\vcRand_1, \ldots, \vcRand_\vcLen)$}
            \State $\CT \gets \emptyset$
            \For{$\vcIdx \in [\vcLen]$}%
                    \Comment{Compute leaf hashes}
                \State $\CT.h_{\operatorname{leaf}(\vcIdx)} \gets \roH(\operatorname{leaf}(\vcIdx) \| \vcVal_{\vcIdx} \| \vcRand_{\vcIdx})$%
                    \label{alg:merkle-leaf}
            \EndFor
            \For{$\ell \in [d-1, ..., 0]$}%
                    \Comment{Compute interior hashes}
                \For{each interior node at level $\ell$ with hash identifier $b$ (see \cref{fig:hiding-merkle})}
                    \State $b_{\text{L}} \gets$ left child of $b$
                    \State $b_{\text{R}} \gets$ right child of $b$
                    \State $\CT.h_b \gets \roH(b \| \CT.h_{b_{\text{L}}} \| \CT.h_{b_{\text{R}}})$%
                        \label{alg:merkle-interior}
                \EndFor
            \EndFor
            \State \Return $\CT$
        \EndProcedure
        
        \medskip
        \Procedure{$\vcLongCommit$}{$\vcVals = (\vcVal_1, \ldots, \vcVal_\vcLen), \vcRands = (\vcRand_1, \ldots, \vcRand_\vcLen)$}
            \State $\CT \gets \operatorname{MerkleTree}(\vcVals, \vcRands)$%
                \label{alg:commit-build}
            \State \Return $\vcCom \gets \CT.h_{\mathtt{{*}\cdots{*}}}$%
                \Comment{Merkle root}%
                \label{alg:commit-root}
        \EndProcedure
        
        \medskip
        \Procedure{$\vcLongOpen$}{$\vcVals, \vcRands, \vcIdx$}
            \State $\CT \gets \operatorname{MerkleTree}(\vcVals, \vcRands)$%
                \label{alg:open-build}
            \State $\vcWit \gets \emptyset$
            \For{$b \in \operatorname{ancestors}(\operatorname{leaf}(\vcIdx))$}%
                    \Comment{Merkle path as witness}%
                    \label{alg:open-path}
                \State $b' \gets \operatorname{sibling}(b)$
                \State $\vcWit.h_{b'} \gets \CT.h_{b'}$%
                    \label{alg:open-sibling}
            \EndFor
            \State \Return $\vcWit$
        \EndProcedure
        
        \medskip
        \Procedure{$\vcLongVerify$}{$\vcCom, \vcIdx, \vcVal, \vcRand, \vcWit$}
            \State $h \gets \roH(\operatorname{leaf}(\vcIdx) \| \vcVal \| \vcRand)$%
                \Comment{Re-compute leaf hash from opening}%
                \label{alg:verify-leaf}
            \For{$b \in \operatorname{ancestors}(\operatorname{leaf}(\vcIdx))$}%
                    \Comment{Re-compute hashes from leaf to root using Merkle path}%
                    \label{alg:verify-traverse}
                \State $b_0 \gets \operatorname{parent}(b)$
                \State $b' \gets \operatorname{sibling}(b)$
                \If{rightmost non-${*}$ bit in $b$ is $0$}%
                    \label{alg:verify-check-bit}
                    \State $h \gets \roH(b_0 \| h \| \vcWit.h_{b'})$%
                        \label{alg:verify-hash-left}
                \Else
                    \State $h \gets \roH(b_0 \| \vcWit.h_{b'} \| h)$%
                        \label{alg:verify-hash-right}
                \EndIf
            \EndFor
            \If{$h = \vcCom$}%
                    \Comment{Check Merkle root}%
                    \label{alg:verify-check}
                \State \Return{$1$}
            \EndIf
            \State \Return{$0$}
        \EndProcedure
    \end{algorithmic}
\end{algorithm}

To obtain the hiding property of \cref{def:vc}, classical (deterministic) Merkle trees~\cite[Sec.~6.6.2]{katzlindellv3} can be augmented with masking randomness at each leaf.
\Cref{alg:hiding-merkle} presents the resulting scheme for vectors of length $\vcLen = 2^d$, for some fixed $d \in \IN$.
\Cref{fig:hiding-merkle} illustrates the construction for $\vcLen = 8$.
Lengths that are not a power of two can be handled via padding.
The construction uses a random oracle $\roH: \{0,1\}^* \to \{0,1\}^\kappa$, where $\kappa$ is the security parameter.

For commitment, $\vcLongCommit$ takes a vector $\vcVals = (\vcVal_1, \ldots, \vcVal_{\vcLen})$ and randomness $\vcRands = (\vcRand_1, \ldots, \vcRand_{\vcLen})$, and constructs a Merkle tree. Each Merkle tree leaf hash incorporates randomness: $h_{\operatorname{leaf}(\vcIdx)} = \roH(\operatorname{leaf}(\vcIdx) \| \vcVal_{\vcIdx} \| \vcRand_{\vcIdx})$ for $\vcIdx \in [\vcLen]$ (\alglocref{alg:hiding-merkle}{alg:merkle-leaf}). The tree builds bottom-up, with hashes at internal nodes computed as $h_b = \roH(b \| h_{b_{\text{L}}} \| h_{b_{\text{R}}})$ from the hashes of their children nodes $h_{b_{\text{L}}}$ and $h_{b_{\text{R}}}$, respectively (\alglocref{alg:hiding-merkle}{alg:merkle-interior}). The commitment is the root hash.
In the example of \cref{fig:hiding-merkle}, the leaf at index $3$ (red-shaded) is computed as $h_{\mathtt{010}} = \roH(\mathtt{010} \| \vcVal_3 \| \vcRand_3)$ (note that $\mathtt{010}$ is the $d=3$-bit binary representation of $3-1=2$). The tree builds upward with internal nodes (blue-shaded) like 
$h_{\mathtt{01{*}}} = \roH(\mathtt{01{*}} \| h_{\mathtt{010}} \| h_{\mathtt{011}})$,
$h_{\mathtt{0{*}{*}}} = \roH(\mathtt{0{*}{*}} \| h_{\mathtt{00{*}}} \| h_{\mathtt{01{*}}})$,
and 
$h_{\mathtt{{*}{*}{*}}} = \roH(\mathtt{{*}{*}{*}} \| h_{\mathtt{0{*}{*}}} \| h_{\mathtt{1{*}{*}}})$ (note the progressive truncation of the binary hash identifiers), 
ultimately producing root $\vcCom = h_{\mathtt{{*}{*}{*}}}$.

For opening, $\vcLongOpen$ returns the standard Merkle path as witness, but the opening includes both value $\vcVal_{\vcIdx}$ and randomness $\vcRand_{\vcIdx}$. The witness $\vcWit$ contains the sibling hashes along the path from leaf to root (\alglocref{alg:hiding-merkle}{alg:open-path}--\alglocref{alg:hiding-merkle}{alg:open-sibling}).
In the example of \cref{fig:hiding-merkle}, opening index $3$ provides, besides $v_3$ and $r_3$, the green-shaded siblings ($h_{\mathtt{011}}$, $h_{\mathtt{00{*}}}$, and $h_{\mathtt{1{*}{*}}}$) of the blue-shaded path from leaf to root ($h_{\mathtt{010}}$, $h_{\mathtt{01{*}}}$, and $h_{\mathtt{0{*}{*}}}$).

For verification, $\vcLongVerify$ receives the claimed opening $(\vcVal, \vcRand)$ as well as the Merkle path as witness $\vcWit$, and recomputes the root. It reconstructs the leaf hash using the opening: $h = \roH(\operatorname{leaf}(\vcIdx) \| \vcVal \| \vcRand)$ (\alglocref{alg:hiding-merkle}{alg:verify-leaf}).
Then follows standard Merkle tree traversal (\alglocref{alg:hiding-merkle}{alg:verify-traverse}--\alglocref{alg:hiding-merkle}{alg:verify-check}), recomputing the hashes from the leaf to the root using the opening and the Merkle path.
In the example of \cref{fig:hiding-merkle},
verification of index $3$ starts by computing $h_{\mathtt{010}}$ from the claimed $(\vcVal_3, \vcRand_3)$, then uses the green-shaded witness to reconstruct the blue-shaded path through $h_{\mathtt{01{*}}}$ and $h_{\mathtt{0{*}{*}}}$ to obtain $h_{\mathtt{{*}{*}{*}}}$ and finally verify against $\vcCom$.

\begin{lemma}
    \label{lem:hiding-merkle-security}
    The construction of \cref{alg:hiding-merkle} satisfies \cref{def:vc}.
\end{lemma}
\deferredproof{uninterestingproofs}{lem:hiding-merkle-security}
\begin{defer}{uninterestingproofs}
\begin{proof}[Proof of \cref{lem:hiding-merkle-security}]
    \begin{itemize}
        \item \textbf{Correctness.}
              By construction, for honestly generated commitments, openings, and witnesses, the verification algorithm
              recomputes the identical hash chain from leaf to root. Specifically, $\vcLongOpen$ extracts the
              authentication path traversed during commitment, and $\vcLongVerify$ reconstructs this path
              deterministically, yielding the same root hash $\vcCom$.

        \item \textbf{Binding.} 
              Assume for contradiction that a PPT adversary $\Adv$ outputs $(\vcVals, \vcVals', \vcRands, \vcRands')$
              with $\vcVals \neq \vcVals'$ or $\vcRands \neq \vcRands'$, but $\vcLongCommit(\vcVals, \vcRands) = \vcLongCommit(\vcVals', \vcRands') = \vcCom$.
              First, observe that both Merkle trees for $(\vcVals, \vcRands)$ and $(\vcVals', \vcRands')$ have the same root hash $\vcCom$.
              Second, since $\vcVals \neq \vcVals'$ or $\vcRands \neq \vcRands'$, there must exist $\vcIdx^* \in [\vcLen]$ where $\vcVal_{\vcIdx^*} \neq \vcVal'_{\vcIdx^*}$ or $\vcRand_{\vcIdx^*} \neq \vcRand'_{\vcIdx^*}$,
              so the leaf hashes at position $\vcIdx^*$ are computed from different inputs,
              $h_{\operatorname{leaf}(\vcIdx^*)} = \roH(\operatorname{leaf}(\vcIdx^*) \| \vcVal_{\vcIdx^*} \| \vcRand_{\vcIdx^*})$
              and
              $h'_{\operatorname{leaf}(\vcIdx^*)} = \roH(\operatorname{leaf}(\vcIdx^*) \| \vcVal'_{\vcIdx^*} \| \vcRand'_{\vcIdx^*})$.
              Third, consider the paths from these leaves to the root in both trees. Since both paths reach the same root $\vcCom$,
              but start from different openings, there must exist some hash node $b$ along these paths where
              the outputs of $\roH$ are the same between the two Merkle trees, $h_b = h'_b$,
              even though the inputs differ.
              Namely, either $b$ is the leaf node itself,
              or $b$ is an interior node with
              $\roH(b \| h_{b_{\text{L}}} \| h_{b_{\text{R}}}) = h_b = h'_b = \roH(b \| h'_{b_{\text{L}}} \| h'_{b_{\text{R}}})$
              even though $b \| h_{b_{\text{L}}} \| h_{b_{\text{R}}} \neq b \| h'_{b_{\text{L}}} \| h'_{b_{\text{R}}}$,
              for $b_{\text{L}}$ and $b_{\text{R}}$ the children of $b$.
              Fourth, in the random oracle model, finding such distinct inputs to $\roH$ that produce the same output 
              occurs with probability at most $q^2/2^\kappa$ (\cf birthday paradox) for an adversary making $q$ queries to $\roH$.
              This probability is negligible in $\kappa$
              since $\Adv$ is PPT and thus $q$ is polynomially bounded in $\kappa$.

        \item \textbf{Position binding.}
              The argument is similar to the binding proof.
              Suppose a PPT adversary $\Adv$ outputs $(\vcCom, \vcIdx, \vcVal, \vcVal', \vcRand, \vcRand', \vcWit, \vcWit')$ with 
              $\vcVal \neq \vcVal'$ or $\vcRand \neq \vcRand'$, where both 
              $\vcLongVerify(\vcCom, \vcIdx, \vcVal, \vcRand, \vcWit) = 1$ and
              $\vcLongVerify(\vcCom, \vcIdx, \vcVal', \vcRand', \vcWit') = 1$.
              First, both verification paths successfully reach the same root $\vcCom$ from position $\vcIdx$.
              Second, since $\vcVal \neq \vcVal'$ or $\vcRand \neq \vcRand'$, the verification algorithm computes leaf hashes
              $h = \roH(\operatorname{leaf}(\vcIdx) \| \vcVal \| \vcRand)$
              and
              $h' = \roH(\operatorname{leaf}(\vcIdx) \| \vcVal' \| \vcRand')$
              from different inputs since $\vcVal \neq \vcVal'$ or $\vcRand \neq \vcRand'$.
              Third, consider the paths from these leaves to the roots in the Merkle trees using witnesses $\vcWit$ and $\vcWit'$.
              Since both paths reach the same root $\vcCom$ but start from different inputs,
              there must exist some node $b$ along these paths where the outputs of $\roH$ are equal, $h_b = h'_b$,
              even though the inputs differ (this holds even if $\vcWit = \vcWit'$).
              Namely, either $b$ is the leaf node itself with different inputs from the openings producing the same hash,
              or $b$ is an interior node where the hash computation yields the same output from different child hashes (even if $\vcWit = \vcWit'$).
              Fourth, in the random oracle model, finding such distinct inputs to $\roH$ that produce the same output
              occurs with probability at most $q^2/2^\kappa$ (\cf birthday paradox) for an adversary making $q$ queries to $\roH$.
              This probability is negligible in $\kappa$ since $\Adv$ is PPT.

        \item \textbf{Hiding.}
              Consider the hiding game where adversary $\Adv$ provides $(\vcVals^{(0)}, \vcVals^{(1)})$.
              The challenger samples $b \getsRandom \{0,1\}$ and $\vcRands \getsRandom \vcRandSet^{\vcLen}$ uniformly at random,
              then gives $\vcCom \gets \vcLongCommit(\vcVals^{(b)}, \vcRands)$ to $\Adv$.
              The adversary may adaptively query openings at indices $\vcIdx_j$ where $\vcVal^{(0)}_{\vcIdx_j} = \vcVal^{(1)}_{\vcIdx_j}$
              and then receives $(\vcVal^{(b)}_{\vcIdx_j}, \vcRand_{\vcIdx_j}, \vcWit_j)$ with $\vcWit_j \gets \vcLongOpen(\vcVals^{(b)}, \vcRands, \vcIdx_j)$.
              To simplify the analysis, we assume that the adversary learns the randomness $\vcRand_{\vcIdx}$ 
              for all indices $\vcIdx$ where $\vcVal^{(0)}_{\vcIdx} = \vcVal^{(1)}_{\vcIdx}$ as well as all the hashes in the Merkle tree.
              We show that the adversary's views for $b=0$ and $b=1$ are indistinguishable, and thus the adversary has no information about $b$ and can only make a random guess $\hat{b}$.

              Since the challenger chooses $\vcRands$ uniformly from $\vcRandSet^{\vcLen}$ where $|\vcRandSet| \geq 2^{\kappa}$,
              and the adversary is PPT (making at most polynomially many queries $q$ to $\roH$),
              the probability that the adversary queries $\roH$ on an input of the form 
              $\operatorname{leaf}(\vcIdx) \| \vcVal^{(b)}_{\vcIdx} \| \vcRand_{\vcIdx}$
              for any position $\vcIdx$ where $\vcVal^{(0)}_{\vcIdx} \neq \vcVal^{(1)}_{\vcIdx}$ 
              is at most $q/2^{\kappa}$, which is negligible.
              Conditioning on this event not occurring, the adversary never queries $\roH$ 
              on the actual inputs used to compute leaf hashes at indices where the vectors differ.
              
              Under this conditioning, the distributions of all Merkle tree hashes are identical for $b=0$ and $b=1$.
              For indices $\vcIdx$ where $\vcVal^{(0)}_{\vcIdx} = \vcVal^{(1)}_{\vcIdx}$,
              the leaf hash $h_{\operatorname{leaf}(\vcIdx)}$ is identical for both values of $b$
              since both the value and randomness are the same.
              For indices $\vcIdx$ where $\vcVal^{(0)}_{\vcIdx} \neq \vcVal^{(1)}_{\vcIdx}$,
              since the adversary has not queried $\roH$ on the corresponding inputs $\operatorname{leaf}(\vcIdx) \| \vcVal^{(0)}_{\vcIdx} \| \vcRand_{\vcIdx}$ and $\operatorname{leaf}(\vcIdx) \| \vcVal^{(1)}_{\vcIdx} \| \vcRand_{\vcIdx}$,
              the random oracle returns uniform and independent hashes $h_{\operatorname{leaf}(\vcIdx)} \getsRandom \{0,1\}^{\kappa}$,
              irrespective of $\vcVal^{(b)}_{\vcIdx}$.
              
              Since the leaf hashes are identically distributed for both values of $b$,
              and the Merkle tree construction deterministically combines leaf hashes to produce hashes at interior nodes,
              all hashes in the tree (including the root) are identically distributed irrespective of $b$.
              
              Thus, in the overwhelmingly probable case where the adversary does not query $\roH$ on the actual inputs used to compute leaf hashes at indices where the vectors differ, even if the adversary knows all hashes of the entire Merkle tree of $\vcVals^{(b)}$ with $\vcRand$, and the randomness $\vcRand_{\vcIdx}$ for all indices $\vcIdx$ where $\vcVal^{(0)}_{\vcIdx} = \vcVal^{(1)}_{\vcIdx}$,
              the missing randomness at indices where the vectors differ prevents the adversary from distinguishing 
              which vector was committed by the challenger.
    \end{itemize}
\end{proof}
\end{defer}

\subsubsection{Hiding Erasure-Correcting Codes}

\begin{algorithm}[tbp]
    \caption{Hiding erasure-correcting code (HECC) instantiation (see \cref{def:hecc-syntax,def:hecc-properties})}
    \label{alg:hecc-impl1}
    \begin{algorithmic}[1]
        \LineComment{Parameters: integers $\heccK, \heccT, \heccN$ with $\heccN \geq \heccK + \heccT$, finite field $\heccF$ with $|\heccF| \geq \heccN$.}
        \LineComment{Fix distinct field elements $\alpha_1, ..., \alpha_{\heccN} \in \heccF$ as \emph{evaluation points}.}
        
        \medskip
        \Procedure{$\heccLongEnc$}{$\heccMsg = (m_1, ..., m_{\heccK}) \in \heccF^{\heccK}, \heccRand = (r_1, ..., r_{\heccT}) \in \heccF^{\heccT}$}
            \State $f(X) \gets \sum_{i=1}^{\heccK} m_i X^{i-1} + \sum_{j=1}^{\heccT} r_j X^{\heccK-1+j}$%
                \Comment{Define polynomial $f(X)$}%
                \label{loc:hecc-impl1-f}
            \For{$i = 1,...,\heccN$}
                \State $\heccShred{i} \leftarrow f(\alpha_i)$%
                    \Comment{Evaluate polynomial $f(X)$ at $\alpha_i$}
            \EndFor
            \State \Return $\heccShreds \gets (\heccShred{1}, ..., \heccShred{\heccN})$%
        \EndProcedure
        
        \medskip
        \Procedure{$\heccLongDec$}{$(\heccShred{i_1}, i_1), ..., (\heccShred{i_{\heccK+\heccT}}, i_{\heccK+\heccT})$}
            \State \Assert{$\forall j \neq k: i_j \neq i_k$}%
                \Comment{Decoding requires indices to be distinct}
            \State $\hat{f}(X) \gets \sum_{j=1}^{\heccK+\heccT} \heccShred{i_j} \prod_{\substack{k=1\\k \neq j}}^{\heccK+\heccT} \frac{X - \alpha_{i_k}}{\alpha_{i_j} - \alpha_{i_k}}$%
                \Comment{Compute polynomial $\hat{f}(X)$ of degree at most $\heccK + \heccT - 1$ using Lagrange interpolation of $((\alpha_{i_1}, \heccShred{i_1}), ..., (\alpha_{i_{\heccK+\heccT}}, \heccShred{i_{\heccK+\heccT}}))$}%
                \label{loc:hecc-impl1-hatf}
            \For{$i = 1,...,\heccK$}
                \State $\hat{m}_i \gets \text{coefficient of $X^{i-1}$ in $\hat{f}(X)$}$%
                    \Comment{Extract message coefficients}
            \EndFor
            \For{$j = 1,...,\heccT$}
                \State $\hat{r}_j \gets \text{coefficient of $X^{\heccK-1+j}$ in $\hat{f}(X)$}$%
                    \Comment{Extract randomness coefficients}
            \EndFor
            \State \Return $(\hat{\heccMsg}, \hat{\heccRand}) \gets ((\hat{m}_1, ..., \hat{m}_{\heccK}), (\hat{r}_1, ..., \hat{r}_{\heccT}))$
        \EndProcedure
    \end{algorithmic}
\end{algorithm}

Leveraging the similarity between Reed--Solomon erasure-correcting codes~\cite{reedsolomoncodes} and secret sharing~\cite{shamirsecretsharing,Blakley1979,ChorGMA85,FranklinY92}, allows us to implement the following \emph{hiding erasure-correcting code} (HECC) primitive, where a \emph{message} is encoded into \emph{shreds} in such a way that (i)~``few'' shreds do not reveal any information about the encoded message, while (ii)~from ``enough''/``many'' shreds, the message can reliably be reconstructed:
\begin{definition}[HECC Syntax]
    \label{def:hecc-syntax}
    For non-negative integers $\heccN, \heccK, \heccT$
    with $\heccN \geq \heccK + \heccT$,
    and a finite field $\heccF$ with $|\heccF| \geq \heccN$,
    a hiding erasure-correcting code $\heccLong$ provides two algorithms:
    \begin{itemize}
        \item
              \textbf{Encoding $\heccLongEnc\colon (\heccMsg \in \heccF^\heccK, \heccRand \in \heccF^\heccT) \to (\heccShreds = (\heccShred{1}, ..., \heccShred{\heccN}) \in \heccF^\heccN)$.}
              Given a \emph{message} $\heccMsg$, represented as a vector of $\heccK$ field elements, and \emph{randomness} $\heccRand$, represented as a vector of $\heccT$ field elements, as inputs, $\heccLongEnc$ deterministically outputs a vector $\heccShreds$ of $\heccN$ field elements $\heccShred{1}, ..., \heccShred{\heccN} \in \heccF$, called \emph{shreds}.

        \item
              \textbf{Decoding $\heccLongDec\colon ((\heccShred{i_j}, i_j)_{j=1}^{\heccK+\heccT} \in (\heccF \times \IN)^{\heccK+\heccT}) \to (\hat\heccMsg \in \heccF^\heccK, \hat\heccRand \in \heccF^\heccT)$.}
              Given a set of $\heccK+\heccT$ pairs of shreds $\heccShred{i_j}$ and their corresponding indices $i_j$ as inputs, $\heccLongDec$ outputs a message $\hat\heccMsg$ and randomness $\hat\heccRand$.
    \end{itemize}
\end{definition}
\begin{definition}[HECC Properties]
    \label{def:hecc-properties}
    A proper HECC satisfies the following two properties:
    \begin{itemize}
        \item
              \textbf{Erasure-correction.}
              For every message $\heccMsg \in \heccF^\heccK$ and every randomness $\heccRand \in \heccF^\heccT$,
              let $\heccShreds = (\heccShred{1}, ..., \heccShred{\heccN}) \gets \heccLongEnc(\heccMsg, \heccRand)$.
              Then, for any set of $\heccK+\heccT$ pairs $(\heccShred{i_j}, i_j)$
              where the indices are distinct,
              $\heccLongDec((\heccShred{i_j}, i_j)_{j=1}^{\heccK+\heccT}) = (\heccMsg, \heccRand)$.

        \item
              \textbf{Hiding.}
              For every two messages $\heccMsg, \heccMsg' \in \heccF^\heccK$,
              let $\heccRand, \heccRand' \getsRandom \heccF^\heccT$,
              and let
              $\heccShreds = (\heccShred{1}, ..., \heccShred{\heccN}) \gets \heccLongEnc(\heccMsg, \heccRand)$,
              and $\heccShreds' = (\heccShred{1}', ..., \heccShred{\heccN}') \gets \heccLongEnc(\heccMsg', \heccRand')$.
              Then, for any set of $\heccT$ distinct indices $i_j$,
              $(\heccShred{i_1}, ..., \heccShred{i_{\heccT}})$ and $(\heccShred{i_1}', ..., \heccShred{i_{\heccT}}')$ are identically distributed.
              This implies that an adversary with at most $\heccT$ shreds cannot learn any information about the underlying message.
    \end{itemize}
\end{definition}

Specifically, we can construct an HECC from a traditional maximum-distance separable (MDS) $(\heccK+\heccT, \heccN$)  erasure-correcting code such as Reed--Solomon,
by sampling $\heccT$ random field elements and appending them to the original message $\heccMsg \in \heccF^\heccK$ before applying the code's encoding.
This construction is carried out explicitly, with Reed--Solomon codes, in \cref{alg:hecc-impl1}.

\begin{lemma}
    \label{lem:hecc-impl1-hecc}
    \Cref{alg:hecc-impl1} provides a HECC as per \cref{def:hecc-syntax,def:hecc-properties}.
\end{lemma}
\deferredproof{uninterestingproofs}{lem:hecc-impl1-hecc}
\begin{defer}{uninterestingproofs}
\begin{proof}[Proof of \cref{lem:hecc-impl1-hecc}]
    \begin{itemize}
        \item \textbf{Erasure-correction.}
              The polynomial $f(X)$ of \alglocref{alg:hecc-impl1}{loc:hecc-impl1-f} has degree at most $\heccK + \heccT - 1$, so it is uniquely determined by evaluations at any $\heccK + \heccT$ distinct points, and thus for $\hat{f}(X)$ of \alglocref{alg:hecc-impl1}{loc:hecc-impl1-hatf}, $f = \hat{f}$.
              Therefore, both $\hat\heccMsg = \heccMsg$ (from coefficients of $X^0, ..., X^{\heccK-1}$ in $\hat{f}$) and $\hat\heccRand = \heccRand$ (from coefficients of $X^{\heccK}, ..., X^{\heccK+\heccT-1}$ in $\hat{f}$).

        \item \textbf{Hiding.}
              We show that for any fixed message $\heccMsg$, any $\heccT$ evaluations (at distinct points) of the polynomial $f(X)$ of \alglocref{alg:hecc-impl1}{loc:hecc-impl1-f} 
              are distributed independently and uniformly random over $\heccF$, if $\heccRand$ is sampled randomly.
              This implies that shreds, at any $\heccT$ distinct indices, of two messages $\heccMsg$ and $\heccMsg'$ are identically distributed for independently random $\heccRand, \heccRand'$, and thus establishes hiding.

              So consider any fixed message $\heccMsg = (m_1, ..., m_{\heccK})$ and set of $\heccT$ distinct indices $i_j$.
              Then,
              \[
                  \forall j \in [\heccT]:
                  \quad
                  \heccShred{i_j} = f(\alpha_{i_j}) = \sum_{k=1}^{\heccK} m_k \alpha_{i_j}^{k-1} + \sum_{j=1}^{\heccT} r_j \alpha_{i_j}^{\heccK-1+j}.
              \]
              Thus, we can write the shreds at the distinct indices $i_j$ as a sum of two matrix-vector products:
              \[
                  \begin{bmatrix}
                      \heccShred{i_1} \\ \vdots \\ \heccShred{i_{\heccT}}
                  \end{bmatrix}
                  =
                  \begin{bmatrix}
                      \alpha_{i_1}^0        & ...    & \alpha_{i_1}^{\heccK-1}        \\
                      \vdots                & \ddots & \vdots                         \\
                      \alpha_{i_{\heccT}}^0 & ...    & \alpha_{i_{\heccT}}^{\heccK-1}
                  \end{bmatrix}
                  \begin{bmatrix}
                      m_1 \\ \vdots \\ m_{\heccK}
                  \end{bmatrix}
                  +
                  \underbrace{
                      \begin{bmatrix}
                          \alpha_{i_1}^{\heccK}        & ...    & \alpha_{i_1}^{\heccK+\heccT-1}        \\
                          \vdots                       & \ddots & \vdots                                \\
                          \alpha_{i_{\heccT}}^{\heccK} & ...    & \alpha_{i_{\heccT}}^{\heccK+\heccT-1}
                      \end{bmatrix}
                  }_{\triangleq M_{\mathrm{masking}}}
                  \begin{bmatrix}
                      r_1 \\ \vdots \\ r_{\heccT}
                  \end{bmatrix}.
              \]
              Since $M_{\mathrm{masking}}$ is a $\heccT \times \heccT$ Vandermonde matrix, it is full-rank over $\heccF$.
              Furthermore, $r_1, ..., r_{\heccT}$ are sampled independently and uniformly random over $\heccF$.
              As a result, 
              the term $M_{\mathrm{masking}} \heccRand$ is uniformly random over $\heccF^\heccT$,
              and
              the shreds $(\heccShred{i_1}, ..., \heccShred{i_{\heccT}})$ are distributed independently and uniformly random over $\heccF$, \emph{independent of the message $\heccMsg$}.
              Thus, at most $\heccT$ shreds provide no information about the message $\heccMsg$.
    \end{itemize}
\end{proof}
\end{defer}
It is clear that the HECC of \cref{alg:hecc-impl1} can trivially be ``vectorized''/``batched'' (see \cite[Fig.~5]{semiavidpr}), in the spirit of interleaved Reed--Solomon codes~\cite{interleavedrscodes}, to accommodate long messages.

\subsubsection{Combining Vector Commitments and Hiding Erasure-Correcting Codes}

Our protocol requires vector commitments with $\vcValSet = \vcRandSet = \IF$ and $\vcLen = \heccN$ that,
even if the masking randomness is chosen as a length-$\heccN$ HECC encoding of a uniformly random length-$\heccK$ message with uniformly random length-$\heccT$ randomness (both in $\IF$, and with $\heccK, \heccT > 0$),
remains hiding to any PPT adversary that knows at most $\heccT$ openings of the vector commitment.
We show that this is indeed the case for \cref{alg:hiding-merkle,alg:hecc-impl1}.

\begin{lemma}
    \label{lem:hecc-impl1-KTuniform}
    Consider \cref{alg:hecc-impl1}.
    Let $\heccMsg \getsRandom \heccF^\heccK$,
    $\heccRand \getsRandom \heccF^\heccT$,
    $\heccShreds = (\heccShred{1}, ..., \heccShred{\heccN}) \gets \heccLongEnc(\heccMsg, \heccRand)$.
    Then, for any set of $\heccK+\heccT$ distinct indices $i_j$,
    $(\heccShred{i_1}, ..., \heccShred{i_{\heccK+\heccT}})$ is uniformly random over $\heccF^{\heccK+\heccT}$.
\end{lemma}
\deferredproof{uninterestingproofs}{lem:hecc-impl1-KTuniform}
\begin{defer}{uninterestingproofs}
\begin{proof}[Proof of \cref{lem:hecc-impl1-KTuniform}]
    The proof is analogous to the proof of the hiding property for \cref{lem:hecc-impl1-hecc}.
    For any set of $\heccK+\heccT$ distinct indices $i_j$, we can write the shreds encoded according to \cref{alg:hecc-impl1} as:
    \[
        \begin{bmatrix}
            \heccShred{i_1} \\ 
            \vdots \\ 
            \heccShred{i_{\heccK+\heccT}}
        \end{bmatrix}
        =
        \underbrace{
            \begin{bmatrix}
                \alpha_{i_1}^0        & \cdots & \alpha_{i_1}^{\heccK+\heccT-1}        \\
                \vdots                & \ddots & \vdots                                \\
                \alpha_{i_{\heccK+\heccT}}^0 & \cdots & \alpha_{i_{\heccK+\heccT}}^{\heccK+\heccT-1}
            \end{bmatrix}
        }_{\triangleq M_{\mathrm{encoding}}}
        \begin{bmatrix}
            m_1 \\ 
            \vdots \\ 
            m_{\heccK} \\ 
            r_1 \\ 
            \vdots \\ 
            r_{\heccT}
        \end{bmatrix}
    \]
    The matrix $M_{\mathrm{encoding}}$ is a $(\heccK+\heccT) \times (\heccK+\heccT)$ Vandermonde matrix with distinct $\alpha_{i_j}$, hence invertible over $\heccF$.
    Since $\heccMsg \getsRandom \heccF^{\heccK}$ and $\heccRand \getsRandom \heccF^{\heccT}$ are sampled independently and uniformly at random, the concatenated vector $(m_1, ..., m_{\heccK}, r_1, ..., r_{\heccT})$ is uniformly random over $\heccF^{\heccK+\heccT}$.
    Because $M_{\mathrm{encoding}}$ is an invertible linear transformation and the input vector is uniformly random over $\heccF^{\heccK+\heccT}$, the output vector $(\heccShred{i_1}, ..., \heccShred{i_{\heccK+\heccT}})$ is also uniformly random over $\heccF^{\heccK+\heccT}$.
\end{proof}
\end{defer}

\begin{lemma}
    \label{lem:hiding-merkle-hecc}
    Consider \cref{alg:hecc-impl1} 
    with $\heccK, \heccT > 0$,
    and \cref{alg:hiding-merkle} with $\vcLen = \heccN$, $\vcValSet = \vcRandSet = \IF$.
    Every PPT adversary $\Adv$ succeeds in the following game with probability negligible in $\kappa$:
    The adversary $\Adv$ provides $(\vcVals^{(0)}, \vcVals^{(1)})$.
    The challenger flips an unbiased coin $b \getsRandom \{0,1\}$, samples $\heccMsg' \getsRandom \IF^\heccK$ and $\heccRand' \getsRandom \IF^\heccT$,
    computes $\vcRands = (\vcRand_1, ..., \vcRand_{\heccN}) \gets \heccLongEnc(\heccMsg', \heccRand')$,
    and gives $\vcCom \gets \vcLongCommit(\vcVals^{(b)}, \vcRands)$ to $\Adv$.
    The adversary $\Adv$ may make up to $\heccT$ queries $\vcIdx_j$ for distinct indices where $\vcVal^{(0)}_{\vcIdx_j} = \vcVal^{(1)}_{\vcIdx_j}$ and receives openings $(\vcRand^{(j)}, \vcWit^{(j)})$ where $\vcRand^{(j)} \gets \vcRand_{\vcIdx_j}$ and $\vcWit^{(j)} \gets \vcLongOpen(\vcVals^{(b)}, \vcRands, \vcIdx_j)$.
    Finally, $\Adv$ outputs a guess $\hat{b} \in \{0,1\}$.
    The adversary succeeds iff $\hat{b} = b$.
\end{lemma}
\deferredproof{uninterestingproofs}{lem:hiding-merkle-hecc}
\begin{defer}{uninterestingproofs}
\begin{proof}[Proof of \cref{lem:hiding-merkle-hecc}]
    The proof follows a similar approach as the proof of the hiding property for \cref{lem:hiding-merkle-security}.
    Consider the hiding game where adversary $\Adv$ provides $(\vcVals^{(0)}, \vcVals^{(1)})$.
    The challenger samples $b \getsRandom \{0,1\}$, $\heccMsg' \getsRandom \IF^\heccK$ and $\heccRand' \getsRandom \IF^\heccT$,
    computes $\vcRands = (\vcRand_1, ..., \vcRand_{\heccN}) \gets \heccLongEnc(\heccMsg', \heccRand')$,
    and gives $\vcCom \gets \vcLongCommit(\vcVals^{(b)}, \vcRands)$ to $\Adv$.
    The adversary may query openings at up to $\heccT$ distinct indices $\vcIdx_j$ where $\vcVal^{(0)}_{\vcIdx_j} = \vcVal^{(1)}_{\vcIdx_j}$
    and receives $(\vcRand_{\vcIdx_j}, \vcWit_j)$ with $\vcWit_j \gets \vcLongOpen(\vcVals^{(b)}, \vcRands, \vcIdx_j)$.
    
    Let $Q$ denote the set of indices queried by the adversary for openings, with $|Q| \leq \heccT$.
    So the adversary learns $(\vcRand_{\vcIdx}, \vcWit_\vcIdx)$ with $\vcWit_\vcIdx \gets \vcLongOpen(\vcVals^{(b)}, \vcRands, \vcIdx)$
    for all $\vcIdx \in Q$.
    Without loss of generality, assume 
    that the adversary learns all hashes of the Merkle tree 
    computed as part of $\vcLongCommit(\vcVals^{(b)}, \vcRands)$.
    We show that the adversary's views for $b=0$ and $b=1$ are indistinguishable.
    
    First, observe that the adversary learns $|Q|$ shreds $\{\vcRand_{\vcIdx} \mid \vcIdx \in Q\}$ from the HECC encoding
    $\vcRands = (\vcRand_1, ..., \vcRand_{\heccN}) \gets \heccLongEnc(\heccMsg', \heccRand')$.
    By \cref{lem:hecc-impl1-KTuniform}, since $\heccMsg'$ and $\heccRand'$ are uniformly random,
    any $\heccK+\heccT$ shreds are jointly uniform over $\IF^{\heccK+\heccT}$.
    Since $\heccK > 0$ and $|Q| \leq \heccT$, this implies that any $|Q|+1$ shreds are jointly uniform.
    Therefore, for any index $\vcIdx \notin Q$, the value $\vcRand_{\vcIdx}$ is uniformly distributed over $\IF$
    even conditioned on the observed shreds $\{\vcRand_{i} \mid i \in Q\}$.
    
    Now consider any index $\vcIdx$ where $\vcVal^{(0)}_{\vcIdx} \neq \vcVal^{(1)}_{\vcIdx}$.
    Since the adversary can only query indices for opening where $\vcVal^{(0)}_{\vcIdx} = \vcVal^{(1)}_{\vcIdx}$, we have $\vcIdx \notin Q$.
    By the above, $\vcRand_{\vcIdx}$ is uniform over $\IF$ from the adversary's perspective \emph{a priori}.
    The leaf hash $h_{\operatorname{leaf}(\vcIdx)}$ at index $\vcIdx$ is computed as $\roH(\operatorname{leaf}(\vcIdx) \| \vcVal^{(b)}_{\vcIdx} \| \vcRand_{\vcIdx})$.
    Since the adversary is PPT, it can make at most $q = \poly(\kappa)$ queries to the random oracle $\roH$.
    Since $|\IF| \geq 2^\kappa$, the probability of the event (denote the event by $E$) that the adversary ever queries $\roH$ on $\operatorname{leaf}(\vcIdx) \| \vcVal^{(b)}_{\vcIdx} \| \vcRand_{\vcIdx}$ for any index $\vcIdx$ where $\vcVal^{(0)}_{\vcIdx} \neq \vcVal^{(1)}_{\vcIdx}$
    is at most $q/|\IF| \leq q/2^{\kappa} = \negl(\kappa)$.

    Conditioned on $\lnot E$, the distribution of all Merkle tree hashes is identical for $b=0$ and $b=1$:
    At indices $\vcIdx$ where $\vcVal^{(0)}_{\vcIdx} = \vcVal^{(1)}_{\vcIdx}$, the leaf hash $h_{\operatorname{leaf}(\vcIdx)}$ is identical for both values of $b$
    (same value and randomness).
    At indices $\vcIdx$ where $\vcVal^{(0)}_{\vcIdx} \neq \vcVal^{(1)}_{\vcIdx}$, the adversary cannot learn $\vcRand_{\vcIdx}$ through opening, and since the adversary cannot have queried $\roH(\operatorname{leaf}(\vcIdx) \| \vcVal^{(b)}_{\vcIdx} \| \vcRand_{\vcIdx})$,
    the leaf hash $h_{\operatorname{leaf}(\vcIdx)}$ is uniform and independent, regardless of $b$.
    Since interior hashes are computed deterministically from leaf hashes,
    all Merkle tree hashes including $\vcCom$ have identical distribution for both values of $b$.
\end{proof}
\end{defer}

\subsection{Setting, Adversary, Network}
\label{sec:model-model}

For ease of exposition, we consider a standard \emph{permissioned} setting~\cite{lewispyeroughgardenpermissionless}
with a computationally bounded adversary and a \emph{partially synchronous} network~\cite{model-psync}, which we recall below.

\paragraph{Setting}
We consider a setting with $n$ \emph{nodes}, denoted by a set $\CPah$.
Each node has a cryptographic identity that is known to all nodes.
\emph{Time} in the considered environment proceeds continuously, indexed by $t$, and nodes are assumed to have synchronized clocks.
In contrast to this notion of ``real time'' provided by the environment,
the protocols we consider proceed in ``logical steps'' called \emph{slots}, indexed by $s$.
The environment has two more features which we detail below:
a PPT \emph{adversary} that attempts to subvert the proper functioning of the protocol,
and a \emph{network} functionality, with which nodes can send each other messages as instructed by the protocol.

\paragraph{Adversary}
Part of every execution is a PPT adversary that attempts to disturb the proper functioning of the protocol (specifically, it seeks to prevent the protocol from guaranteeing the consensus security properties outlined in \cref{sec:model-consensus-basics}).
For this purpose, for ease of exposition, we assume that the adversary chooses a set $\CPa \subseteq \CPah$ of $f \triangleq |\CPa|$ nodes to \emph{corrupt} at the beginning of the execution before any of the protocol's randomness is drawn (\emph{static corruption}).
The adversary can cause these so-called \emph{Byzantine} nodes to deviate arbitrarily from the protocol for the entire execution.
Recall that in all its actions, the adversary is computationally bounded to be probabilistic polynomial-time (PPT).
Non-adversary \emph{honest} nodes, denoted by $\CPh \triangleq \CPah \setminus \CPa$, follow the protocol as specified.
The adversary also determines the delays incurred by messages sent among honest nodes, as detailed next.

\paragraph{Network}
Nodes can send messages to each other via a fully-connected network of private point-to-point links.
The network is \emph{partially synchronous}~\cite{model-psync},
meaning there is a delay upper-bound~$\Delta$, known to all nodes,
and a ``global stabilization time'' $\GST$, chosen adaptively by the adversary during execution,
such that before $\GST$, the adversary can cause arbitrary message delays,
while after $\GST$, the message delay is adversarial but at most $\Delta$.
In particular, this means that every message sent by an honest node by $t$ is delivered to
all honest nodes by $\max(t, \GST) + \Delta$.

\subsection{Consensus Basics}
\label{sec:model-consensus-basics}

\subsubsection{Safety and Liveness of Atomic Broadcast}
\label{sec:model-consensus-basics-ab}

We are interested in constructing protocols that solve the \emph{atomic broadcast} variant of \emph{consensus}, where, continuously over time, nodes are input \emph{payloads} by the environment, and they output sequences of payloads considered \emph{confirmed}, in so-called \emph{logs}.
Intuitively, the consensus protocol should ensure that
(i)~logs output by honest nodes are \emph{consistent} across nodes and across time (\emph{safety}),
and (ii)~every input payload eventually makes it into the log of every honest node (\emph{liveness}).
More formally, if we denote by $L_t^p$ the log output by node $p$ at time $t$, 
and if $L \preceq L'$ denotes that the sequence $L$ is a prefix of (or equal to) the sequence $L'$,
then we require:
\begin{definition}[Safety]
    \label{def:safety}
    For all honest nodes $p,p'$ and times $t,t'$,
    we have $L_t^p \consistent L_{t'}^{p'}$,
    where $L_t^p \consistent L_{t'}^{p'}$ denotes
    that either $L_t^p \preceq L_{t'}^{p'}$ or $L_{t'}^{p'} \preceq L_t^p$.
\end{definition}
\begin{definition}[Liveness]
    \label{def:liveness}
    For every payload $\tx$ input to an honest node, there exists a time~$t_0$, such that for all times $t \geq t_0$, for every honest node $p$, we have $\tx \in L_{t}^p$.
\end{definition}
\begin{definition}[Security]
    \label{def:security}
    An atomic broadcast protocol is \emph{secure}
    with \emph{resilience} $\tau$
    iff in every partially synchronous execution with $f/n \leq \tau$,
    except with probability negligible in some cryptographic security parameter $\kappa$,
    the protocol is safe and live.
\end{definition}
Note that safety implies that the logs output by honest nodes are all prefixes of one ``ground truth'' log.
It is therefore meaningful to speak of \emph{the log} output by a protocol.
Specifically, for a safe protocol producing output logs $L_t^p$, we denote by $\TheLOG$ ``the log of the protocol'', where $\TheLOG$ is the shortest log with the property that $\forall p, t: L_t^p \preceq \TheLOG$.

\subsubsection{Slightly Enriched Interface}
\label{sec:model-consensus-basics-enriched}

Our protocol will be constructed using, as an opaque-box, an off-the-shelf partially synchronous consensus protocol $\PIsmr$.
For both our protocol and for $\PIsmr$, we assume an interface that is slightly enriched over the vanilla atomic broadcast primitive recalled above,
namely in the following way.
We assume that the protocol internally proceeds in ``logical steps'' called \emph{slots} (terminology borrowed from Alpenglow~\cite{alpenglow}).
Each slot has an associated \emph{leader} node $\Leader{s}$ that gets to propose a payload for inclusion in the log,
and each slot has an associated fixed \emph{slot proposal time} $\SlotTime{s}$,
where the leader is expected to produce and broadcast its proposal.
Payloads confirmed in the log are annotated with a slot number (the payload is said to be \emph{confirmed at that slot}) (similar to~\cite[Def.~9]{timelinesssupersafety}).
If an honest leader proposes a payload that is subsequently confirmed,
this payload will be annotated with the slot in which it was proposed.
In any case, honest nodes agree on the slot numbers of confirmed payloads, and the slot numbers are non-decreasing along the log.

Slightly more formally, we consider consensus protocols $\Pi$ with the following interface:
\begin{itemize}
    \item Protocol $\Pi$ is parameterized by a leader sequence $\Leader{s}$.
    \item Protocol $\Pi$ comes with a fixed slot proposal time sequence $\SlotTime{s}$.
    \item By time $\SlotTime{s}$, leader $\Leader{s}$ of slot $s$ invokes $\Pi.\smrInput(s, \txs)$ to propose payload $\txs$ in slot $s$.
    \item As payloads get confirmed, the protocol emits events $\Pi.\smrOutput(s, \txs)$, consecutively for $s=1,2,...$.
    Here, $\txs$ is the payload confirmed at slot $s$, or $\bot$ (if no payload is confirmed at slot $s$, for instance because $\Leader{s}$ did not invoke $\Pi.\smrInput(s, \txs)$ in time, or before $\GST$).
\end{itemize}

It is important to highlight that the above is primarily a \emph{syntactic} variation of consensus that will subsequently be convenient,
rather than a change to the consensus security \emph{semantics}.
(Virtually) all partially synchronous candidate protocols for $\PIsmr$ we are aware of are easily modified to satisfy this refined interface.
They already have a corresponding notion of slots:
PBFT~\cite{pbft}, Tendermint~\cite{tendermint}, or HotStuff~\cite{hotstuff} calls them \emph{views},
Simplex~\cite{simplex} calls them \emph{heights}, Streamlet~\cite{streamlet} calls them \emph{epochs}.
They are easily modified to have a fixed regular cadence of slot proposal times, for instance every~$2\Delta$ time,
and to annotate confirmed payloads with the slot number in which they were proposed (over which these protocols also establish consensus).

For any log $L$ annotated with slot numbers, we denote by $L[s]$ the payloads in $L$ confirmed \emph{at slot $s$}, and we denote by $L[:s]$ the prefix of $L$ of payloads confirmed \emph{at any slot $s' \leq s$}.

Most partially synchronous candidate protocols for $\PIsmr$, like PBFT, Tendermint, (non-chained) HotStuff, or Simplex, readily satisfy the following liveness guarantee:
\begin{definition}[Leader-Driven Liveness]
    \label{def:honleader-liveness}
    There exists $c\geq0$ such that 
    for every slot $s$ with $\SlotTime{s} \geq \GST + c\Delta$,
    if $\Leader{s} \in \CPh$ and $\Leader{s}$ invokes $\Pi.\smrInput(s, \txs)$ no later than $\SlotTime{s}$, then $\txs \subseteq \TheLOG[:s]$.
\end{definition}

\subsection{Economic Consensus Properties}
\label{sec:model-consensus-econ}

To achieve the goals outlined in \cref{sec:introduction}, we require, in addition to traditional safety and liveness, two new \emph{economically-motivated} consensus properties that we define below.
First, we need a stronger liveness-like notion, called \emph{selective-censorship resistance}, that guarantees that transactions are not just \emph{eventually} confirmed, but are confirmed at the next possible opportunity.
Second, we need a privacy-like notion, called \emph{hiding}, that guarantees that the adversary cannot learn about pending transactions before it is ``too late'' for the adversary to act on the information in the transactions.

\subsubsection{Selective-Censorship Resistance}

Intuitively, the leader $\Leader{s}$ drives consensus during slot $s$.
If $\Leader{s}$ does not follow the protocol---for instance, if $\Leader{s}$ crashes during the slot---it becomes hard to satisfy that transactions input by $\SlotTime{s}$ are confirmed at slot $s$.
We thus only require that \emph{either} all transactions input to honest nodes by $\SlotTime{s}$ are confirmed at slot $s$, \emph{or} \emph{no} transactions are confirmed at slot $s$ at all.

\begin{definition}[Selective-Censorship Resistance]
    \label{def:scr}
    A safe consensus protocol
    with output logs $L_t^p$
    is \emph{selective-censorship resistant} iff
    for every slot $s$ with $\SlotTime{s} \geq \GST$,
    it is the case that
    either
    $\TheLOG[s] = \emptyset$,
    or
    for every transaction $\tx$ input to some honest node by some time $t \leq \SlotTime{s}$,
    $\tx \in \TheLOG[:s]$ (or both).
\end{definition}

This means that in order for the adversary to censor \emph{any} transactions in a given slot, the adversary has to censor \emph{all} transactions for that slot.
\Cref{def:scr} is similar to the notion of \emph{short-term censorship resistance} defined in~\cite{Alpos_CRoverview}.

\subsubsection{Hiding}

In addition to selective-censorship resistance,
we require that the adversary cannot learn information about pending transactions input at honest nodes, before the transactions' eventual inclusion (and positions) in the protocol's log is decided.
To make this notion precise, we build on the well-known concept of \emph{valency}~\cite{attiyaimpossibilities,flpimpossibility}.
The valency of a consensus protocol at a particular point in the protocol's execution
captures consensus decisions that are inevitable (no matter the adversary's future actions), given the state of honest nodes and of the environment at that point in the execution.
Importantly,
the valency captures not only consensus decisions that honest nodes have recognized (and externalized in their output logs) yet,
but also comprises consensus decisions that have been ``effectively made'' by the protocol but not yet detected by honest nodes.
\begin{definition}[Valency]
    \label{def:valency}
    The \emph{valency} $\ValencyLOG{t}$ of a consensus protocol
    at a particular point $t$ in an execution
    is the longest annotated log $L$ such that,
    for all adversarial strategies for the continuation of the execution,
    it holds that $L \preceq \TheLOG$.
\end{definition}

A consensus protocol is hiding if the adversary cannot learn any information about honest nodes' input transactions before the transactions enter the valency of the protocol.
\begin{definition}[Hiding]
    \label{def:hiding}
    A consensus protocol is \emph{hiding} iff
    every PPT adversary $\Adv$ has no more than a negligible advantage over a random guess in winning the following game:
    The adversary $\Adv$ selects an honest node $p$ and two transactions $\tx_0, \tx_1$.
    The challenger flips an unbiased local coin $b \getsRandom \{0,1\}$,
    and provides node $p$ with input $\tx_b$.
    The protocol's execution commences.
    Let $t^*$ be the first time such that $\tx_b \in \ValencyLOG{t^*}$,
    where $\ValencyLOG{t^*}$ denotes the valency of the protocol at time $t^*$.
    At this point $t^*$, the adversary $\Adv$ is asked to produce $\hat{b}$,
    and $\Adv$ wins the game iff $\hat{b} = b$.
\end{definition}

\subsubsection{Practical Considerations of Selective-Censorship Resistance and Hiding}

To discuss practical considerations, it is important to recognize how the abstraction used for the formal parts of this paper matches with the intended application setting.
In a blockchain like Solana or Ethereum, users submit transactions to infrastructure providers such as RPCs or block builders.
Our abstraction does not capture users or RPCs/block builders.
Instead, it treats nodes as both the entities where transactions are input, and the entities that run the consensus protocol.

As a result, \cref{def:hiding} differs slightly from the hiding notions used for instance in encrypted mempools, in a practically relevant way:
Encrypted mempools guarantee to users the ``privacy'' of their transactions from all nodes. 
The result is inefficient use of the scarce resource ``block space'',
because nodes cannot filter spam or ensure that transactions are able to pay fees.
In contrast, \cref{def:hiding} does not guarantee ``privacy'' to users, it only guarantees that nodes cannot learn the private transaction flow of other nodes before the corresponding transactions are confirmed.
This allows RPCs/block builders to select high-value transactions and allocate block space efficiently.

In practice already today, users effectively enjoy ``privacy'' due to the trust relationships and repeated-game dynamics with their corresponding RPCs/block builders (this is not captured in our model).
An advantage of a hiding protocol
is that there are multiple nodes a user could route its transaction flow to.
A user can ensure to only employ nodes that respect the user's privacy.

Note that a protocol being hiding without also being selective-censorship resistant does not help with the economic properties we desire.
If a protocol is not selective-censorship resistant,
an adversary can selectively censor all transactions that are not input to the adversary.
Thereby, the adversary effectively coerces users to submit transactions through the adversary, so that de-facto all input transactions are revealed to the adversary (undermining the hiding property).

\section{Multiple Concurrent Proposers Protocol}
\label{sec:protocol}

\subsection{Philosophy \& Overview}
\label{sec:protocol-philosophy}

\begin{figure}[tbp]
    \centering
    \begin{tikzpicture}[%
        x=2cm,
        y=-2cm,
        entity/.style={font=\footnotesize,anchor=center,align=center,inner sep=1pt},
        arrow/.style={-latex},
        arrow_shred/.style={arrow,myA16zTeal},
        arrow_attestation/.style={arrow,myA16zAquamarine},
        arrow_block/.style={arrow,myA16zMagenta},
        entity_box/.style={draw=none,fill=gray!5},
        entity_box_tl/.style={yshift=1em,xshift=-1em},
        entity_box_br/.style={yshift=-1em,xshift=1em},
        phase_label/.style={font=\footnotesize,align=left,anchor=west,xshift=-7cm},
        role_label/.style={font=\footnotesize,align=left,anchor=west,xshift=5cm},
    ]

        \def\inputY{0.25}
        \def\inputList{1/-2,2/-1,3/0,4/1,5/2}

        \def\proposerY{1}
        \def\proposerList{1/-1.5,2/-0.5,3/0.5,4/1.5}

        \def\relayY{2}
        \def\relayList{1/-1.5,2/-0.5,3/0.5,4/1.5}

        \def\consensusY{3}

        \def\reconstructY{4}
        \def\reconstructList{1/-1.5,2/-0.5,3/0.5,4/1.5}

        \def\userList{1/-2/50/0.6,2/-1/50/0.5,3/0/50/0.5,4/1/50/0.5,5/2/90/0.5}
        \def\userY{5}

        \def\outputY{5.6}

        \pgfdeclarelayer{background}
        \pgfdeclarelayer{foreground}
        \pgfsetlayers{background,main,foreground}

        \foreach \i/\x in \inputList {
            \node [entity] (tx\i) at (\x,\inputY) {\tangoicon[height=1.3em]{mimetypes_text-x-generic} $\tx_{\i}$};
        }

        \foreach \i/\x in \proposerList {
            \node [entity] (p\i) at (\x,\proposerY) {\tangoicon[height=1.3em]{devices_computer} $p_{\i}$};
        }

        \foreach \i/\x in \relayList {
            \node [entity] (r1\i) at (\x,\relayY) {\tangoicon[height=1.3em]{places_network-server} $r_{\i}$};
        }

        \node [entity] (l) at (-0.5,\consensusY-0.25) {\tangoicon[height=1.3em]{emblem_person_red} $\Leader{s}$};

        \node [entity,minimum width=8cm,minimum height=0.5cm,draw=black,fill=gray!10] (c) at (0,\consensusY+0.25) {Consensus $\PIsmr$};

        \foreach \i/\x in \reconstructList {
            \node [entity] (r2\i) at (\x,\reconstructY) {\tangoicon[height=1.3em]{places_network-server} $r_{\i}$};
        }

        \foreach \i/\x/\angle/\looseness in \userList {
            \node [entity] (u\i) at (\x,\userY) {\tangoicon[height=1.3em]{emblem_person_blue} $u_{\i}$};
        }

        \foreach \i/\x/\angle/\looseness in \userList {
            \node [entity] (o\i) at (\x, \outputY) {\tangoicon[height=1.3em]{mimetypes_text-x-generic}\tangoicon[height=1.3em]{mimetypes_text-x-generic} $\txs$};
        }

        \begin{pgfonlayer}{background}
            \foreach \i/\x in \proposerList {
                \draw [entity_box] ([entity_box_tl]p\i.north west) rectangle ([entity_box_br]r2\i.south east);
            }
        \end{pgfonlayer}

        \draw [arrow] (tx1) -- (p1);
        \draw [arrow] (tx2) -- (p1);
        \draw [arrow] (tx3) -- (p2);
        \draw [arrow] (tx4) -- (p2);
        \draw [arrow] (tx4) -- (p3);
        \draw [arrow] (tx4) -- (p4);
        \draw [arrow] (tx5) -- (p4);

        \foreach \i/\x in \proposerList {
            \foreach \j/\y in \relayList {
                \draw [arrow_shred] (p\i) -- (r1\j);
            }
        }

        \foreach \i/\x in \relayList {
            \draw [arrow_attestation] (r1\i) -- (l);
        }

        \draw [arrow_block] (l) -- (c);
        \foreach \i/\x in \relayList {
            \draw [arrow_block] (c) -- (r2\i);
        }

        \foreach \j/\y in \reconstructList {
            \foreach \i/\x/\angle/\looseness in \userList {
                \draw [arrow_shred] (r2\j) -- (u\i);
            }
        }
        \foreach \i/\x/\angle/\looseness in \userList {
            \draw [arrow_block] (c) to [out=-20,in=90,in looseness=0.75] (2,\reconstructY) to [out=-90,in=\angle,out distance=3em,in distance=3em] (u\i);
        }

        \foreach \i/\x/\angle/\looseness in \userList {
            \draw [arrow] (u\i) -- (o\i);
        }

        \node [phase_label] at (0,\proposerY*0.5+\relayY*0.5) {$\SlotTime{s}-2\Delta$:\\\textsc{Proposer}\\\textsc{phase}};
        \node [phase_label] at (0,\relayY*0.5+\consensusY*0.5) {$\SlotTime{s}-\Delta$:\\\textsc{Relay}\\\textsc{phase}};
        \node [phase_label] at (0,\consensusY*0.5+\reconstructY*0.5) {$\SlotTime{s}$:\\\textsc{Consensus}\\\textsc{phase}};
        \node [phase_label] at (0,\reconstructY*0.5+\userY*0.5) {After consensus:\\\textsc{Reconstruction}\\\textsc{phase}};

        \node [role_label] at (0,\proposerY) {Proposers $\Proposers{s}$};
        \node [role_label] at (0,\relayY) {Relays $\Relays{s}$};
        \node [role_label] at (0,\consensusY) {Leader $\Leader{s}$\\All nodes};
        \node [role_label] at (0,\reconstructY) {Relays $\Relays{s}$};
        \node [role_label] at (0,\userY) {Users};

        \node [role_label] at (0,\proposerY*0.5+\relayY*0.5) {\textcolor{myA16zTeal}{Shreds}};
        \node [role_label] at (0,\relayY*0.6+\consensusY*0.4) {\textcolor{myA16zAquamarine}{Attestations}};
        \node [role_label] at (0,\consensusY*0.4+\reconstructY*0.6) {\textcolor{myA16zMagenta}{Block}};
        \node [role_label] at (0,\reconstructY*0.5+\userY*0.5) {\textcolor{myA16zMagenta}{Block} \& \textcolor{myA16zTeal}{shreds}};

    \end{tikzpicture}
    \caption{Overview of the Multiple Concurrent Proposers (MCP) protocol
    (\cref{alg:mech2-main,alg:mech2-recover,sec:protocol}), showing one slot $s$. 
    The protocol operates in four main phases: 
    (1)~\textsc{Proposer phase}: Proposers $\Proposers{s}$ collect transactions into batches,
    encode the batches into shreds,
    and send 
    the shreds
    to relays $\Relays{s}$ privately.
    (2)~\textsc{Relay phase}: Relays $\Relays{s}$ verify and store shreds, and send attestations (about shreds they have received) to the consensus leader $\Leader{s}$.
    (3)~\textsc{Consensus phase}: The leader $\Leader{s}$ aggregates relay attestations into a block, which is input to the consensus protocol~$\PIsmr$ run by all nodes.
    (4)~\textsc{Reconstruction phase}: After consensus, the relays broadcast stored shreds. All nodes decode available batches from the shreds, and output the transaction batches.
    }
    \label{fig:mcp-overview}
\end{figure}
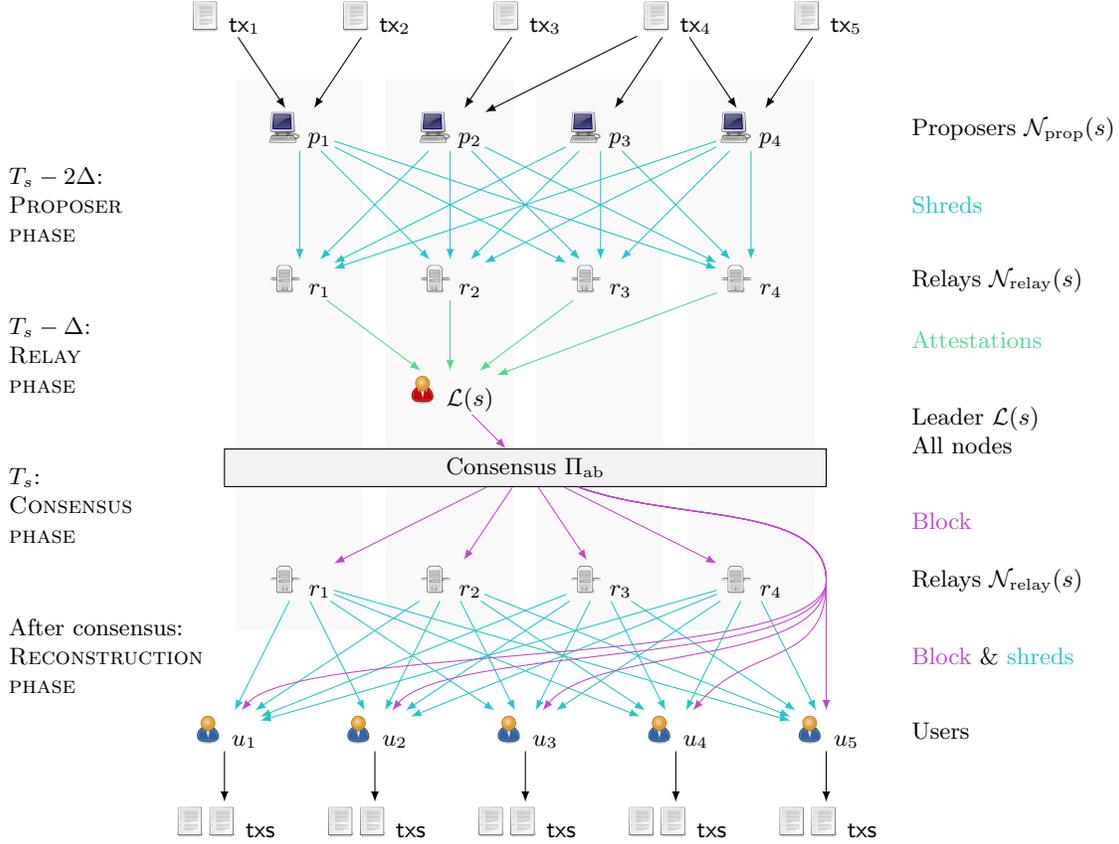

In traditional consensus protocols, a single \emph{leader} per slot collects transactions into a block for proposal as the next consensus decision.
This leaves the leader with outsized power over transaction inclusion, which is detrimental to the economic properties of the system.
Our multiple concurrent proposers (MCP) protocol modifies this approach
(see \cref{fig:mcp-overview}): 
instead of one leader proposing all the transactions for a slot, multiple \emph{proposers} each submit a \emph{batch} of transactions for the same slot.
Then, a committee of \emph{relays} mediates between proposers and the consensus leader. 
Relays receive \emph{shreds}, HECC-encoded pieces of batches, and create \emph{attestations} that certify the timely receipt of shreds and vouch for the \emph{availability} of the batches to the leader.
As a result, the leader can incorporate batches by reference into its block, \emph{even before} learning the batches' contents, and with confidence that once consensus is reached, these batches can be reconstructed from the shreds stored at relays.
In addition, the relays constrain the leader's degrees of freedom in choosing which batches to include in the block, by requiring that the leader include attestations from ``many'' relays to form a valid block, which in turn forces the leader to incorporate the batches of honest proposers.

It is important to note that proposers, relays, and consensus nodes are separate roles and the corresponding nodes could be chosen from different sets. For ease of exposition, however, we describe here the scenario where each of the system-wide set of $n$ are eligible for each role. We also assume that all of the $n$ nodes act as proposers for every slot. In practice, the set of proposers would be subsampled from the set of all nodes. Obviously, then our selective-censorship resistance property applies only to transactions that are submitted to honest proposers.

In more detail, our MCP protocol operates in four distinct phases per slot.
In the \emph{proposer phase}, each proposer collects transactions into a batch, encodes it using HECC into multiple shreds, and distributes a distinct shred to each relay.
During the \emph{relay phase}, relays validate and store received shreds without learning the contents of the underlying batches, then create attestations that certify timely receipt and availability of batches' shreds to the consensus leader. 
The \emph{consensus phase} sees the leader aggregate these attestations into a consensus block. 
Batches are implicitly confirmed if sufficiently many relay attestations attest to the batches' shreds being available.
Then, the underlying consensus protocol runs, and, once the block is confirmed, the \emph{reconstruction phase} begins.
In this phase, relays broadcast their stored shreds so that nodes can reconstruct all the confirmed transaction batches. Nodes then take the union of transactions included in the batches and orders them by a deterministic rule (e.g., by priority fee) to determine the order they are added to those nodes' logs.

\paragraph{Selective-Censorship Resistance}
Since the consensus leader must include attestations from many relays to obtain a valid block, the leader is forced to include attestations from a significant number of honest relays. Attestations from honest relays will reference all honest proposals (under synchrony). We choose the threshold for the maximum amount of attestations the leader can exclude such that we are guaranteed sufficiently many honest relay attestations to be able to confirm all honest proposals (during synchrony). Thus the leader's only way to censor a target batch is to not propose any block at all.

\paragraph{Hiding}
Proposers encode batches using HECC with parameters that ensure that shreds stored at adversarial relays reveal no information about batch contents. Additionally, honest relays only broadcast their shreds after consensus has been reached on transaction batch inclusion.  
As a result, no node (other than the originating proposer) can observe or adapt to transaction contents before batches are confirmed.\footnote{Notably, our protocol does not solve the problem of obscuring transactions from the proposer. The concern regarding a proposer's potential adverse selection against a user after seeing their bid in an auction is mitigated by the repeated nature of the interaction between proposers and users. In a competitive environment with multiple proposers, users may switch proposer if one proposer repeatedly acts against their interests.}

\subsection{Detailed Description}
\label{sec:protocol-description}

\begin{algorithm}[tbp]
    \caption{Basic multiple concurrent proposers (MCP) gadget $\PImcp$ to augment any consensus protocol $\PIsmr$ for combined protocol $\PImcp(\PIsmr)$ to satisfy \cref{def:hiding,def:scr},
    with target resilience $\tau$.
    \textbf{Part~1/2:} \textsc{Proposer}, \textsc{Relay}, \textsc{Consensus} phases.
    See \cref{alg:mech2-recover} for part~2/2: \textsc{Reconstruction} phase.
    Pseudocode described from the perspective of node $p$.}
    \label{alg:mech2-main}
    \begin{algorithmic}[1]
        \LineComment{Parameters:
            coding rate $\gamma\in[0,1]$,
            relay threshold $\mu\in[0,1]$,
            availability threshold $\varphi\in[0,1]$;
            $\heccLong$ uses parameters 
            $\heccN=\NumRelays$, $\heccK=(\gamma-\tau)\NumRelays$, $\heccT=\tau\NumRelays$.
        }%
            \label{loc:init-params}%
            \label{loc:init}
        \LineComment{System-wide setup: 
            every node $p$ has $(\sigSK_p, \sigPK_p) \gets \sigLongGen(1^\kappa)$; 
            all $\sigPK_p$ are common knowledge; 
            subsequently, in the algorithm below, $\sigSK \triangleq \sigSK_p$.
        }%
            \label{loc:init-setup}
        \LineComment{System-wide random lottery: 
            for every slot $s$, 
            sample uniformly 
            \emph{vectors} of
            $\NumProposers$ nodes as proposers $\Proposers{s}$,
            and
            $\NumRelays$ nodes as relays $\Relays{s}$.
            Note that $\Proposers{s}$ and $\Relays{s}$ are \emph{ordered},
            so that for $q\in\Relays{s}$, $\Relays{s}^{-1}[q]$
            is the position of $q$ in $\Relays{s}$.
            Analogously for $\Proposers{s}$.
        }%
            \label{loc:init-lottery}
        \LineComment{Interaction with consensus protocol:
            for every slot $s$,
            one node is sampled independently and uniformly as leader $\Leader{s}$,
            and there is a slot proposal time $\SlotTime{s}$.
        }%
            \label{loc:init-consensus}
            
        \smallskip
        \State $L \gets []$%
            \Comment{Initialize empty annotated output log (default value for uninitialized entries: $\bot$)}
        \State $S \gets \emptyset$%
            \Comment{Dictionary to store shreds received when acting as a relay (default value for uninitialized entries: $\bot$)}
        
        \For{slot $s = 1, 2, ...$}
            \At{$\SlotTime{s} - 2\Delta$}
                \Comment{\textsc{Proposer phase}}
                \label{loc:proposer}
                \If{$p \in \Proposers{s}$}%
                    \Comment{Node $p$ is a proposer for slot $s$}
                    \State Collect pending transactions (from invocations of $\PImcp(\PIsmr).\smrInput(s, \txs)$) into batch $\heccMsg$%
                        \label{loc:proposer-collect}
                    \State $\heccRand \getsRandom \heccF^{\heccT}$%
                        \Comment{Sample randomness for HECC}%
                        \label{loc:proposer-sample}
                    \State $\heccShreds = (\heccShred{1}, ..., \heccShred{\NumRelays}) \gets \heccLongEnc(\heccMsg, \heccRand)$%
                        \label{loc:proposer-encode}
                    \State $\heccMsg' \getsRandom \heccF^{\heccK}$, $\heccRand' \getsRandom \heccF^{\heccT}$%
                        \Comment{Sample randomness for commitment hiding}%
                        \label{loc:proposer-sample-vc}
                    \State $\vcRands = (\vcRand_1, ..., \vcRand_{\NumRelays}) \gets \heccLongEnc(\heccMsg', \heccRand')$%
                        \label{loc:proposer-encode-vc}
                    \State $\vcCom \gets \vcLongCommit(\heccShreds, \vcRands)$%
                        \label{loc:proposer-commit}
                    \For{$r \in \Relays{s}$}
                        \State $\vcWit_r \gets \vcLongOpen(\heccShreds, \vcRands, \Relays{s}^{-1}[r])$%
                            \Comment{Recall, $\Relays{s}^{-1}[q]$ is the position of $q$ in $\Relays{s}$.}%
                            \label{loc:proposer-open}
                        \State $\sigSig_r \gets \sigLongSign(\sigSK, \vcCom)$%
                            \label{loc:proposer-sign}
                        \State Privately send $(\vcCom, \heccShred{\Relays{s}^{-1}[r]}, \vcRand_{\Relays{s}^{-1}[r]}, \vcWit_r, \sigSig_r)$ to relay $r$%
                            \label{loc:proposer-send}
                    \EndFor
                \EndIf
            \EndAt
            
            \At{$\SlotTime{s} - \Delta$}%
                \Comment{\textsc{Relay phase}}
                \label{loc:relay}
                \If{$p \in \Relays{s}$}%
                    \Comment{Node $p$ is a relay for slot $s$}
                    \State $A \gets \emptyset$%
                        \Comment{Initialize empty dictionary}
                    \For{$q \in \Proposers{s}$}
                        \If{$(\vcCom_q, \heccShred{q}, \vcRand_q, \vcWit_q, \sigSig_q)$ privately received from $q$}
                            \State \Assert{$\sigLongVerify(\sigPK_{q}, \vcCom_q, \sigSig_q) \land \vcLongVerify(\vcCom_q, \Relays{s}^{-1}[p], \heccShred{q}, \vcRand_q, \vcWit_q)$}%
                                \label{loc:relay-verify}
                            \State $S[s, q] \gets (\vcCom_q, \heccShred{q}, \vcRand_q, \vcWit_q, \sigSig_q)$%
                                \Comment{Store shred and randomness for proposer $q$ in slot $s$}%
                                \label{loc:relay-store}
                            \State $A[q] \gets (\vcCom_q, \sigSig_q)$%
                                \Comment{Attest to having stored shred for proposer $q$ in slot $s$}%
                                \label{loc:relay-attest}
                        \EndIf
                    \EndFor
                    \State $\sigSig \gets \sigLongSign(\sigSK, A)$%
                        \label{loc:relay-sign}
                    \State Send $(A, \sigSig)$ to consensus leader $\Leader{s}$%
                        \label{loc:relay-send}
                \EndIf
            \EndAt
            
            \At{$\SlotTime{s}$}%
                \Comment{\textsc{Consensus phase}}
                \label{loc:consensus}
                \If{$p = \Leader{s}$}%
                    \Comment{Node $p$ is the consensus leader for slot $s$}
                    \State $B \gets \emptyset$
                        \Comment{Initialize empty dictionary}
                    \For{$r \in \Relays{s}$}
                        \If{$(A_r, \sigSig_r)$ received from $r$}
                            \State \Assert{$\sigLongVerify(\sigPK_r, A_r, \sigSig_r)$}%
                                \label{loc:consensus-verify-relay}
                            \For{$(q \mapsto (\vcCom_q, \sigSig_q)) \in A_r$}
                                \State \Assert{$(q \in \Proposers{s}) \land \sigLongVerify(\sigPK_q, \vcCom_q, \sigSig_q)$}%
                                    \label{loc:consensus-verify-proposer}
                            \EndFor
                            \State $B[r] \gets (A_r, \sigSig_r)$%
                                \label{loc:consensus-include}
                        \EndIf
                    \EndFor
                    \State $\sigSig \gets \sigLongSign(\sigSK, B)$%
                        \label{loc:consensus-sign}
                    \State Invoke $\PIsmr.\smrInput(s, (B, \sigSig))$%
                        \Comment{Input to consensus protocol}%
                        \label{loc:consensus-propose}
                \EndIf
            \EndAt

            \LineComment{See \cref{alg:mech2-recover} for part~2/2: \textsc{Reconstruction} phase.}
        \EndFor
    \end{algorithmic}
\end{algorithm}

\begin{algorithm}[tbp]
    \caption{Basic multiple concurrent proposers (MCP) gadget $\PImcp$ to augment any consensus protocol $\PIsmr$ for combined protocol $\PImcp(\PIsmr)$ to satisfy \cref{def:hiding,def:scr},
    with target resilience $\tau$.
    \textbf{Part~2/2:} \textsc{Reconstruction} phase.
    See \cref{alg:mech2-main} for part~1/2: \textsc{Proposer}, \textsc{Relay}, \textsc{Consensus} phases.
    Pseudocode described from the perspective of node $p$.}
    \label{alg:mech2-recover}
    \begin{algorithmic}[1]
        \LineComment{See \cref{alg:mech2-main} for parameters and initialization.}
        \LineComment{Messages received as part of the \textsc{Reconstruction} phase are automatically re-broadcast to all nodes.}
        
        \For{slot $s = 1, 2, ...$}
            \LineComment{See \cref{alg:mech2-main} for part~1/2: \textsc{Proposer}, \textsc{Relay}, \textsc{Consensus} phases.}

            \Upon{$\PIsmr.\smrOutput(s, \bot)$ triggered and $L[s-1] \neq \bot$}
                \State $L[s] \gets \emptyset$, triggering $\PImcp(\PIsmr).\smrOutput(s, L[s])$%
                    \label{loc:reconstruction-output-empty1}
            \EndUpon

            \Upon{$\PIsmr.\smrOutput(s, (B, \sigSig))$ triggered and $L[s-1] \neq \bot$}%
                \Comment{\textsc{Reconstruction phase}}%
                \label{loc:reconstruction}
                \Try
                    \State \Assert{$\sigLongVerify(\sigPK_{\Leader{s}}, B, \sigSig) \land (|B| \geq \mu \NumRelays)$}%
                        \label{loc:reconstruction-verify-leader}
                    \For{$(r \mapsto (A_r, \sigSig_r)) \in B$}
                        \State \Assert{$(r \in \Relays{s}) \land \sigLongVerify(\sigPK_r, A_r, \sigSig_r)$}%
                            \label{loc:reconstruction-verify-relay}
                        \For{$(q \mapsto (\vcCom_q, \sigSig_q)) \in A_r$}
                            \State \Assert{$(q \in \Proposers{s}) \land \sigLongVerify(\sigPK_q, \vcCom_q, \sigSig_q)$}%
                                \label{loc:reconstruction-verify-proposer}
                        \EndFor
                    \EndFor
                \Catch{assertion failure}
                    \State $L[s] \gets \emptyset$, triggering $\PImcp(\PIsmr).\smrOutput(s, L[s])$%
                        \label{loc:reconstruction-output-empty2}
                    \State \BreakOutOf{\cref{loc:reconstruction} ``upon'' block}
                \EndTry

                \State $P \gets \emptyset$%
                    \Comment{Set of proposers whose batches to reconstruct for $L[s]$}%
                    \label{loc:reconstruction-init-available}
                \For{$q \in \Proposers{s}$}
                    \State $A' \gets \{ (r, \vcCom) \mid \exists A, \sigSig, \sigSig': (B[r] = (A, \sigSig)) \land (A[q] = (\vcCom, \sigSig')) \}$%
                        \Comment{Relay attestations for $q$}%
                        \label{loc:reconstruction-collect-attestations-per-proposer}
                    \If{$(|A'| \geq \varphi \NumRelays) \land \exists \vcCom: \forall (r, \vcCom') \in A': \vcCom' = \vcCom$}%
                        \Comment{No equivocation, enough shreds available}%
                        \label{loc:reconstruction-check-available}
                        \State $P \gets P \cup \{ (q, \vcCom) \}$%
                            \label{loc:reconstruction-mark-available}
                    \EndIf
                \EndFor
                
                \If{$p \in \Relays{s}$}%
                    \Comment{Node $p$ is a relay for slot $s$}
                    \For{$(q, \vcCom) \in P$}
                        \State Broadcast $(s, p, q, S[s, q])$%
                            \Comment{Broadcast shred stored for proposer $q$ in slot $s$ (if any)}%
                            \label{loc:reconstruction-shred-broadcast}
                    \EndFor
                \EndIf

                \State $S_s \gets \emptyset$%
                    \Comment{Initialize empty dictionary}
                \Upon{$(s, p, q, (\vcCom, \heccShred{}, \vcRand, \vcWit, \sigSig))$ received}%
                    \Comment{Collect shreds for proposers whose batches to reconstruct for $L[s]$}
                    \State \Assert{$(q, \vcCom) \in P \land \sigLongVerify(\sigPK_{q}, \vcCom, \sigSig) \land \vcLongVerify(\vcCom, \Relays{s}^{-1}[p], \heccShred{}, \vcRand, \vcWit)$}%
                        \label{loc:reconstruction-shred-verify}
                    \State $S_s[p, q] \gets (\heccShred{}, \vcRand)$%
                        \label{loc:reconstruction-shred-store}
                \EndUpon

                \Upon{$\forall (q, \vcCom) \in P: |\{ p \mid S_s[p, q] \neq \bot \}| \geq \gamma \NumRelays$}%
                    \Comment{Enough shreds collected to reconstruct batches}%
                    \label{loc:reconstruction-threshold}
                    \State $\heccMsg \gets \emptyset$%
                        \Comment{Initialize empty dictionary}
                    \For{$(q, \vcCom) \in P$}%
                            \label{loc:reconstruction-batch}
                        \State $(\heccMsg[q], \heccRand[q]) \gets \heccLongDec(\{ (S_s[p, q][0],\Relays{s}^{-1}[p]) \mid S_s[p, q] \neq \bot \})$%
                            \label{loc:reconstruction-decode}
                        \State $(\heccMsg'[q], \heccRand'[q]) \gets \heccLongDec(\{ (S_s[p, q][1],\Relays{s}^{-1}[p]) \mid S_s[p, q] \neq \bot \})$%
                            \label{loc:reconstruction-decode-vc}
                        \If{$\vcLongCommit(\heccLongEnc(\heccMsg[q], \heccRand[q]), \heccLongEnc(\heccMsg'[q], \heccRand'[q])) \neq \vcCom$}%
                            \label{loc:reconstruction-verify-decode}
                            \State $\heccMsg[q] \gets \bot$%
                                \label{loc:reconstruction-discard}
                        \EndIf
                    \EndFor
                    \State $L[s] \gets \operatorname{orderAndConcat}(\heccMsg)$, triggering $\PImcp(\PIsmr).\smrOutput(s, L[s])$%
                        \Comment{Order transactions according to application logic; output annotated log}%
                        \label{loc:reconstruction-output}
                \EndUpon
            \EndUpon
        \EndFor
    \end{algorithmic}
\end{algorithm}

We now describe the MCP protocol construction in detail.
Pseudo-code is provided in \cref{alg:mech2-main,alg:mech2-recover}. 
The protocol consists of a gadget $\PImcp$ that can be applied to any standard consensus protocol $\PIsmr$ (that satisfies the enriched interface described in \cref{sec:model-consensus-basics}) so that the composite $\PImcp(\PIsmr)$ satisfies the economic properties described in \cref{sec:model-consensus-econ}.
After setup, the protocol iterates through slots,
each of which consists of four phases, which we detail below.

\subsubsection{Setup (\alglocref{alg:mech2-main}{loc:init})} 
\label{sec:protocol-description-setup}

The protocol construction has three main parameters (\alglocref{alg:mech2-main}{loc:init-params}): 
\emph{coding rate} $\gamma \in [0,1]$, 
\emph{relay threshold} $\mu \in [0,1]$, 
and \emph{availability threshold} $\varphi \in [0,1]$. 
The construction is also parameterized by the target resilience $\tau$ of $\PImcp(\PIsmr)$.
The relay threshold $\mu$ determines the fraction of relays from which an attestation must be included to form a valid consensus block (\alglocref{alg:mech2-recover}{loc:reconstruction-verify-leader}).
This ensures that adversarial consensus leaders cannot get away without including attestations from ``many'' honest relays (and, thereby, about ``many'' honest proposers' batches, when the network is synchronous).
The availability threshold $\varphi$ determines the fraction of relays that must have attested for a proposer's batch for said batch to be considered available and required to be included in the protocol's output log
(\alglocref{alg:mech2-recover}{loc:reconstruction-check-available}).
This ensures that only batches for which it is guaranteed that they can be reconstructed from the relays' shreds are considered confirmed. This threshold is higher than the coding rate to account for the fact that adversarial relays might attest to a batch but never release the corresponding shreds.

All nodes share cryptographic parameters established during setup (\alglocref{alg:mech2-main}{loc:init-setup}). 
Each node $p$ has a key pair $(\sigSK_p, \sigPK_p)$ for digital signatures, where all public keys are common knowledge.
For each slot $s$, the protocol randomly samples (\alglocref{alg:mech2-main}{loc:init-lottery}) uniformly $\NumProposers$ nodes as \emph{proposers} $\Proposers{s}$ and $\NumRelays$ nodes as \emph{relays} $\Relays{s}$. For a given relay $r\in \Relays{s}$, we use $\Relays{s}^{-1}[r]$ to denote the index of relay $r$ for slot $s$. This is relevant, as proposers send specific information to each relay based on that relay's index in a slot. 
The protocol operates over an underlying consensus protocol $\PIsmr$ which follows the enriched interface described in \cref{sec:model-consensus-basics}.
The consensus protocol comes with a \emph{leader} $\Leader{s}$ (which we assume to be sampled independently and uniformly) and a \emph{slot proposal time} $\SlotTime{s}$ for each slot $s$. Given the schedule $\SlotTime{s}$ for $\PIsmr$, we denote the start of slot $s$ for $\PImcp$ by $\Tilde{\SlotTime{s}} = T_s-2\Delta$.

\subsubsection{Proposer Phase (\alglocref{alg:mech2-main}{loc:proposer})}
\label{sec:protocol-description-proposer}

At time $\SlotTime{s} - 2\Delta$, each proposer $p \in \Proposers{s}$ assembles pending transactions into a \emph{batch} $\heccMsg$ (\alglocref{alg:mech2-main}{loc:proposer-collect}).\footnote{For the purposes of this writeup, we ignore block size limits and transaction fees. We assume batches can be arbitrarily large and include all transactions a proposer hears about. See \Cref{sec:discussion-blockspace} for an example of how to address this.} 
The proposer then HECC encodes (using fresh randomness) this batch into~$\NumRelays$ \emph{shreds}~$\heccShreds$, where each shred~$\heccShred{i}$ is intended for relay~$i$ (\alglocref{alg:mech2-main}{loc:proposer-encode}). 
This encoding preserves hiding, since no information can be learned about $\heccMsg$ from no more than $\tau \NumRelays$ shreds, while $\gamma \NumRelays$ shreds allow reconstruction of $\heccMsg$.
To ensure consistency during reconstruction (\alglocref{alg:mech2-recover}{loc:reconstruction-verify-decode}), the proposer computes a vector commitment $\vcCom$ over the entire shred vector $\heccShreds$ using additional HECC-encoded randomness (\alglocref{alg:mech2-main}{loc:proposer-sample-vc,loc:proposer-encode-vc,loc:proposer-commit}). This ensures that the vector commitment remains hiding given the openings known to adversarial relays, but given the openings known to honest relays, both the underlying vector and randomness can be reconstructed. 
For each relay $r \in \Relays{s}$, the proposer generates an opening proof~$\vcWit_r$ demonstrating that shred $\heccShred{\Relays{s}^{-1}[r]}$ appears at index $\Relays{s}^{-1}[r]$ within the committed vector (\alglocref{alg:mech2-main}{loc:proposer-open}) together with the commitment randomness $\vcRand_{\Relays{s}^{-1}[r]}$. The proposer signs the commitment~$\vcCom$ producing the signature $\sigSig_r$ (\alglocref{alg:mech2-main}{loc:proposer-sign}), and sends the shred packet $(\vcCom, \heccShred{\Relays{s}^{-1}[r]}, \vcRand_{\Relays{s}^{-1}[r]}, \vcWit_r, \sigSig_r)$ privately to relay $r$ (\alglocref{alg:mech2-main}{loc:proposer-send}).

\subsubsection{Relay Phase (\alglocref{alg:mech2-main}{loc:relay})}
\label{sec:protocol-description-relay}

At time $\SlotTime{s} - \Delta$, each relay $p \in \Relays{s}$ verifies shreds received from proposers.
Specifically, for each proposer $q$ that transmitted a shred packet $(\vcCom_q, \heccShred{q}, \vcRand_q, \vcWit_q, \sigSig_q)$, the relay verifies both the proposer's signature $\sigSig_q$ on commitment $\vcCom_q$ and the opening proof $\vcWit_q$ demonstrating that $\heccShred{q}$ appears at the expected position $\Relays{s}^{-1}[p]$ in the committed shred vector, associated with randomness $\vcRand_q$ (\alglocref{alg:mech2-main}{loc:relay-verify}). 
This ensures that adversarial nodes cannot forge shred packets,
and enables consistency during reconstruction (\alglocref{alg:mech2-recover}{loc:reconstruction-verify-decode}).
The relay stores (\alglocref{alg:mech2-main}{loc:relay-store}) each verified shred packet in local memory $S[s, q]$ for potential later broadcast (\alglocref{alg:mech2-recover}{loc:reconstruction-shred-broadcast}), 
but, crucially, does not reveal the shred at this point. 
Instead, the relay constructs an \emph{attestation}~$A$, a dictionary mapping $q$ to $(\vcCom_q, \sigSig_q)$ (\alglocref{alg:mech2-main}{loc:relay-attest}). This attestation certifies that the relay received, verified, and stored a shred from proposer $q$, while revealing only the commitment $\vcCom_q$ and the signature $\sigSig_q$, which reveal no information about the individual shred $\heccShred{q}$ due to the hiding property of the vector commitment scheme. The relay signs attestation~$A$ (\alglocref{alg:mech2-main}{loc:relay-sign}), and sends the signed attestation $(A, \sigSig)$ to the consensus leader $\Leader{s}$ of slot $s$ (\alglocref{alg:mech2-main}{loc:relay-send}).

\subsubsection{Consensus Phase (\alglocref{alg:mech2-main}{loc:consensus})}
\label{sec:protocol-description-consensus}

At time $\SlotTime{s}$, the \emph{leader} $\Leader{s}$ for slot $s$ of the consensus protocol $\PIsmr$ collects valid attestations from the relays, and assembles a \emph{block} $B$ for input to $\PIsmr$. 
Specifically, the leader validates each attestation $(A_r, \sigSig_r)$ received from relay $r$ by verifying the relay's signature $\sigSig_r$ on $A_r$ (\alglocref{alg:mech2-main}{loc:consensus-verify-relay}), and by verifying the proposer's signature within each entry $(q \mapsto (\vcCom_q, \sigSig_q)) \in A_r$ (\alglocref{alg:mech2-main}{loc:consensus-verify-proposer}). 
Successfully validated attestations are incorporated into block $B$ as $B[r]$ (\alglocref{alg:mech2-main}{loc:consensus-include}).
The leader signs $B$ (\alglocref{alg:mech2-main}{loc:consensus-sign}), and submits the signed block as input $(B, \sigSig)$ for slot $s$ to the underlying consensus protocol $\PIsmr$ (\alglocref{alg:mech2-main}{loc:consensus-propose}). 

\subsubsection{Reconstruction Phase (\alglocref{alg:mech2-recover}{loc:reconstruction})}
\label{sec:protocol-description-reconstruction}

Upon receiving a signed block $(B, \sigSig)$ from $\PIsmr$ as consensus decision for slot $s$, all nodes initiate reconstruction (\alglocref{alg:mech2-recover}{loc:reconstruction}). 
Nodes first validate $(B, \sigSig)$ by verifying the leader's signature $\sigSig$ and confirming that $B$ contains attestations from at least $\mu \NumRelays$ relays (\alglocref{alg:mech2-recover}{loc:reconstruction-verify-leader})---a threshold that will be tuned in \cref{sec:analysis} appropriately to ensure selective-censorship resistance. 
Each embedded attestation $(A_r, \sigSig_r)$ undergoes signature verification, 
as do the proposers' signatures on $(\vcCom_q, \sigSig_q)$ within each attestation~$A_r$ (\alglocref{alg:mech2-recover}{loc:reconstruction-verify-relay}, \alglocref{alg:mech2-recover}{loc:reconstruction-verify-proposer}).
Note that all checks are deterministic and thus all honest nodes reach the same verdict.
If any check fails, the block is rejected and an empty consensus decision is returned by $\PImcp(\PIsmr)$ for slot $s$ (\alglocref{alg:mech2-recover}{loc:reconstruction-output-empty2}).

Nodes determine which proposers' batches are available for reconstruction and required to be included in the consensus decision of $\PImcp(\PIsmr)$ for slot $s$, by analyzing the relays' attestations. 
For each proposer $q \in \Proposers{s}$, nodes collect all attestations $A_r[q]$  where $A_r$ is relay $r$'s attestation, for all relays $r$ where relay $r$ attested to having received a valid shred for proposer $q$ (\alglocref{alg:mech2-recover}{loc:reconstruction-collect-attestations-per-proposer}). 
The batch of proposer $q$ is deemed \emph{available} 
(\alglocref{alg:mech2-recover}{loc:reconstruction-mark-available})
if at least $\varphi \NumRelays$ relays attested to the same commitment $\vcCom$ and no relay attested to a conflicting commitment (\alglocref{alg:mech2-recover}{loc:reconstruction-check-available}). 
The availability threshold $\varphi$ will be tuned appropriately in \cref{sec:analysis} to ensure that the proposer's batch can be reconstructed from the shreds stored by honest relays for that batch.
Note again that all computations are deterministic, and honest nodes have consensus on $(B, \sigSig)$, and thus all honest nodes reach the same verdict regarding which proposers' batches are available.

Note that only after consensus has been reached on which proposers' batches are to be included in the consensus decision of $\PImcp(\PIsmr)$ for slot $s$, do relays broadcast their stored shred packets $S[s, q]$ for each proposer $q$ whose batch was deemed available (\alglocref{alg:mech2-recover}{loc:reconstruction-shred-broadcast}). 
This order of events preserves hiding, as the adversary learns no information about honest proposers' batches from the shreds stored at the roughly $f\NumRelays/n$ adversarial relays, due to the hiding property of the HECC and of the vector commitment. 
Nodes collect broadcast shred packets, and, upon receiving at least $\gamma \NumRelays$ valid shreds and associated randomness values for proposer $q$ (\alglocref{alg:mech2-recover}{loc:reconstruction-threshold}), 
apply HECC decoding to reconstruct both the original batch $\heccMsg_q$ and its encoding randomness $\heccRand_q$, as well as the commitment randomness (\alglocref{alg:mech2-recover}{loc:reconstruction-decode,loc:reconstruction-decode-vc}). 
Consistency of reconstruction is ensured by re-encoding the recovered batch and commitment randomness, and checking that the resulting vector commitment matches the attested $\vcCom$ (\alglocref{alg:mech2-recover}{loc:reconstruction-verify-decode}). 
If not, which happens exclusively if an adversary proposer does not follow the HECC encoding and vector commitment scheme during the proposer phase, the batch is discarded (\alglocref{alg:mech2-recover}{loc:reconstruction-discard}).
Valid reconstructed batches are added to the annotated consensus output log $L[s]$ of $\PImcp(\PIsmr)$ for slot $s$ according to a specific deterministic transaction ordering rule (\alglocref{alg:mech2-recover}{loc:reconstruction-output}).

\section{Analysis}
\label{sec:analysis}

\subsection{Parameter Choices}
\label{sec:analysis-parameters}

Regardless of our parameter choices for $\gamma, \varphi, \mu$ in \cref{alg:mech2-main}, as long as $\PIsmr$ is safe against $f$ Byzantine nodes, $\PImcp(\PIsmr)$ is also safe against $f$ Byzantine nodes.
However, the other three properties of liveness, selective-censorship resistance, and hiding only hold under specific choices of 
$\mu$ (relay threshold: minimum relay attestations required for a valid consensus block), 
$\varphi$ (availability threshold: minimum attestations to attempt batch reconstruction), 
$\gamma$ (coding rate: fraction of shreds needed for reconstruction),
in relation to the resilience $\tau \triangleq f/n$ of $\PIsmr$,
which also determines $\heccT$ ($\heccLong$ parameter such that at most $\heccT$ shreds reveal no information about batches' contents).
Let $f_{\mathrm{relay}}$ be the number of Byzantine relay nodes in a given slot.
We list the inequalities required so that our subsequent analyses of the security properties of $\PImcp(\PIsmr)$ hold. 

\textbf{Liveness.} 
$f_{\mathrm{relay}} \leq (\varphi-\gamma)\NumRelays$.
$\tau \leq \varphi - \gamma$.
For a liveness violation to occur, it must be that a batch is considered available (received $\varphi\NumRelays$ relay attestations) but less than $\gamma\NumRelays$ relays release shreds for the batch, so that nodes are unable to complete reconstruction.\footnote{See \cref{sec:discussion-livenessgadget} for an additional gadget that allows to recover liveness if such an event occurs.}
The condition ensures that batches deemed available will eventually be reconstructed by all honest nodes. 

\textbf{Selective-Censorship Resistance.} 
$f_{\mathrm{relay}} \leq (\mu-\varphi)\NumRelays$.
$\tau \leq \mu - \varphi$.
For an adversary to selectively censor a transaction, they must control enough relays that do not attest to the corresponding batches in which that transaction is included, so that none of those batches are considered available, even though a valid block is proposed by a potentially Byzantine leader.
The condition ensures that a valid block must contain enough honest relay attestations to deem all batches of honest proposers available.

\textbf{Hiding.}  
$f_{\mathrm{relay}} \leq \heccT$.
$\tau \leq \heccT/\NumRelays$.
Honest relays only release shreds as part of $\PImcp(\PIsmr)$ after a batch is considered available and confirmed by $\PIsmr$.
Thus, the only information an adversary has about transactions included in batches deemed available, is from the shreds received by Byzantine relays.
The condition ensures that from those shreds, an adversary cannot learn any information about batches of honest proposers.

Note that increasing $\mu$, and decreasing $\gamma$ towards $\heccT$, weakly improves the resilience of all three of these properties.
At the same time, $f_{\mathrm{relay}} \leq (1-\mu)\NumRelays$ must hold, otherwise adversarial relays withholding attestations can keep honest consensus leaders from proposing valid blocks.
Furthermore, smaller $\gamma$ means less of $\heccLong$ encoding is used for payload vs.\ randomness, 
causing net transaction throughput to go down.
Furthermore, $\gamma\NumRelays = \heccK+\heccT$, so $\gamma\NumRelays > \heccT$ is required.

Concretely, we instantiate $\PImcp(\PIsmr)$ with parameters (\alglocref{alg:mech2-main}{loc:init-params}) $\mu=4/5$,
$\varphi=3/5$,
and $\gamma = 2/5$, 
for a target resilience of $\tau=1/5$.
We assume
$\NumProposers=n$ and $\NumRelays = \Theta(n)$.
That is, we assume that, in each slot, all nodes are eligible to submit batches of transactions,
and a constant fraction of the nodes are relays sampled uniformly  (\alglocref{alg:mech2-main}{loc:init-lottery}).
Consequently, $\heccLong$ is instantiated with $\heccN = \NumRelays$, $\heccK=(\gamma-\tau)\NumRelays=\NumRelays/5$, and $\heccT=\tau\NumRelays=\NumRelays/5$,  
so that at most $\heccT$ shreds reveal no information about batches' contents, 
and $\heccK+\heccT=2\NumRelays/5=\gamma\NumRelays$ shreds suffice to reconstruct batches (\alglocref{alg:mech2-recover}{loc:reconstruction-decode}).
We also assume that the component protocol $\PIsmr$ is secure (\cref{def:security}) with resilience $1/5$ and satisfies the enriched consensus interface of \cref{sec:model-consensus-basics-enriched} (\alglocref{alg:mech2-main}{loc:init-consensus}) and leader-driven liveness (\cref{def:honleader-liveness}).

\begin{definition}%
    \label{def:wop}
    A sequence of events $\{E_n\}_{n\in\IN}$ occurs \emph{with overwhelming probability} (w.o.p.) iff the probability of the sequence of complementary events $\{\overline{E_n}\}_{n\in\IN}$ is negligible in $n$.
\end{definition}

We start by showing that w.o.p.,\ for every slot $s$, at least $4\NumRelays/5$ of the relays are honest. 

\begin{lemma}
    \label{lem:wop}
    Over execution horizons of $\poly(n)$ slots,
    w.o.p.,\ $\forall s: |\Relays{s} \cap \CPa| < \NumRelays/5$.
\end{lemma}
\begin{proof}[Proof of \cref{lem:wop}]
    Let $X_s$ denote the number of Byzantine relays in slot $s$.
    Since the adversary controls at most $f = \beta n$ nodes where $\beta < 1/5$, and relays are randomly selected, we have $\Expe{X_s} = \beta\NumRelays$.
    For any fixed slot $s$, applying a Chernoff bound yields
    \[
    \Prob{X_s \geq \frac{\NumRelays}{5}} 
    = \Prob{X_s \geq (1+\delta)\Expe{X_s}}
    \leq \exp\left( -\frac{\beta\NumRelays\delta^2}{2+\delta} \right)
    = \exp\left( -\frac{\NumRelays(1-5\beta)^2}{5(1+5\beta)} \right)
    \]
    where $\delta = \frac{1}{5\beta} - 1$ results from $\Expe{X_s} = \beta\NumRelays$ and setting $(1+\delta)\beta\NumRelays = \NumRelays/5$.
    
    Since $\beta < 1/5$ is constant and $\NumRelays = \Theta(n)$, this probability is $\exp(-\Theta(n))$.
    Taking a union bound over $\poly(n)$ slots, the probability that any slot has $\NumRelays/5$ or more Byzantine relays is at most $\poly(n) \cdot \exp(-\Theta(n)) = \negl(n)$.
\end{proof}

For the rest of the analysis, we assume that every slot has at least $4\NumRelays/5$ honest relays, except for the safety of the protocol, which is satisfied even when this assumption is not met.

\subsection{Safety}
\label{sec:analysis-safety}

We start by showing that $\PImcp(\PIsmr)$ is safe.
Intuitively, the safety of the underlying consensus protocol $\PIsmr$ implies that every honest node has the same view of what batches are deemed available and of the vector commitments to the shreds each proposer produced for their batches.
We show that by using these vector commitments, either all honest nodes give up on a particular proposer's batch, or they reconstruct the same batch.
From this we get that all honest nodes must reconstruct the same set of batches for every slot.
Since the output log of transactions for a slot is a deterministic function of the reconstructed batches (\alglocref{alg:mech2-recover}{loc:reconstruction-output}), this implies honest nodes always have consistent logs.

Note that it is possible (albeit under appropriate parameters, very unlikely) for the protocol to subsample a set of relays where more than $\NumRelays/5$ relays are Byzantine.
In this case, the adversarial relays can withhold shreds they attested to receiving, causing honest nodes to be unable to reconstruct the batches deemed available.
In this case, honest nodes are caught waiting for these shreds, and liveness is violated.
However, safety is not harmed by this.
\Cref{sec:discussion-livenessgadget} discusses a technique to resolve consensus stalls in those cases.

\begin{theorem}
    \label{thm:mcp-safety}
    Assuming $\PIsmr$ is secure with resilience $n/5$,\ $\PImcp(\PIsmr)$ is safe (\cref{def:safety}).
\end{theorem}
\begin{proof}[Proof sketch of \cref{thm:mcp-safety}]
    Fix any slot $s$ and any two honest nodes $p_1,p_2$.
    We show that if $p_1$ eventually assigns $L^{p_1}[s]=m$, then, if $p_2$ assigns $L^{p_2}[s]$, then $L^{p_2}[s]=m$.
    Observe from \cref{alg:mech2-recover} that honest nodes assign a slot $L[s]$ only once, and only in strictly increasing order after $L[s-1]$ is assigned.
    Let $B$ be the output of $\PIsmr$ for slot $s$.
    By safety of $\PIsmr$, all honest nodes, especially $p_1$ and $p_2$, receive the same $B$ for slot $s$ from $\PIsmr$. 
    
    If $B = \bot$, or $B$ does not pass verification (\alglocref{alg:mech2-recover}{loc:reconstruction-verify-leader,loc:reconstruction-verify-relay,loc:reconstruction-verify-proposer}), 
    then every honest node $p$ assigns $L^p[s] = \emptyset$ (\alglocref{alg:mech2-recover}{loc:reconstruction-output-empty1,loc:reconstruction-output-empty2}), as desired.
    Thus consider the case that $B$ is a valid block passing verification.
    Let $P$ be the set of proposers whose batches are deemed available by $B$ according to the availability condition (\alglocref{alg:mech2-recover}{loc:reconstruction-check-available}).
    Note that $P$ is a deterministic function of $B$, so all honest nodes have the same view of $P$.
    We show that for any proposer $q$ such that $(q,C)\in P$, regardless of what set of shreds any honest node receives, either all nodes reconstruct the same batch $\heccMsg[q]$, or all nodes fail to verify that the $\heccLong$ encoding of the batch they reconstruct matches the vector commitment found in $B$, and hence  set $\heccMsg[q]=\bot$.

    Consider a pair $(q,C)\in P$ 
    in \alglocref{alg:mech2-recover}{loc:reconstruction-batch},
    and the two honest nodes $p_1$ and $p_2$.
    Note that this implies that 
    both $p_1$ and $p_2$ each have received $\gamma\NumRelays$ valid shreds for the proposer $q$ from the relays
    (\alglocref{alg:mech2-recover}{loc:reconstruction-threshold}).
    Note that for every shred packet $(s,p,q,(\vcCom,\heccShred{},\vcRand,\vcWit,\sigSig))$ received by an honest node
    (\alglocref{alg:mech2-recover}{loc:reconstruction-shred-verify}), 
    the node checks $\vcWit$ is a valid opening of $\vcCom$ to $(\heccShred{},\vcRand)$ at index $\Relays{s}^{-1}[p]$.
    Let $\CI_1$ and $\CI_2$ denote the sets of relays from which $p_1$ and $p_2$, respectively, received valid shreds about the proposal of $q$.
    Let $S_1$ and $S_2$ denote the dictionaries $p_1$ and $p_2$ store for these shreds for slot $s$, and let $(\heccMsg_1[q], \heccRand_1[q],\heccMsg_1'[q], \heccRand_1'[q]),(\heccMsg_2[q], \heccRand_2[q],\heccMsg_2'[q], \heccRand_2'[q])$ denote the messages $p_1$ and $p_2$ decode, respectively, from $S_1$ and $S_2$ as defined in \alglocref{alg:mech2-recover}{loc:reconstruction-decode,loc:reconstruction-decode-vc}.

    Let
    \[
    \tilde\vcVals_k = (\tilde\vcVal_{k,1}, ..., \tilde\vcVal_{k,\heccN}) \triangleq \heccLongEnc(\heccMsg_k[q], \heccRand_k[q])
    \]
    and
    \[
    \tilde\vcRands_k = (\tilde\vcRand_{k,1}, ..., \tilde\vcRand_{k,\heccN}) \triangleq \heccLongEnc(\heccMsg_k'[q], \heccRand_k'[q])
    \]
    for $k \in \{1,2\}$.
    We first show that
    $\vcLongCommit(\tilde\vcVals_1, \tilde\vcRands_1) = \vcCom$
    iff
    $\vcLongCommit(\tilde\vcVals_2, \tilde\vcRands_2) = \vcCom$
    in \alglocref{alg:mech2-recover}{loc:reconstruction-verify-decode}.
    Without loss of generality, assume
    $\vcLongCommit(\tilde\vcVals_1, \tilde\vcRands_1) = \vcCom$.
    By the position binding property of
    $\vcLong$ (\cref{def:vc}),
    since $\vcLongCommit(\tilde\vcVals_1, \tilde\vcRands_1) = \vcCom$, and since $p_2$ has verified the openings of the shreds it received,
    it must be the case that
    for all $p \in \CI_2$,
    $S_2[p,q][0] = \tilde\vcVal_{1,\Relays{s}^{-1}[p]}$ and
    $S_2[p,q][1] = \tilde\vcRand_{1,\Relays{s}^{-1}[p]}$.
    Then, by the erasure-correction property of
    $\heccLong$ (\cref{def:hecc-properties}),
    it must be the case that
    \[
    (\heccMsg_1[q], \heccRand_1[q],\heccMsg_1'[q], \heccRand_1'[q]) = (\heccMsg_2[q], \heccRand_2[q],\heccMsg_2'[q], \heccRand_2'[q]).
    \]
    Since $\heccLongEnc$ is deterministic, this implies that
    $\tilde\vcVals_1 = \tilde\vcVals_2$ and
    $\tilde\vcRands_1 = \tilde\vcRands_2$,
    and since $\vcLongCommit$ is deterministic, this implies that
    $\vcLongCommit(\tilde\vcVals_1, \tilde\vcRands_1) = \vcLongCommit(\tilde\vcVals_2, \tilde\vcRands_2) = \vcCom$, as desired.

    It is clear that $p_1$ overwrites $\heccMsg[q] \gets \bot$ in \alglocref{alg:mech2-recover}{loc:reconstruction-discard} iff
    $p_2$ does the same.
    It remains to show that if neither $p_1$ nor $p_2$ overwrite $\heccMsg[q] \gets \bot$ in \alglocref{alg:mech2-recover}{loc:reconstruction-discard}, then
    $\heccMsg_1[q] = \heccMsg_2[q]$.
    This follows from observing that under those circumstances,
    $\vcLongCommit(\tilde\vcVals_1, \tilde\vcRands_1) = \vcCom = \vcLongCommit(\tilde\vcVals_2, \tilde\vcRands_2)$,
    so that by the binding property of $\vcLong$ (\cref{def:vc}),
    $\tilde\vcVals_1 = \tilde\vcVals_2$ and
    $\tilde\vcRands_1 = \tilde\vcRands_2$,
    and thus by the erasure-correction property of
    $\heccLong$ (\cref{def:hecc-properties}),
    $\heccMsg_1[q] = \heccMsg_2[q]$, as desired.

    Since every honest node waits to reconstruct the batches from every proposer in $P$ before assigning anything to $L[s]$ in \alglocref{alg:mech2-recover}{loc:reconstruction-output}, and for every proposer the corresponding batch is reconstructed identically across honest nodes, and the resulting transactions from the reconstructed batches are ordered deterministically, we have that any two honest nodes that assign anything to $L[s]$ in \alglocref{alg:mech2-recover}{loc:reconstruction-output}, must assign the same contents.
\end{proof}

\subsection{Selective-Censorship Resistance}
\label{sec:analysis-censorship}

Before showing that $\PImcp(\PIsmr)$ is live, we first show that $\PImcp(\PIsmr)$ is selective-censorship resistant.
Liveness then follows from $\PImcp(\PIsmr)$ being selective-censorship resistant and $\PIsmr$ additionally satisfying leader-driven liveness.
Intuitively, there are two ways for an adversary to potentially selectively censor a transaction.
The adversary can control enough relays and have them claim not to have seen a batch from a certain proposer; or a Byzantine consensus leader can purposefully leave out relay attestations corresponding to a target batch.
We show that given the adversary controls fewer than $\NumRelays/5$ relays, neither attack (nor any other attack) works.

As a result, an adversary's only choice, if they wish to censor any target transactions, is to control the consensus leader for that slot \emph{and} either assemble an invalid block or not assemble any block at all.
While this gives an adversary \emph{some} ability to censor, an adversary cannot include any of their own transactions either, and everyone has the ability to resubmit fresh transactions in the next slot.
In other words, while the adversary can censor \emph{indiscriminately}, it cannot do so \emph{selectively}, which is precisely what \cref{def:scr} asserts.

\begin{theorem}
    \label{thm:mcp-censorship}
    Assuming $\PIsmr$ is secure with resilience $n/5$, 
    w.o.p.,\ $\PImcp(\PIsmr)$ is selective-censorship resistant (\cref{def:scr}).
\end{theorem}
\begin{proof}[Proof sketch of \cref{thm:mcp-censorship}]
    Following \cref{lem:wop}, we consider only executions where 
    $\forall s: f_{\mathrm{relay}} \triangleq |\Relays{s} \cap \CPa| < \NumRelays/5$,
    and, following \cref{thm:mcp-safety}, 
    we may assume $\PImcp(\PIsmr)$ is safe.
    Consider a slot $s$ for which $\Tilde{\SlotTime{s}} \geq \GST$.
    Recall $\Tilde{\SlotTime{s}} = \SlotTime{s}-2\Delta$ where $\SlotTime{s}$ refers to the time when slot $s$ starts under $\PIsmr$.
    Let a transaction $\tx$ arrive at an honest node $p$ before time $\Tilde{\SlotTime{s}}$. We show that then either $\tx\in L^p_\infty[s]$ or $L^p_\infty[s] = \emptyset$ (and so, by safety of $\PImcp(\PIsmr)$, either $\tx \in \TheLOG[s]$ or  $\TheLOG[s] = \emptyset$, as required).

    If $\tx$ has not yet been included in any confirmed slot, then honest node $p$ includes $\tx$ in its batch $\heccMsg$ for slot $s$ (\alglocref{alg:mech2-main}{loc:proposer-collect}).
    Then $p$ samples randomness $\heccRand$ (\alglocref{alg:mech2-main}{loc:proposer-sample}), encodes the batch into shreds $\heccShreds = \heccLongEnc(\heccMsg, \heccRand)$ (\alglocref{alg:mech2-main}{loc:proposer-encode}), and distributes these shreds with vector commitments and opening proofs to the relays (\alglocref{alg:mech2-main}{loc:proposer-send}).
    Since honest nodes broadcast their shreds at $\Tilde{\SlotTime{s}}$, 
    and $\Tilde{\SlotTime{s}} \geq \GST$,
    every honest relay $r$ will receive their shred and corresponding vector commitment randomness along with valid opening proofs by $\SlotTime{s}-\Delta$.
    It follows that every honest relay attests to the proposal of $p$ being available (\alglocref{alg:mech2-main}{loc:relay-attest}) and gets their attestation to the consensus leader $\Leader{s}$ by $\SlotTime{s}$ (\alglocref{alg:mech2-main}{loc:relay-send}).
    
    By the liveness of $\PIsmr$, eventually, all nodes receive either $\PIsmr.\smrOutput(s,\bot)$ or $\PIsmr.\smrOutput(s, (B, \sigSig))$.
    In the former case, every honest node $p$ outputs $L^p_\infty[s]=\emptyset$ (\alglocref{alg:mech2-recover}{loc:reconstruction-output-empty1}).
    In the latter case, if $B$ is an invalid block, all honest nodes also output $L^p_\infty[s]=\emptyset$ (\alglocref{alg:mech2-recover}{loc:reconstruction-output-empty2}).
    Otherwise, if $B$ is a valid block, it must contain at least $\mu \NumRelays$ unique relay attestations (\alglocref{alg:mech2-recover}{loc:reconstruction-verify-leader}).
    Since at least $(\mu - 1/5)\NumRelays \geq \varphi\NumRelays$ of these attestations are from honest relays, we get that the proposal of $p$ must be deemed available by $B$ (\alglocref{alg:mech2-recover}{loc:reconstruction-check-available}).
    By the safety of $\PIsmr$, all honest relays have the same view of $B$ 
    and thus of $P$ (\alglocref{alg:mech2-recover}{loc:reconstruction-mark-available}),
    and hence will all release their shreds for the proposal of $p$ to all other nodes (\alglocref{alg:mech2-recover}{loc:reconstruction-shred-broadcast}).
    Thus, all honest nodes receive at least $\varphi\NumRelays \geq \gamma\NumRelays$ 
    shreds for the proposal of $p$ (\alglocref{alg:mech2-recover}{loc:reconstruction-threshold}),
    and will successfully reconstruct $\heccMsg$ from decoding the shreds (\alglocref{alg:mech2-recover}{loc:reconstruction-decode}).
    Since the union of all transactions in reconstructed batches for slot $s$ gets added to the output log for slot $s$ (\alglocref{alg:mech2-recover}{loc:reconstruction-output}), we have $\tx\in L^p_\infty[s]$ for every honest node $p$.
\end{proof}

\subsection{Liveness}
\label{sec:analysis-liveness}

The liveness of MCP follows similarly, except we consider a slot $s$ with $\Tilde{\SlotTime{s}} \geq \GST + c\Delta$, where $c$ is the constant of the leader-driven liveness guarantee (\cref{def:honleader-liveness}) of $\PIsmr$, and $\Leader{s} \in \CPh$.

\begin{theorem}
    \label{thm:mcp-liveness}
    Assuming $\PIsmr$ is secure with resilience $n/5$
    and satisfies leader-driven liveness (\cref{def:honleader-liveness}), then, 
    w.o.p.,\ $\PImcp(\PIsmr)$ is live (\cref{def:liveness}) (and even satisfies leader-driven liveness).
\end{theorem}
\begin{proof}[Proof sketch of \cref{thm:mcp-liveness}]
    Following \cref{lem:wop}, we consider only executions where $\forall s: f_{\mathrm{relay}} \triangleq |\Relays{s} \cap \CPa| < \NumRelays/5$.

    We first show that for every $s$, every honest node $p$ eventually completes the reconstruction phase
    (\cref{alg:mech2-recover})
    and assigns $L^p[s]$.
    We proceed by induction.
    The claim is vacuously true for $s=0$.
    Assume $p$ eventually completes the reconstruction phase for slot $k-1$.
    Then, for slot $k$, $p$ is only blocked on receiving $\PIsmr.\smrOutput(k, \bot)$ or $\PIsmr.\smrOutput(k, (B, \sigSig))$ to start reconstruction.
    Since $\PIsmr$ satisfies leader-driven liveness, $p$ eventually witnesses one of these events.
    If $\PIsmr.\smrOutput(k, \bot)$
    (\alglocref{alg:mech2-recover}{loc:reconstruction-output-empty1})
    or 
    $\PIsmr.\smrOutput(k, (B, \sigSig))$
    where $B$ is deemed invalid
    (\alglocref{alg:mech2-recover}{loc:reconstruction-verify-leader,loc:reconstruction-verify-relay,loc:reconstruction-verify-proposer}), 
    then $p$ assigns $L^p[k] = \emptyset$, as desired.

    Otherwise, let $P$ be the set of proposers whose batches are deemed available by $B$,
    according to the availability condition of \alglocref{alg:mech2-recover}{loc:reconstruction-check-available}.
    Note that $P$ is a deterministic function of $B$, so all honest nodes have identical $P$, as shown in the proof of \cref{thm:mcp-safety}.
    Since any $(q,C)\in P$ has $\varphi\NumRelays$ corresponding attestations in $B$, we have that at least $\varphi\NumRelays-f_{\mathrm{relay}}\geq \gamma\NumRelays$ honest relays attested to $(q,C)$ and custody valid shreds for the batch of proposer $q$.
    Thus, each of these honest relays will eventually release their shred, and for every $(q,C)\in P$, every honest node $p$ will receive at least $\gamma\NumRelays$ valid shreds, allowing them to complete the reconstruction phase and assign $L^p[s]$
    (\alglocref{alg:mech2-recover}{loc:reconstruction-output}). 

    Now, consider a transaction $\tx$ submitted to an honest node $p$ at time $t$.
    Then, let $s$ be the earliest slot such that $\Tilde{\SlotTime{s}} \geq t$, $\Tilde{\SlotTime{s}} \geq \GST + c\Delta$, and $\Leader{s}$ is an honest node.
    (To conclude leader-driven liveness of $\PImcp(\PIsmr)$, consider $p=\Leader{s}$.)
    Then, at time $\Tilde{\SlotTime{s}}$, if $\tx\notin L^p_{\Tilde{\SlotTime{s}}}$, node $p$ includes $\tx$ in its batch $\heccMsg$ 
    (\alglocref{alg:mech2-main}{loc:proposer-collect}) 
    and gets her shreds to each relay by time $\SlotTime{s}-\Delta$ (\alglocref{alg:mech2-main}{loc:proposer-send}).
    All honest relays will then attest to the proposal of $p$ being available (\alglocref{alg:mech2-main}{loc:relay-attest}), and get their attestations to $\Leader{s}$ by $\SlotTime{s}$ (\alglocref{alg:mech2-main}{loc:relay-send}).
    Since $\Leader{s}$ is honest, they will include all timely attestations in their block $B$ (\alglocref{alg:mech2-main}{loc:consensus-include}).

    Since $\Tilde{\SlotTime{s}} \geq \GST$, $\Leader{s}$ will hear the attestations from all honest relays in time.
    Since $\NumRelays - f_{\mathrm{relay}} \geq \mu\NumRelays$, $\Leader{s}$ hears enough attestations to input a valid block $B$ to $\PIsmr$ (\alglocref{alg:mech2-main}{loc:consensus-propose}).
    Then, since $\PIsmr$ satisfies leader-driven liveness, we have that, eventually,
    $B$ gets confirmed by $\PIsmr$ with output $\PIsmr.\smrOutput(s, (B, \sigSig))$.
    From here, the reconstruction phase for slot $s$ is guaranteed to terminate, as discussed above, and the proof follows identically to the proof for showing $\PImcp(\PIsmr)$ is selective-censorship resistant (see proof of \cref{thm:mcp-censorship}), 
    in that $B$ deems the proposal of $p$ available, sufficiently many honest relays release shreds for the proposal of $p$ (\alglocref{alg:mech2-recover}{loc:reconstruction-shred-broadcast}), and subsequently all honest nodes reconstruct $\heccMsg$ (\alglocref{alg:mech2-recover}{loc:reconstruction-decode}), and confirm the corresponding transactions in their logs (\alglocref{alg:mech2-recover}{loc:reconstruction-output}).
\end{proof}

\subsection{Hiding}
\label{sec:analysis-hiding} 

Intuitively, since we use a HECC (\cref{def:hecc-syntax}) with parameter $\heccT=\tau\NumRelays$, along with a hiding vector commitment (\cref{def:vc}), for any slot where fewer than $\tau\NumRelays$ relays are Byzantine, via \cref{lem:hiding-merkle-hecc}, we have that the adversary cannot recover any information about what will become the content of $\TheLOG[s]$ until at least one honest relay releases their shreds.
However, since an honest relay only releases its shreds once the block for that slot is confirmed in $\PIsmr$, an adversary only sees the contents of transactions in a slot once that slot's valency of $\PImcp(\PIsmr)$ has already been determined.

\begin{theorem}
    \label{thm:mcp-hiding}
    Assuming $\PIsmr$ is secure with resilience $n/5$
    and satisfies leader-driven liveness,
    and $\heccLong$ is instantiated with $K=T=\NumRelays/5$ and $N=\NumRelays$ as in \cref{sec:analysis-parameters}, w.o.p.,\ $\PImcp(\PIsmr)$ is hiding (\cref{def:hiding}).
\end{theorem}
\begin{proof}[Proof sketch of \cref{thm:mcp-hiding}]
    Following \cref{lem:wop}, we consider only executions where 
    $\forall s: f_{\mathrm{relay}} \triangleq |\Relays{s} \cap \CPa| < \NumRelays/5$.
    Consider the hiding game of \cref{def:hiding}.
    The adversary $\CA$ chooses an honest node $p$ and transactions $\tx_0,\tx_1$.
    The challenger flips a local coin $b \getsRandom \{0,1\}$ uniformly, and inputs $\tx_b$ to $p$.
    Let $t^*$ be the first time such that $\tx_b\in\ValencyLOG{t^*}$, where $\ValencyLOG{t^*}$ denotes the valency of $\PImcp(\PIsmr)$ at $t^*$ (\cref{def:valency}).

    Since $p$ is honest and $\PImcp(\PIsmr)$ is live (by \cref{thm:mcp-liveness}), $\tx_b$ will eventually be confirmed, and thus $t^*$ is well-defined and finite.
    Let $\heccMsg_{p,s}$ denote the batch $p$ submits for slot $s$ (\alglocref{alg:mech2-main}{loc:proposer-collect}).
    Note that $p$ might have to submit batches including $\tx_b$ across multiple slots, either during periods of asynchrony before $\GST$, or in the case that the adversary controls the consensus leaders for these slots and causes them to not propose valid blocks.

    Then, let $s^*$ be the first slot such that $\tx_b \in \heccMsg_{p,s^*}$ and such that honest nodes trigger the output of $\PIsmr.\smrOutput(s^*,(B,\sigma))$,
    where $B$ is a valid block 
    (\alglocref{alg:mech2-recover}{loc:reconstruction-verify-leader,loc:reconstruction-verify-relay,loc:reconstruction-verify-proposer})
    that indicates the batch of proposer $p$ as available
    (\alglocref{alg:mech2-recover}{loc:reconstruction-check-available}).
    Let $t_0$ be the earliest time any honest node triggers the output of $\PIsmr.\smrOutput(s^*,(B,\sigma))$.
    Following the arguments for safety, selective-censorship resistance, and liveness, once $\PIsmr$ confirms $B$, $\tx_b$ position in $\TheLOG$ of $\PImcp(\PIsmr)$ is implicitly confirmed: under the given assumptions, reconstruction of slot $s^*$ is ``deterministic'' in that eventually, slot $s^*-1$ will complete reconstruction, $B$ deems the batch of proposer $p$ as available, sufficiently many relays will release their shreds for $\heccMsg_{p,s^*}$, all honest nodes reconstruct $\heccMsg_{p,s^*}$, and finally all honest nodes include the transactions from $\heccMsg_{p,s^*}$ into their logs for $s^*$.
    It follows that $\tx_b$ is in the valency log of $\PImcp(\PIsmr)$ by time $t_0$, so $t_0 \geq t^*$.

    The earliest time any honest node broadcasts a shred for a batch $\heccMsg_{p,s}$ containing $\tx_b$, is after $\PIsmr$ confirms a block attesting to $\heccMsg_{p,s}$ as available, which is no earlier than $t_0$.
    Thus, since $t_0\geq t^*$, until $t^*$, the only information honest relays expose about the batches $p$ submits, are signatures on the commitments given to them by $p$ (\alglocref{alg:mech2-main}{loc:relay-attest}).
    Since these commitments are already known to the adversary, these signatures reveal no information about the batches, and we can ignore them going forward.
    Crucially, the adversary does not learn any information about $\heccMsg_{p,s}$ from honest shreds until $t_0$, which is after $t^*$.
    Let $\CS$ be the set of slots that elapse between $t^*$ and the time $\tx_b$ is submitted to $p$.
    Note that while the adversary has access to all the information stretching back to the first slot, no slots before the time $\tx_b$ is submitted, contain information to help the adversary determine $b$.

    Hence, up to time $t^*$, for each slot $s$, the only information the adversary learns about $\heccMsg_{p,s}$ is from the information (commitments and openings) submitted to Byzantine relays (\alglocref{alg:mech2-main}{loc:proposer-send}).
    Namely, let $Q_s$ denote the set of adversary controlled relays for slot $s$.
    Then, the adversary's view with respect to $\heccMsg_{p,s}$ consists of  $V_s = \{(\vcCom, \heccShred{\Relays{s}^{-1}[r]}, \vcRand_{\Relays{s}^{-1}[r]}, \vcWit_r, \sigSig_r) \mid r\in Q_s\}$ (\alglocref{alg:mech2-main}{loc:proposer-send}).
    Since $|Q_s| \leq f_{\mathrm{relay}}$, and $\heccLongEnc$ is instantiated with $\heccT = \tau\NumRelays \geq f_{\mathrm{relay}}$, this is exactly the setting of \cref{lem:hiding-merkle-hecc}.
    Thus, by \cref{lem:hiding-merkle-hecc}, we have that from $V_s$, the adversary cannot discern whether $\tx_0$ or $\tx_1$ is in $\heccMsg_{p,s}$.

    For every slot, $p$ generates fresh randomness to use in $\heccLongEnc$ (\alglocref{alg:mech2-main}{loc:proposer-sample}) and $\vcLongCommit$ (\alglocref{alg:mech2-main}{loc:proposer-sample-vc}).
    As a result,
    from the relevant part of the adversary's view, $V_\CS = \{ V_s \mid s \in \CS \}$, the adversary cannot distinguish if the batches $\{ \heccMsg_{p,s} \}_{s\in\CS}$ contain $\tx_0$ or $\tx_1$. 
\end{proof}

\section{Discussion}
\label{sec:discussion}

\subsection{Tradeoffs in Probabilistic Guarantees}
\label{sec:discussion-probguarantees}

The MCP protocol of \cref{sec:protocol} is always safe.
But, liveness, selective-censorship resistance, and hiding are only satisfied with overwhelming probability.
In the parameterization analyzed in \cref{sec:analysis}, all three properties only hold when the adversary controls fewer than $\NumRelays/5$ relays per slot.
Thus, if $\NumRelays$ is set low, an adversary might get lucky and have an outsized number of its nodes sampled as relays for a given slot, and then threaten these properties of the protocol.

However, a protocol might favor some of these properties over others.
A natural choice is for the protocol to preserve liveness against a stronger adversary, to render the probability of a liveness fault for moderately-strong adversaries extremely low, at the cost of a weaker adversary being able to break selective-censorship resistance or hiding occasionally.

As described in \cref{sec:analysis}, a liveness fault requires $f_{\mathrm{relay}} > (\varphi-\gamma)\NumRelays$, selective-censorship resistance can be violated if $f_{\mathrm{relay}} > (\mu-\varphi)\NumRelays$, and hiding can be violated when $f_{\mathrm{relay}} > \heccT$.
Thus, we can bias MCP towards favoring liveness by increasing the availability threshold $\varphi$, and/or decreasing $\heccT$ (corresponding to a lower $\gamma$).
We give one such choice of these parameters tuned to favor liveness, along with the probabilities of each of the properties being violated for a fixed slot.

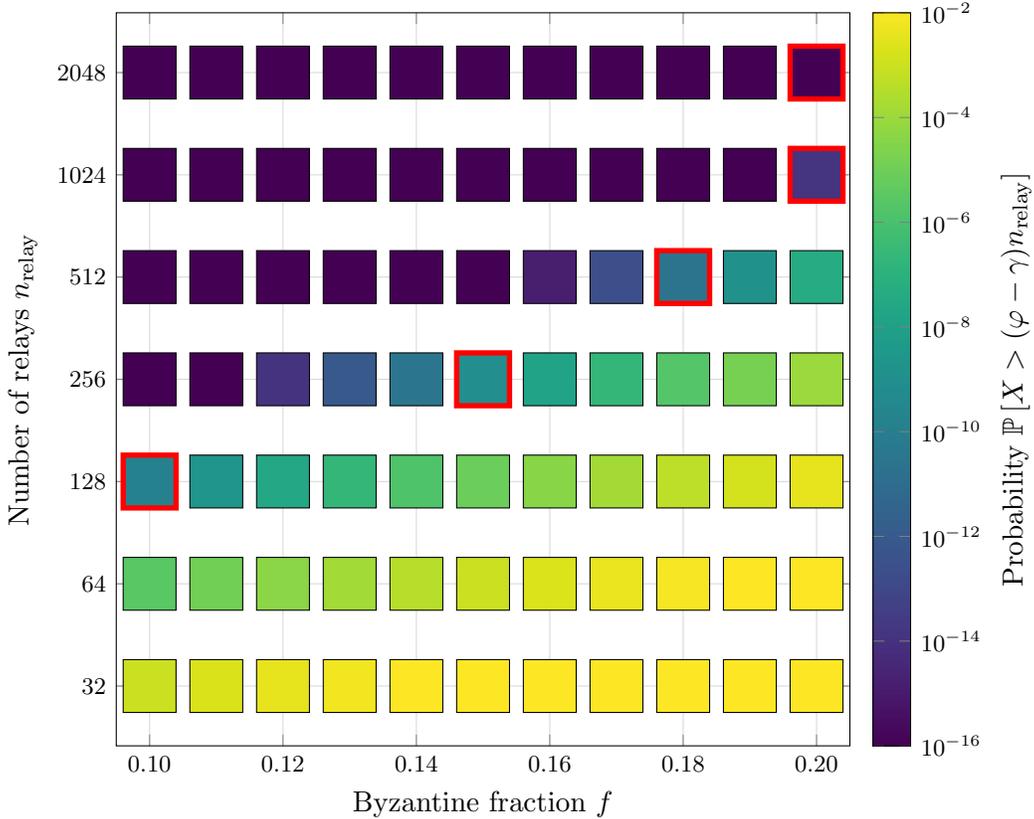
\begin{figure}[tbp]
    \centering
    \begin{tikzpicture}
        \begin{axis}[
            width=0.7\textwidth,
            height=0.7\textwidth,
            xlabel={Byzantine fraction $f$},
            ylabel={Number of relays $\NumRelays$},
            xmin=0.095, xmax=0.205,
            ymin=21.333333333333332, ymax=3072.0,
            xtick={0.10,0.12,0.14,0.16,0.18,0.20},
            xticklabels={0.10,0.12,0.14,0.16,0.18,0.20},
            ytick={32,64,128,256,512,1024,2048},
            yticklabels={32,64,128,256,512,1024,2048},
            ymode=log,
            log basis y=2,
            point meta min=-16,
            point meta max=-2,
            ticklabel style={font=\footnotesize},
            grid=both,
            grid style={gray!30},
            colorbar,
            colormap name=viridis,
            colorbar style={
                ylabel={Probability $\Prob{X > (\varphi - \gamma) \NumRelays}$},
                yticklabel={$10^{\pgfmathprintnumber{\tick}}$}
            },
        ]

        \addplot[
            scatter,
            only marks,
            mark=square*,
            mark size=10pt,
            point meta=explicit,
            scatter/use mapped color={
                draw=none,
                fill=mapped color
            },
        ] table[x=f, y=n, meta expr=log10(\thisrow{prob})] {floats/fig-probability-visualization-3col.csv};

        \addplot[
            scatter,
            only marks,
            mark=square,
            mark size=10pt,
            scatter src=explicit,
            scatter/use mapped color={
                draw=red,
                fill=none,
                line width=2pt
            },
        ] table[x=f, y=n, meta expr=\thisrow{border}, restrict expr to domain={\thisrow{border}}{0.5:1.5}] {floats/fig-probability-visualization-3col.csv};
        
        \end{axis}
    \end{tikzpicture}
    
    \caption{%
        Probability $\Prob{X > (\varphi - \gamma) \NumRelays}$ under different parameter settings for the MCP protocol.
        The heatmap shows how the probability $\Prob{X > (\varphi - \gamma) \NumRelays}$ varies with Byzantine fraction $f$ and number of relays $\NumRelays$,
        for $\varphi = 0.6$ and $\gamma = 0.3$.
        Red rectangles highlight the highest $f$ for a given $\NumRelays$ at which a probability of at most $1:10^9$ is maintained.%
    }
    \label{fig:probability-visualization}
\end{figure}

The number of Byzantine relays sampled in a given slot follows a binomial distribution, $X \sim \operatorname{Binomial}(\NumRelays,f)$.
We can then compute, as a function of $\NumRelays, f, \gamma, \mu, \varphi, \heccT$, the probability of a potential liveness fault, selective-censorship fault, and hiding fault in a given slot.
These probabilities are $\Prob{X > (\varphi - \gamma) \NumRelays}$, $\Prob{X > (\mu - \varphi) \NumRelays}$, and $\Prob{X > \heccT}$, respectively.

For instance, setting $\NumRelays = 512$, $\gamma = 0.3$, $\heccT=\NumRelays \cdot 0.15$, $\heccK=\NumRelays \cdot 0.15$, $\varphi = 0.55$, $\mu=0.8$, and $f = 0.15 \cdot \NumRelays$, we get that both the probability of a potential liveness failure and the probability of a potential selective-censorship failure is roughly $1$ in $10^9$ per slot, which is once every $10$ years given $400$~millisecond slot times.
In this case, since $f=\heccT$, the probability of hiding being broken in every slot is fairly high.
But, we reiterate that $\heccT$ shreds is only the threshold for which it is information-theoretically possible for an adversary to learn \emph{some} information about the batch.
In practice, an adversary would likely need more shreds before they can glean \emph{useful} information about batch contents.
Furthermore, if a proposer desires to hide its transactions against stronger adversaries, it can unilaterally decide to dedicate part of its batch to additional randomness.
Given that the regular operating mode of the protocol is with parameters $\heccK$ and $\heccT$ for encoding batches, a proposer can instead encode their batch with parameters $\heccK',\heccT'$ such that $\heccT' \geq \heccT$ and $\heccK'+\heccT'=\heccK+\heccT$, but a proposer will always be capped by $\gamma$ on the maximum resilience they can get for hiding.

\subsection{Recovering Liveness}
\label{sec:discussion-livenessgadget}

Even in cases where honest nodes are left unable to reconstruct batches deemed available, due to missing/withheld shreds blocking the confirmation of subsequent payloads, the underlying atomic broadcast protocol $\PIsmr$ still makes progress independently (\cref{alg:mech2-main} does not wait for any response from reconstruction before making a new block).
We can use this to construct a liveness recovery gadget, where nodes reach consensus on the ``unavailability of a batch''.
For instance, if nodes are blocked on receiving enough shreds for a given batch, they can submit a vote into $\PIsmr$ indicating a specific batch as being unavailable.
Once a quorum of such votes is gathered, a ``batch-skip certificate'' is produced, certifying it is safe for nodes to skip reconstruction.
Thus, either some honest node will eventually gather enough shreds to reconstruct a batch and forward the contents to all other nodes, or nodes will eventually agree to skip the batch.\footnote{As stated here, such a gadget would make the MCP protocol unsafe in the case where an adversary releases shreds so that some nodes successfully reconstruct a batch (and hence confirm the corresponding transactions) while other nodes vote to skip the batch. This can be addressed, at the cost of additional communication/latency, by having nodes submit additional votes certifying successful reconstruction before confirming batches.}

From here, there are multiple options for how to resume batch reconstruction/confirmation, including:
(a) Full rollback.
All blocks proposed from the failed slot up to the issuance of the skip certificate are not confirmed.
This approach is simple, but may discard batches that could be successfully reconstructed.
(b) Isolated skip.
Only the failed slot is skipped, while subsequent slots that can be reconstructed are preserved.
This could be further refined to operate per-batch rather than per-slot.
A challenge with this approach is 
the bursty bandwidth required to release a backlog of batches deemed available but confirmed only at a later time.

\subsection{Choosing Proposers and Allocating Blockspace}
\label{sec:discussion-blockspace}

A degree of freedom
in implementing MCP is the selection of proposers and the allocation of blockspace among them.
Since execution and networking resources are bounded, individual batches must be capped in size to ensure feasibility of reconstruction and timely confirmation.

An advantage of the MCP construction is that the communication overhead scales with the aggregate blocksize, rather than with the number of proposers.
Specifically, if a batch $u$ is partitioned into $M$ chunks $u_1,...,u_M$, each individually encoded, then a single proposer transmitting $u$ produces the same total number of shreds as $k$ proposers jointly transmitting disjoint sub-batches whose concatenation equals $u$.
Such partitioning is required regardless of the number of proposers, as all transactions for a block are unlikely to fit inside one $\NumRelays$-length codeword.
Thus, the incremental overhead of additional proposers consists only of one extra commitment per proposer, and the associated signature verifications by relays and the leader.
This cost is modest and permits a large number of concurrent proposers.

A natural proposer selection rule is stake-weighted eligibility: every node holding at least a minimum stake fraction (e.g., $0.2\%$) is eligible as a proposer, with blockspace proportional to its stake.
Nodes unwilling to propose themselves, may delegate their blockspace to other participants.
Consequently, the proposer set remains relatively stable, while the barrier to participation remains low enough that clients can reliably route transactions to at least one honest proposer.

\subsection{Localized Relay Optimization}
\label{sec:discussion-localrelay}

The protocol described in \cref{sec:protocol} takes $5$ rounds.
One round can be saved through a localized relay optimization.
Each validator would maintain relays in several locations around the world.
After \alglocref{alg:mech2-main}{loc:relay-send}, the relay privately forwards its shreds to its sibling (relays from the same validator).
Each of the local relays listens for the output of $\PIsmr$, and broadcasts shreds as soon as the broadcast conditions are met (\alglocref{alg:mech2-recover}{loc:reconstruction-shred-broadcast}).

\section*{Acknowledgment}
We thank
Ittai Abraham,
Jeff Bezaire,
Dan Boneh,
Joseph Bonneau,
Yuval Efron,
Quentin Kniep,
Maher Latif,
Kartik Nayak,
Mallesh Pai,
Tom Pointon,
Guru-Vamsi Policharla,
Ling Ren,
Dan Robinson,
Tim Roughgarden,
Jakub Sliwinski,
Alberto Sonnino,
Ertem Nusret Tas,
and
Roger Wattenhofer
for fruitful discussions.
Special thanks to Yuval Efron with whom we have discussed extensions of valency to multi-shot consensus (like \cref{def:valency}) on various occasions.

\bibliographystyle{plainurl}
\bibliography{references}

\appendix

\deferredsection{uninterestingproofs}{Deferred Proofs}

\deferredsection{merkledetails}{Details of Hiding Merkle Tree Vector Commitments}

\clearpage
\tableofcontents

\end{document}